\documentclass[%
 reprint,
 floatfix,
superscriptaddress,
 amsmath,amssymb,
 aps,
]{revtex4-2}

\usepackage{graphicx}
\usepackage{dcolumn}
\usepackage{bm}
\usepackage{hyperref}
\usepackage{color}
\usepackage{url}
\usepackage{subcaption}
\usepackage[labelformat=parens,labelsep=quad,skip=3pt]{caption}
\usepackage[nowatermark]{fixmetodonotes}
\usepackage[margin=1.in]{geometry}
\usepackage{floatrow}

\usepackage{amssymb}
\usepackage{graphicx,amsmath}
\usepackage{enumitem} 
\usepackage{tikz}
\def\checkmark{\tikz\fill[scale=0.4](0,.35) -- (.25,0) -- (1,.7) -- (.25,.15) -- cycle;}
\usepackage{amssymb}
\usepackage{pifont}
\newcommand{\xmark}{\ding{53}}%
\usepackage{cleveref}
\usepackage{floatrow}
\usepackage[printonlyused]{acronym}
\usepackage{rotating}

\usepackage{adjustbox}
\usepackage{isotope}

\begin{document}

\preprint{APS/123-QED}

\title{AGKY Hadronization Model Tuning in GENIE v3}

\author{J\'{u}lia Tena-Vidal}
\email{publications@genie\_mc.org}
\affiliation{University of Liverpool, Dept. of Physics, Liverpool L69 7ZE, UK}
\author{Costas Andreopoulos}
\affiliation{University of Liverpool, Dept. of Physics, Liverpool L69 7ZE, UK}
\affiliation{Science and Technology Facilities Council, Rutherford Appleton Laboratory, Particle Physics Dept., Oxfordshire OX11 0QX, UK}
\author{Christopher Barry}
\affiliation{University of Liverpool, Dept. of Physics, Liverpool L69 7ZE, UK}
\author{Steve Dennis}
\altaffiliation[Now at ]{University of Cambridge}
\affiliation{University of Liverpool, Dept. of Physics, Liverpool L69 7ZE, UK}
\author{Steve Dytman}
\affiliation{University of Pittsburgh, Dept. of Physics and Astronomy, Pittsburgh PA 15260, USA}
\author{Hugh Gallagher}
\affiliation{Tufts University, Dept. of Physics and Astronomy, Medford MA 02155, USA}
\author{Steven Gardiner}
\affiliation{Fermi National Accelerator Laboratory, Batavia, Illinois 60510, USA}
\author{Walter Giele}
\affiliation{Fermi National Accelerator Laboratory, Batavia, Illinois 60510, USA}
\author{Robert Hatcher}
\affiliation{Fermi National Accelerator Laboratory, Batavia, Illinois 60510, USA}
\author{Or Hen}
\affiliation{Massachusetts Institute of Technology, Dept. of Physics, Cambridge, MA 02139, USA}
\author {Igor D. Kakorin}
\affiliation{Joint Institute for Nuclear Research (JINR), Dubna, Moscow region, 141980, Russia}
\author {Konstantin S. Kuzmin}
\affiliation{Joint Institute for Nuclear Research (JINR), Dubna, Moscow region, 141980, Russia}
\affiliation{Alikhanov Institute for Theoretical and Experimental Physics (ITEP) \\
             of NRC ``Kurchatov Institute'', Moscow, 117218, Russia}
\author {Anselmo Meregaglia}
\affiliation{ CENBG, Universit\'e de Bordeaux, CNRS/IN2P3, 33175 Gradignan, France}
\author {Vadim A. Naumov}
\affiliation{Joint Institute for Nuclear Research (JINR), Dubna, Moscow region, 141980, Russia}
\author{Afroditi Papadopoulou}
\affiliation{Massachusetts Institute of Technology, Dept. of Physics, Cambridge, MA 02139, USA}
\author{Marco Roda}
\affiliation{University of Liverpool, Dept. of Physics, Liverpool L69 7ZE, UK   }
\author{Vladyslav Syrotenko}
\affiliation{Tufts University, Dept. of Physics and Astronomy, Medford MA 02155, USA}
\author{Jeremy Wolcott}
\affiliation{Tufts University, Dept. of Physics and Astronomy, Medford MA 02155, USA}

\collaboration{GENIE Collaboration}

\date{\today}
             
\begin{abstract}
The GENIE neutrino Monte Carlo describes neutrino-induced hadronization with an effective model, known as AGKY, which is interfaced with PYTHIA at high invariant mass. Only the low-mass AGKY model parameters were extracted from hadronic shower data from the FNAL 15 ft and BEBC experiments. 
In this paper, the first hadronization tune on averaged charged multiplicity data from deuterium and hydrogen bubble chamber experiments is presented, with a complete estimation of parameter uncertainties.
A partial tune on deuterium data highlights the tensions between hydrogen and deuterium datasets.  

\end{abstract}

\maketitle


\section{Introduction}

The next generation of neutrino oscillation experiments will rely on the precise understanding  of  neutrino interactions at the percent level.
Experiments such as T2K~\cite{T2KObservation}, NOvA~\cite{Nova}, MINERvA~\cite{ALIAGA2014130} and MicroBooNE~\cite{AGUILARAREVALO200928} study neutrino interactions over a broad energy range.
In the few GeV region, $0\pi$ and $1\pi$ contributions dominate the event rate.
Hence, most of the effort has been focused on the theoretical understanding of these interactions ~\cite{NuInteractions6,NuInteractions1,NuInteractions2,NuInteractions4,NuInteractions5} as well as the precise measurement of quasielastic~\cite{1002.2680,0706.0926,Kunxian:2015ymr,PhysRevD.99.012004,PhysRevD.91.112002,uBooNEQE,uBooNECCNp} and pion production cross sections~\cite{1210.4717,1002.2680,0810.3903,MiniBooNEMineral,0805.0186,PhysRevD.95.012010,PhysRevD.98.052002,uBooNECCpi0}.
Pions, before \ac{FSI}, can be produced by either neutrino resonance interactions~\cite{mypaper_1} or hadronization processes.
Hadronization models provide information about the multiplicities and kinematics of the hadrons before \ac{FSI} given the neutrino-nucleon interaction and the event kinematics. 
The knowledge of the exact mixture of hadrons in showers affects the efficiency to distinguish between Neutral-Current (NC) and \ac{CC} events, the event topological characterization~\cite{0706.0926,0805.0186}, impacts the estimation of backgrounds~\cite{PhysRevLett.101.131802} and the calorimetric energy reconstruction. 
\ac{FSI} interaction modeling and detector efficiency corrections are also crucial to avoid confusion in measurements of neutrino-induced hadron production.
Unfortunately, due to the lack of unified models for exclusive hadronic multiparticle production over the energy range of interest for neutrino experiments, one must resort to stitching together different modeling ingredients. 
The GENIE neutrino~\ac{MC} event generator~\cite{Andreopoulos:2009rq} uses the \ac{AGKY} hadronization model~\cite{Yang_2009} whose validity extends down to the inelastic threshold.
At low hadronic invariant mass $W$ the model is based on the \ac{KNO}, while at high-$W$ it is based on the PYTHIA \ac{MC}~\cite{PYTHIAManual}.

\begin{figure}
    \centering
    \begin{subfigure}{\columnwidth}
        \centering\includegraphics[width=\columnwidth]{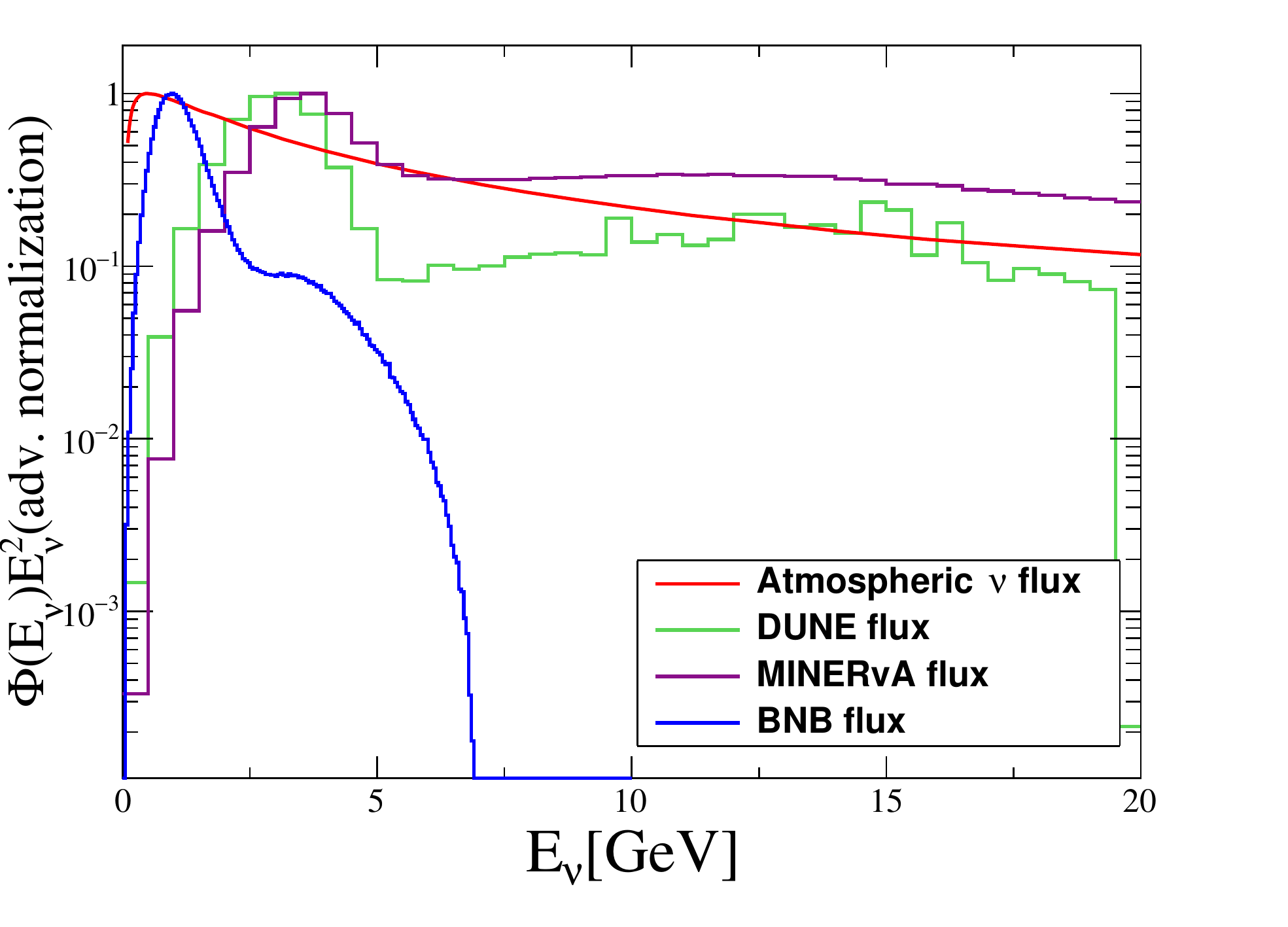}    
    \end{subfigure}
    \begin{subfigure}{\columnwidth}
        \centering\includegraphics[width=\columnwidth]{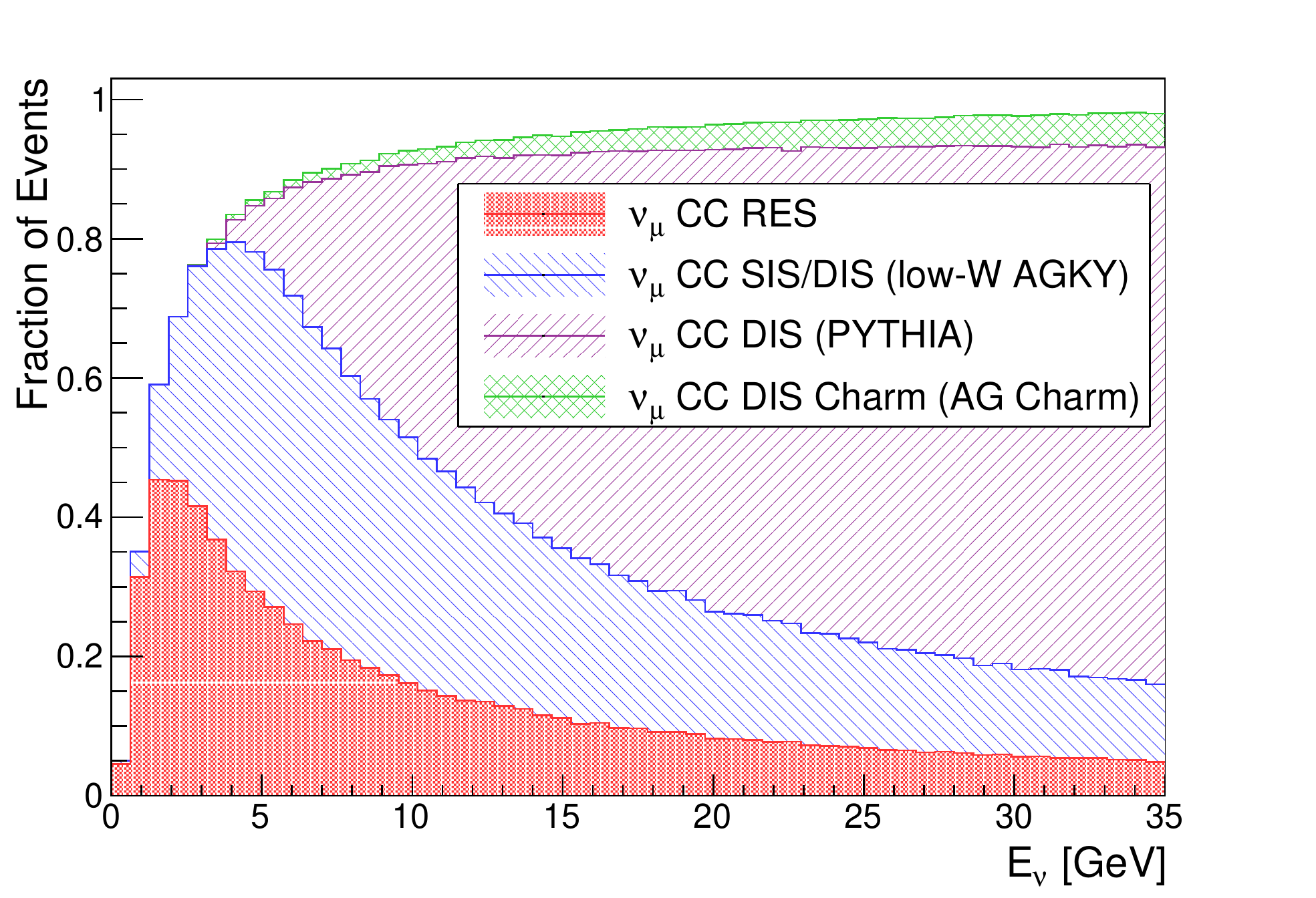}
    \end{subfigure}
    \caption{
    Normalized neutrino fluxes are shown for the atmospheric neutrino flux at Kamioka~\cite{Honda_2015}, DUNE~\cite{DUNEFLUX}, MINERvA~\cite{Valencia_2019} and BNB~\cite{Aguilar_Arevalo_2009} flux predictions (top plot).
    Breakdown of CC events as a function of the neutrino energy from $\nu_{\mu}$ scattering on $^{40}$Ar (bottom plot).
    The plot was obtained with GENIE v3.00.06 using tune \texttt{G18\_02a\_02\_11a}.
    The main components are: resonance (RES), shallow and deep inelastic scattering (SIS/DIS) and deep inelastic charm production (DIS Charm).
    DIS contributions are split according to the hadronization model used: low-$W$ AGKY and PYTHIA. 
    }
\label{fig:fracevents}
\end{figure}

Current and future experiments operate at high energies, where potential biases originating from
hadronization mismodeling become important. 
For instance, DUNE~\cite{dune}, IceCube-Gen2~\cite{ishihara2019icecube,WILLIAMS2020161650} and ORCA~\cite{ORCA} will focus on the $2$ to $20$ GeV energy range where \ac{DIS} events are dominant. 
The neutrino energy dependence on the main inelastic components of the expected event rate for \ac{CC} $\nu_\mu$-\isotope[40]{Ar} scattering is shown in Fig.~\ref{fig:fracevents}. 
Some relevant neutrino fluxes of interest are shown in Fig.~\ref{fig:fracevents} (top).
It is seen that the contribution to the event rate from GeV neutrinos is mainly driven by \ac{CC} RES events as well as \ac{SIS} and \ac{DIS} events from the {low-$W$} \ac{AGKY} model, whereas PYTHIA events dominate at high neutrino energies.

The description of the AGKY hadronization model implementation in GENIE is described in Sec.~\ref{sec:AGKY}.
There is a separate hadronization model to simulate \ac{DIS} charm production, the Aivazis, Olness, and Tung model~\cite{PhysRevD.50.3085}.
Hadronic remnants produced in the interaction are hadronized with PYTHIA.

The \ac{AGKY} model parameters controlling hadronization at low invariant masses were extracted from some of the FNAL 15 ft bubble chamber and the \ac{BEBC} analyses~\cite{PhysRevD.27.47,Barlag1982}.
PYTHIA has never been tuned to low-energy neutrino-induced hadronization data. 
In 2010, GENIE revisited the \ac{AGKY} parameter values and modified a number of PYTHIA parameters using information from the NUX PYTHIA tune~\cite{NUX}, as discussed in Sec.~\ref{sec:AGKY}.
We refer to this parameter set as the \emph{2010} GENIE AGKY tune or \emph{2010} GENIE tune. 
Despite the modifications, several discrepancies between the model and neutrino-induced hadron shower data remained~\cite{TeppeiTUNE,Kuzmin_2013}.

This paper summarises the results of the first tune of the \ac{AGKY} hadronization model against averaged charged multiplicity data on hydrogen and deuterium targets from bubble chamber experiments.
The analysis is performed within the GENIE v3.00.06 global analysis framework \cite{mypaper_1}. 
The base configuration used for all the plots presented here is the \texttt{G18\_02a\_02\_11a}.
This paper is organized as follows: the \ac{AGKY} model specifics relevant for this work are described in Sec.~\ref{sec:AGKY}, followed by an explanation of the analysis procedure applied to the hydrogen and deuterium datasets in Sec.~\ref{sec:dataanalysesBBCH}. Section~\ref{sec:parametrization} discusses the free parameters in the model, and Sec.~\ref{sec:Likelihood} presents the construction of the likelihood function used for fitting. 
The \ac{AGKY} best-fit results are summarised in Sec.~\ref{sec:AGKYTune}.

\section{The AGKY model}
\label{sec:AGKY}

The \ac{AGKY}~\cite{Yang_2009} model is the main hadronization model used in GENIE.
As a function of hadronic invariant mass $W$, three different regimes are defined: an empirical model anchored to bubble chamber data at low-$W$ ($W<W^{\text{tr}}_{\min}$), a pure PYTHIA region for high-$W$ ($W>W^{\text{tr}}_{\max}$) and a transition region that connects them. 
In the transition region, the probability to produce a PYTHIA event increases linearly with $W$, from zero at $W^{\text{tr}}_{\min}$ to 1 at $W^{\text{tr}}_{\max}$.
The values of the transition region limits are $W^{\text{tr}}_{\min}=2.3$~GeV$/c^2$ and $W^{\text{tr}}_{\max}=3.0$~GeV$/c^2$. 
The empirical low-$W$ model and PYTHIA are valid in different mass ranges and they are combined accordingly. 

\begin{figure}
\centering
\includegraphics[width=\columnwidth]{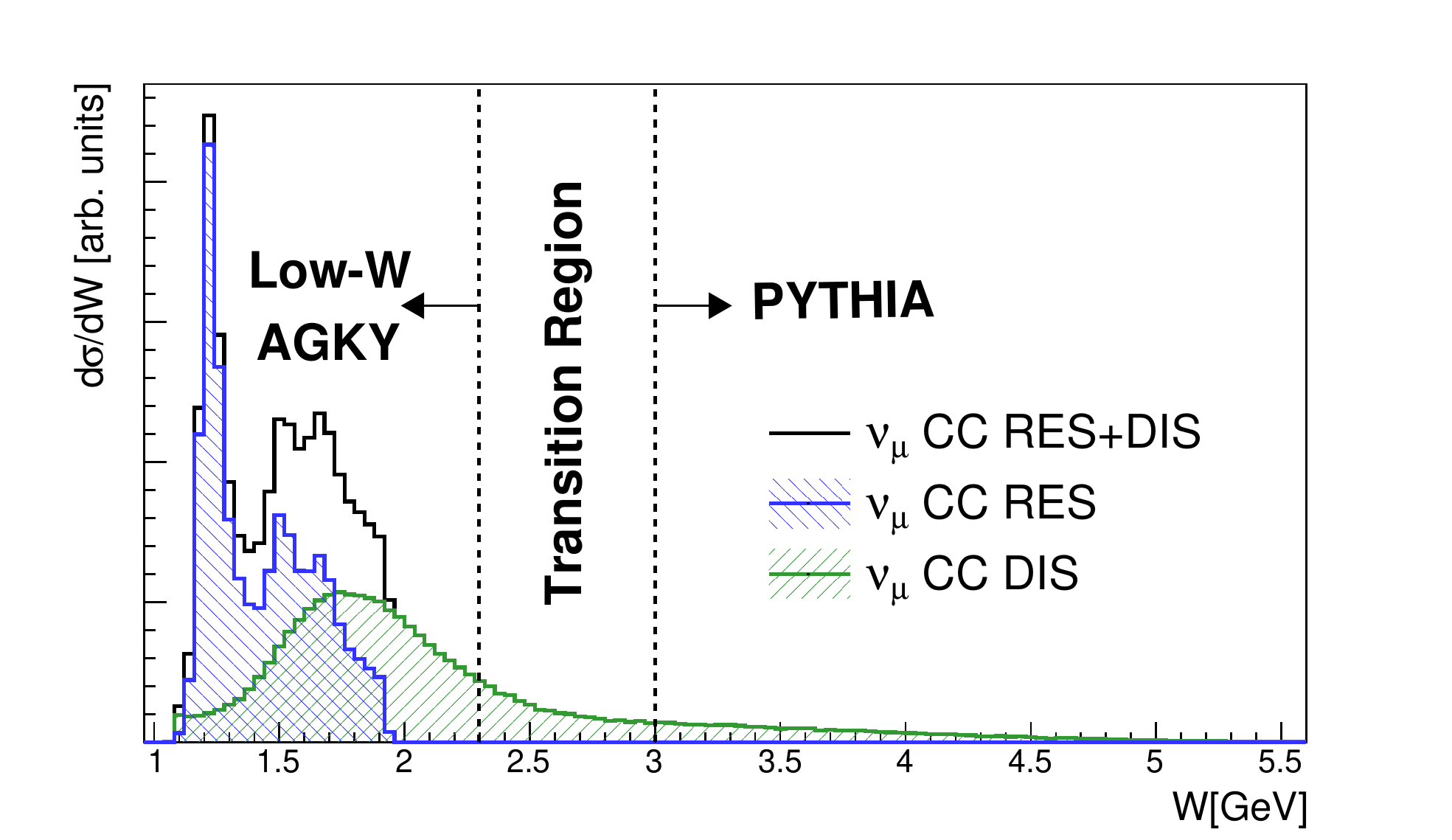}
\caption{Flux integrated CC inelastic differential cross section as a function of the hadronic invariant mass for a DUNE $\nu_\mu$ beam on $^{40}$Ar, obtained with the \texttt{G18\_02a\_02\_11a} tune. 
The distribution is decomposed in RES and DIS contributions.
The DIS contribution to the total number of events is 38\% and 36\% for RES events.
The $\nu_\mu$ flux maximum is between $1$ and $5$ GeV.}
\label{fig:DuneBreakdown}
\end{figure}

The low-$W$ AGKY and PYTHIA algorithms are described in the Sec.~\ref{subsec:lowWAGKY} and Sec.~\ref{subsec:PYTHIA} respectively. 
The contribution of the main inelastic components as a function of $W$ for events generated with the DUNE flux~\cite{DUNEFLUX} is shown in Fig.~\ref{fig:DuneBreakdown}. 
Most of the \ac{DIS}/\ac{SIS} events use the low-$W$ AGKY model while the PYTHIA events are coming from the high-energy tail of the beam.

\subsection{Effective low-$W$ AGKY hadronization model}
\label{subsec:lowWAGKY}

At low-$W$, the showers are made of one baryon and any number of $\pi$ or $K$ consistent with momentum, charge, baryon and strange number, isospin and parity conservation laws:
\begin{eqnarray*}
    \nu_\mu + p & \rightarrow & \mu^- + X^{++}, \label{Eq:nup}\\
    \nu_\mu + n & \rightarrow & \mu^- + X^{+}, \label{Eq:nun}\\
    \bar{\nu}_\mu + p & \rightarrow & \mu^+ + X^{0},  \label{Eq:barnup}\\
    \bar{\nu}_\mu + n & \rightarrow & \mu^+ + X^{-}.  \label{Eq:barnun}
\end{eqnarray*}
For instance, when approaching the pion production threshold, the $\nu_\mu p$ interaction would produce a shower made of a proton and a $\pi^+$. 
In general, the hadron multiplicity at the lowest possible $W$ is 2 as the hadronic final state can only be made of a pion and a nucleon. 

As $W$ increases, more possibilities are available.
The model draws random integer numbers from the simulated hadronic multiplicity distribution to generate the number of particles in the shower, then the particles are labeled so that baryon number, charge, and strangeness are conserved.
The particle content of a shower is selected so that the total mass is not exceeding $W$.
The four-momenta of the hadronic shower particles 
are generated by a weighted phase space decay of a particle of mass $W$ to the selected hadronic-multiparticle state.  
There are many ingredients in the simulation of the hadronic probability distribution: average hadronic multiplicity data, the \ac{KNO} scaling law, particle content rules, phase space weighting and others, as discussed in detail in Ref.~\cite{Yang_2009}.
In this paper we focus on the description of the hadronic multiplicity. 
The hadronic multiplicity probability distribution depends on two ingredients: the measured average as a function of $W$, and an empirical parameterization of multiplicity dispersion. 
Both parameterizations must be extracted from data. 

Empirical observation suggests that the average charged multiplicity is linear with $\ln W^2$:
        \begin{equation}
            \langle n_{\text{ch}} \rangle (W) = \alpha_{\text{ch}} + \beta_{\text{ch}} \ln \left(\frac{W^2}{\text{GeV}^2/c^4}\right).
            \label{Eq:Avnch}
        \end{equation}
The coefficients $\alpha_{\text{ch}}$ and $\beta_{\text{ch}}$ depend on the initial sate and their values can be extracted from neutrino-induced hadronization data, see Sec.~\ref{sec:dataanalysesBBCH}. 
This behaviour has also been proved to be true for heavier nuclear targets~\cite{collaboration2017study,KayisTopaksu:2007pe}. 
From fits to $\pi^0$ production data, it is known that $\langle n_{\text{ch}} \rangle\sim0.5\langle n_{\pi^0}\rangle$~\cite{Wittek1988}.
Therefore, the total hadronic multiplicity is obtained from the charged one as
\begin{equation}
 \langle n\rangle(W) \equiv 1.5\langle n_{\text{ch}} \rangle (W) \,.
 \label{eq:total_multiplicity}
\end{equation}

\begin{figure}
    \centering
    \begin{subfigure}{0.8\columnwidth}
             \includegraphics[width=\linewidth]{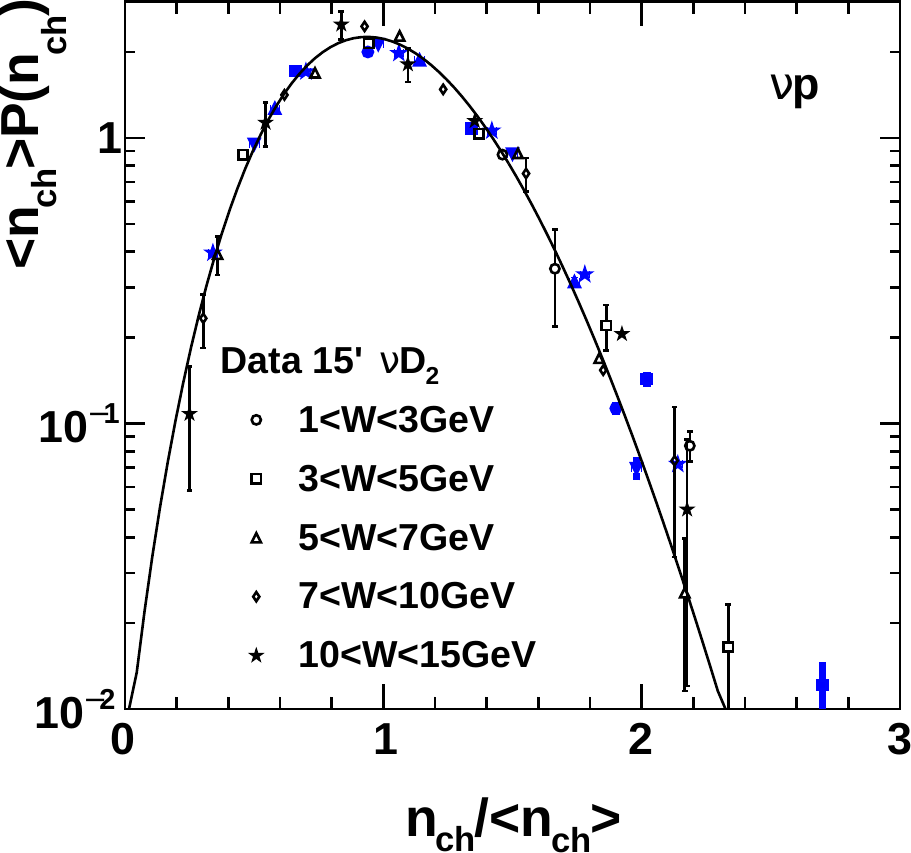}
        \caption{ KNO distribution for $\nu p$ interactions. }   
    \end{subfigure}\\
    \begin{subfigure}{0.8\columnwidth}
        \includegraphics[width=\linewidth]{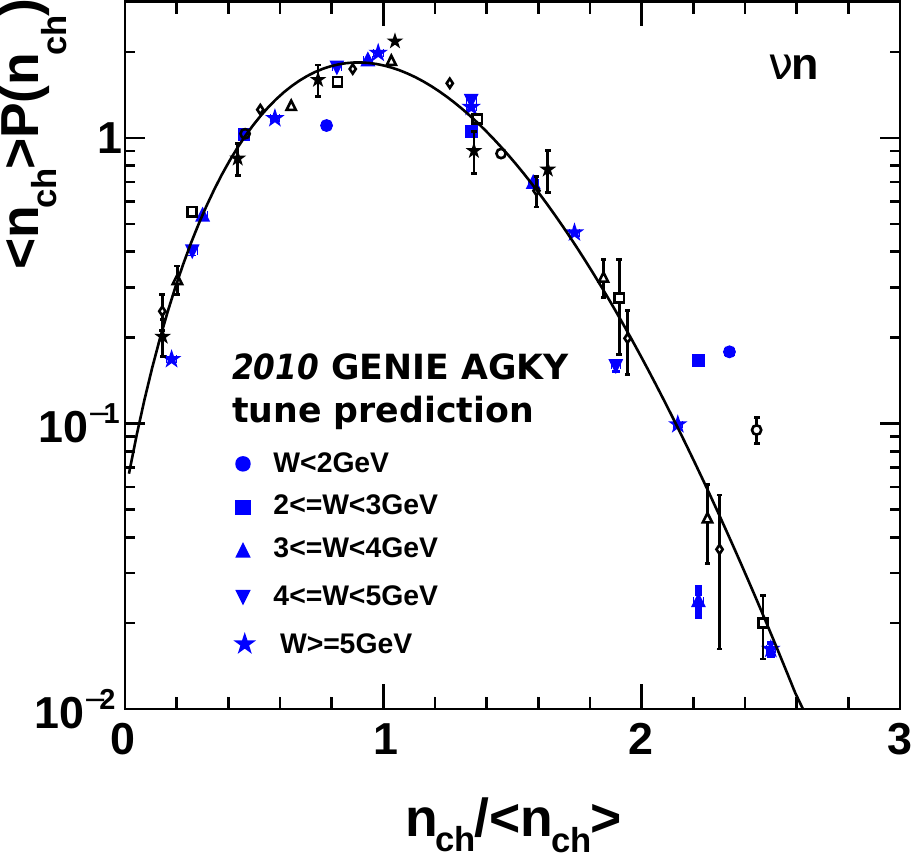}
        \caption{KNO distribution for $\nu n$ interactions.  }   
    \end{subfigure}
\caption{KNO scaling distributions for neutrino interactions on deuterium~\cite{Yang_2009}. The solid line is the best fit result of the Levy function to FNAL 15 ft bubble chamber data~\cite{PhysRevD.27.47}. Blue dots show the \emph{2010} GENIE AGKY prediction for a given $W$ range specified in the legend.}
   \label{fig:KNOData}
\end{figure}

Given the average $\langle n \rangle$, the hadronic multiplicity distribution, $n$, can be obtained from the \ac{KNO} scaling law, which relates the dispersion of hadron multiplicities with a universal scaling function~\cite{KNO}, 
\begin{equation}    
    \langle n \rangle P(n)=f\left(\frac{n}{\langle n \rangle}\right).
    \label{eq:Phadro}
\end{equation}
The scaling function $f(n/\langle n \rangle)$ is parametrized with the Levy function $L(n/\langle n \rangle;\text{c})$
\begin{equation}
    L\left(n/\langle n \rangle;\text{c}\right)=\frac{2e^{-\text{c}}\text{c}^{\text{c}\frac{n}{\langle n \rangle}+1}}{\Gamma\left(\text{c}\frac{n}{\langle n \rangle}+1\right)},
\end{equation}
where $\Gamma$ is the gamma function and $\text{c}$ is the free parameter that has to extracted from data and depends on the interaction isospin. 
By construction, the dispersion of the hadronic multiplicity distribution is independent from the average, see Fig.~\ref{fig:KNOData}.
The \emph{2010} GENIE AGKY values of $\alpha_\text{ch}$, $\beta_\text{ch}$ and c are specified in Tab.~\ref{tab:agkyParameters}.

\begin{table}   
    \centering
    \begin{tabular}{@{\extracolsep\fill} c c c c c c} \hline\hline\noalign{\smallskip}
    Parameter           && $\nu_\mu p$ & $\nu_\mu n$ & $\bar{\nu}_\mu p$ & $\bar{\nu}_\mu n$ \\     \noalign{\smallskip}\hline\hline\noalign{\smallskip}
    $\alpha_\text{ch}$  && 0.40 & -0.20 & 0.02 & 0.80 \\
    $\beta_\text{ch}$   && 1.42 &  1.42 & 1.28 & 0.95 \\
    $\text{c}$          && 7.93 &  5.22 & 5.22 & 7.93 \\   \noalign{\smallskip}\hline\hline
    \end{tabular}
    \caption{ \emph{2010} GENIE tune low-$W$ AGKY parameters. }
    \label{tab:agkyParameters}
\end{table}

\subsection{PYTHIA in GENIE}
\label{subsec:PYTHIA}
The PYTHIA algorithm is well known for its wide use in high-energy collider experiments to simulate the evolution from a few-body hard process to a multi-hadronic final state~\cite{PYTHIAIntroduction,PYTHIAManual}. 
The PYTHIA hadronization model is based on the Lund string fragmentation framework which describes the hadronization process as break-ups in a string throughout production of new $q\bar{q}$ pairs~\cite{LUNDSTRING}. 
Each string represents a color flux which is subject to a linear confined potential.
In the Lund model, the $q\bar{q}$ pairs break by tunneling, which, together with causality, defines the Lund symmetric fragmentation function,
\begin{equation}
    f(z) \propto \frac{(1-z)^a}{z} \exp\left( \frac{-b m^2_\bot}{z} \right)
    \label{eq:fragfunction}
\end{equation}
with the transverse mass of the hadron defined as $m_\bot^2 \equiv m^2+p^2_\bot/c$ and $z$ being quantities that characterise the hadronic shower~\cite{Ferreres-Sole:2018vgo}. 
The transverse momentum is defined as $p_\bot^2=p^2_x+p^2_y$.
$z$ describes the fraction of  available  light cone  momentum $E+p_z$ transferred to the hadrons produced with energy $E$, and it is defined as $z=E/\nu$. 
The parameters $a$ and $b$, known as Lund $a$ and Lund $b$,  are free parameters of the model that are responsible to distribute the longitudinal energy of the hadronic system after the interaction and they should be tuned to reproduce experimental data~\cite{Ferreres-Sole:2018vgo}. 
In terms on the effect on $\langle n_{\text{ch}} \rangle $, as Lund $a$ increases, the multiplicity increases as well, while the opposite is happening for Lund $b$.

In GENIE, PYTHIA is used to simulate the hadronization at high energy invariant masses.
Specifically, GENIE v3.00.06 uses PYTHIA~6.
Future GENIE releases will slowly transition to PYTHIA~8.
In particular, in  v3.00.06, PYTHIA~8 is partially integrated in GENIE and it is fully integrated in the AGKY model.
After the partial integration of PYTHIA~8, simulation outputs remained unchanged. 
Hence, the tune presented in this paper is also valid for PYTHIA~8.
Moreover, different GENIE \ac{CMC}~\cite{mypaper_1} have no impact on the hadronization predictions.

\begin{table*}   
    \centering
    \begin{tabular}{@{\extracolsep\fill} c c c c c c} \hline\hline\noalign{\smallskip}
    Parameter &  Name in PYTHIA\,\, & PYTHIA default\,\, & NUX tune\,\, & HERMES tune\,\, & \emph{2010} GENIE tune\\     \noalign{\smallskip}\hline\hline\noalign{\smallskip}
    $P_{s\bar{s}}$  & \texttt{PARJ(2)} &0.30 & 0.21 &0.25 &  0.30  \\
    $\langle p_\bot^2 \rangle$ [GeV$^2/c^2$]   & \texttt{PARJ(21)} &0.36 & 0.44  & 0.42 & 0.44 \\
    $E_{\text{CutOff}}$ [GeV]  &\texttt{PARJ(33)} &0.80 & 0.20 & 0.47 & 0.20 \\
    Lund $a$ & \texttt{PARJ(41)}& 0.30    & 0.30 & 0.68 & 0.30 \\
    Lund $b$ [$c^4$/GeV$^2$] & \texttt{PARJ(42)}& 0.58 & 0.58  & 0.35 & 0.58 \\    \noalign{\smallskip}\hline\hline
    \end{tabular}
     \caption{{Summary of different PYTHIA parameterizations. The parameter configuration for PYTHIA, NUX, HERMES and \emph{2010} GENIE tunes are specified. The details on the HERMES tune are given in Sec.~\ref{subsec:HERMESTUNE}. }}
    \label{tab:PYTHIAParameters_default}
\end{table*}

\begin{figure}
    \centering
    \includegraphics[width=0.75\columnwidth]{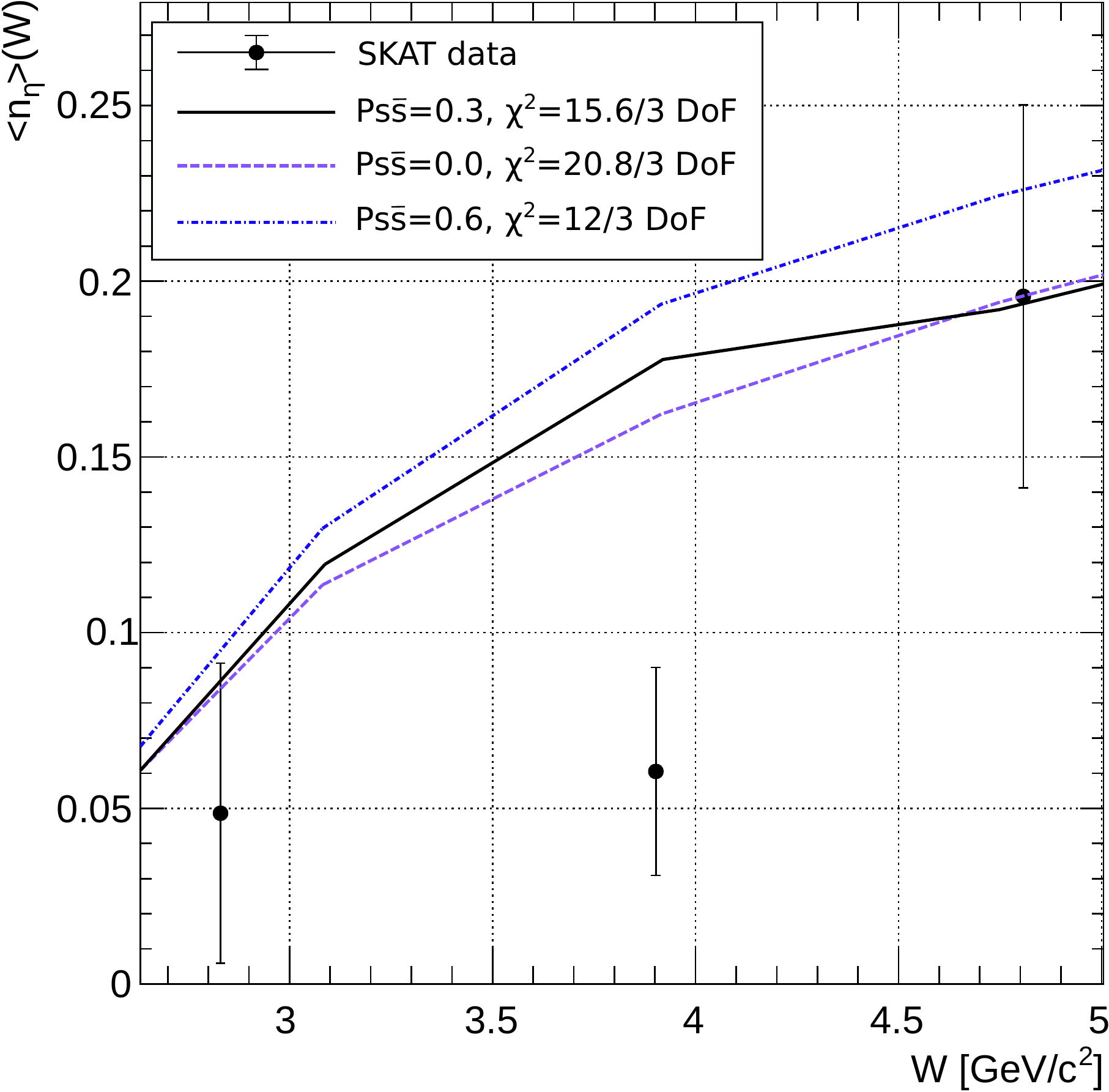}
    \caption{Parameter impact on the averaged $\eta$ production data from SKAT~\cite{Agababyan:2008gg}. }
    \label{fig:ssProd}
\end{figure}

\begin{figure*}
    \centering
        \centering
        \includegraphics[width=0.8\columnwidth]{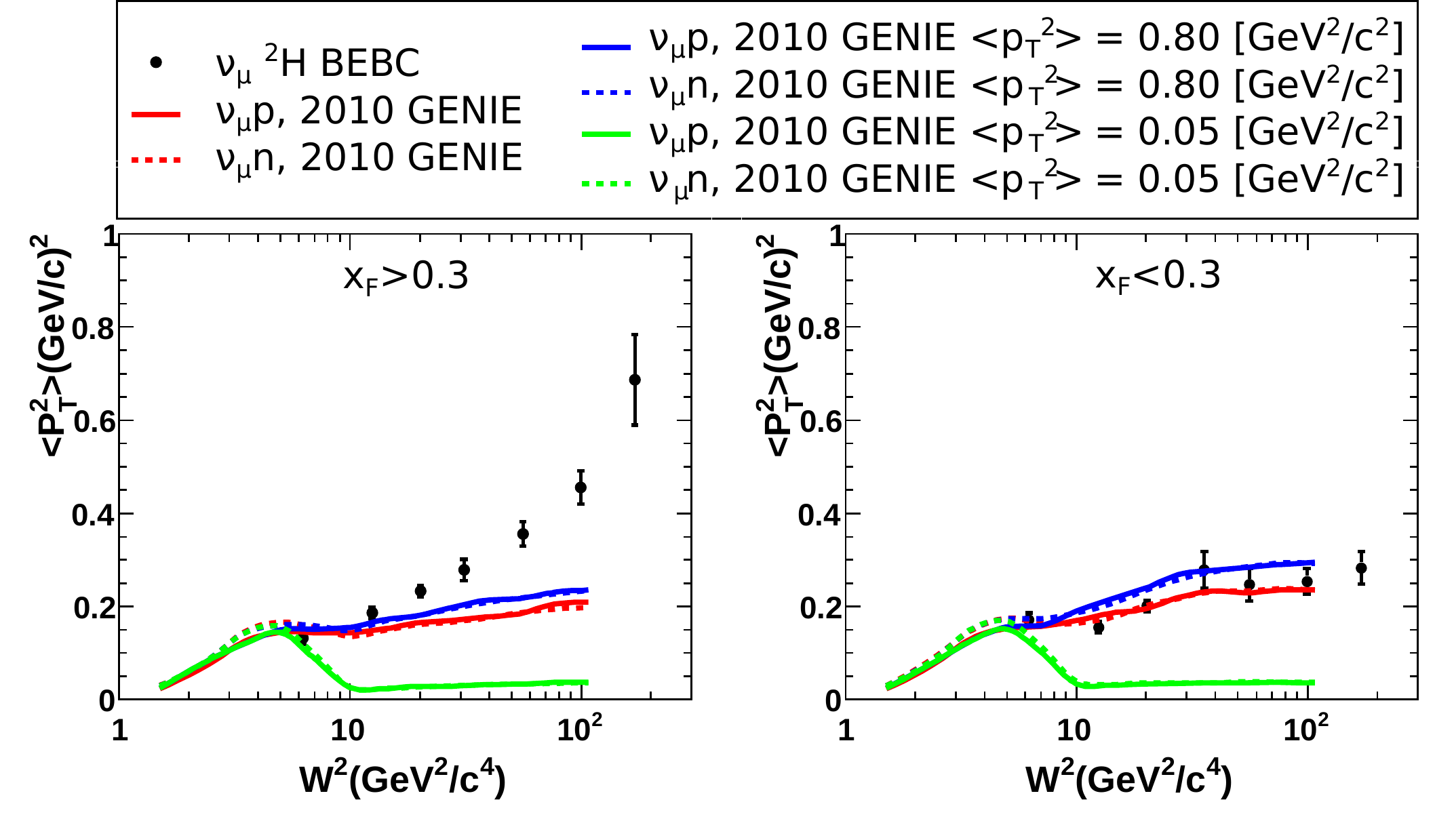}
        \caption{Effect of the $\langle p_{T}^2 \rangle$ parameter on the $\langle p_{T}^2\rangle$ distributions as a function of $W^2$ for $\nu_\mu$ data on $^2$H from the \ac{BEBC} experiment under different Feynman-$x$ ($x_{\text{F}}=p_L/p_{L,\max}$) conditions~\cite{Yang_2009}: $x_{\text{F}}>0.3$ (left) and  $x_{\text{F}}<0.3$ (right). The \emph{2010} GENIE parameter value is ${\langle p_{T}^2 \rangle=0.44~\text{GeV}^2/c^2}$. The validation range used for this plot is specified in the legend.}
        \label{fig:pt2validation}
\end{figure*}

The default PYTHIA parameters shown in Tab.~\ref{tab:PYTHIAParameters_default} come from fits to high energy $e^+ - e^-$ experiments~\cite{RUN1_0, RUN_1,RUN_2,RUN_3,RUN_4,RUN_5,RUN_6,RUN_7,RUN_8,RUN_9,tunes_pythia_list,PYTHIA8_TUNE} (${\sqrt{s} \sim 35~\text{GeV}}$).
PYTHIA's description to data at low energy, such as modern neutrino oscillation experiments ($1-10$~GeV) or even lower energy $e^\pm-p$ experiments such as the HERMES experiment (at $27$ GeV)~\cite{HERMES}, is not accurate, see Sec.~\ref{sec:tune_review}. 
The first attempt to improve this disagreement was in 2010, where some of the PYTHIA parameters were tweaked according to a NUX PYTHIA tune~\cite{NUX}.
The parameters modified by the NUX PYTHIA tune are:
\begin{itemize}
    \item $P_{s\bar{s}}$ controls the $s\bar{s}$ production suppression,
    \item $\langle p_\bot^2 \rangle$ determines the average hadron transverse momentum squared,
    \item $E_{\text{CutOff}}$ is the energy cut-off for the fragmentation process.
\end{itemize}
These parameters are related to important hadron shower characteristics. 
The assumption of tunneling break-ups implies the suppression of heavy-quark production, limiting its production in soft fragmentation processes. 
The suppression factor for heavy quarks is $u\bar{u}$:$d\bar{d}$:$s\bar{s}$:$c\bar{c}\sim1$:1:$0.3$:$10^{-11}$~\cite{Ferreres-Sole:2018vgo}. 
This is supported by $\eta$ production data, Fig.~\ref{fig:ssProd}. 
Previous tunes are in agreement with this fact, see Tab.~\ref{tab:PYTHIAParameters_default}.
Each quark anti-quark pair receive opposite $p_\bot$ kicks at each string breaking point according to a Gaussian distribution.
The $\langle p_\bot^2 \rangle$ parameter controls the variance of the Gaussian distribution used at the breaking point. 
There is different datasets available to constrain this parameter~\cite{Yang_2009}, see for instance Fig.~\ref{fig:pt2validation}.
Finally, $E_{\text{CutOff}}$ determines the minimum energy at which the fragmentation of the parton system can occur, set to $0.8$ GeV in PYTHIA.
\emph{2010} GENIE uses the best-fit-value from the NUX PYTHIA tune, where $E_{\text{CutOff}}=0.20$~GeV.

In GENIE v3 and previous releases, there is only one parameter set configuration for the low-$W$ AGKY model (Tab.~\ref{tab:agkyParameters}) and PYTHIA (Tab.~\ref{tab:PYTHIAParameters_default}) that is common for all \ac{CMC}'s.

\section{Neutrino-induced hadronization data review}
\label{sec:dataanalysesBBCH}
The characterization of the AGKY parameters relies on neutrino-induced hadronization data from \ac{BEBC} and FNAL 15 ft experiments. 
These experiments published a variety of observables related to hadronization. 
This work is based mainly on charged multiplicity data as a function of the hadronic invariant mass, hence, it is what this review is focusing on.
The analyses procedure for both experiments are similar and depend on the target type that can be hydrogen or deuterium.
The different analysis requirements need to be implemented in the GENIE hadronization analyses for a meaningful data/MC comparison, see Sec.~\ref{sec:Likelihood}.
In this section, the analyses of interest for this work are discussed in detail.

\subsection{Hydrogen data}
\label{subsec:dataanalysesBBCH_H}
The bubble chamber at Fermilab (FNAL 15 ft) and \ac{BEBC} at CERN follow similar analyses procedures.
The data considered in this work are those listed in Tab.~\ref{tab:summary_data_H}.

\begin{table*}   
    \centering
    \begin{tabular}{@{\extracolsep\fill}c c c c c c c} \hline\hline\noalign{\smallskip}
    \textbf{Experiment}& $\mathbf{N_{p}}\,\,\,\,$ & 
    \textbf{$\mathbf{W^2}$ [GeV$\mathbf{^2/c^4}$]} & \textbf{Cuts} & 
    \textbf{Syst.}      & \textbf{In Fit} & \textbf{Ref.} \\     \noalign{\smallskip}\hline\hline\noalign{\smallskip}
    \multicolumn{7}{c}{$\nu_\mu + p \rightarrow \mu^- + X^{++}$} \\ 
    \noalign{\smallskip}\hline\hline\noalign{\smallskip}
                        &      &            & $E_\nu^{\text{reco}} \geq 15$ GeV           &             &                           &                           \\
     FNAL 15 ft (1976)  & 25   & [1.5, 150] & $p^{\text{visible}}_L \geq 10$ GeV$/c$      & \emph{Included}    & $W^2<20$ GeV$^2/c^4$      & \cite{PhysRevLett.36.124} \\
                        &      &            & $p^\mu \geq 5$ GeV$/c$                      &             &                           &                           \\
                        &      &            & $p_{T}^{\mu} \geq 1$ GeV$/c$                &             &                           &                           \\
    \noalign{\smallskip}\hline\noalign{\smallskip}
                        &      &            & $p^\mu \geq 3$ GeV$/c$                      &             &                           &                           \\
     BEBC (1983)        & 11   & [9, 121]   & $E^{\text{visible}} \geq 5$ GeV             & $3-5$\%     & \xmark                    & \cite{GRASSLER1983269}    \\
                        &      &            & $W^2 \geq 9$ GeV$^2$/c$^4$                  &             &                           &                           \\
    \noalign{\smallskip}\hline\noalign{\smallskip}
                        &      &            & $Q^2 \geq 1$ (GeV$/c$)$^2$                  &             &                           &                           \\
     BEBC (1990)        &  6   & [6, 150]   & $p^\mu \geq 3$ GeV$/c$                      & \emph{Statistical} & $W^2<9$ GeV$^2/c^4$       & \cite{Jones1990}          \\
                        &      &            & $W^2 \geq 4$ GeV$^2/c^4$                    &             &                           &                           \\
    \noalign{\smallskip}\hline\noalign{\smallskip}
     BEBC (1992)        &  5   & [12, 144]  &  $p^\mu \geq 3$ GeV$/c$                     & \emph{Included}    & \checkmark                & \cite{BEBC1992}           \\
    \noalign{\smallskip}\hline\hline\noalign{\smallskip}
    \multicolumn{7}{c}{$\bar{\nu}_\mu + p \rightarrow \mu^+ + X^{0}$} \\ 
    \noalign{\smallskip}\hline\hline\noalign{\smallskip}
                        &      &            & $p_{\text{ch}} \geq 5$ GeV$/c$              &             &                           &                           \\
                        &      &            & $p^{\text{tot}}_{\text{FW}} \geq 2$ GeV$/c$ &             &                           &                           \\
     FNAL 15 ft (1981)  & 10   & [16, 100]  & $y_{\text{B}}\geq0.1$                                & \emph{Statistical} & $W^2<30~\text{GeV}^2/c^4$ & \cite{PhysRevD.25.624}    \\
                        &      &            & $y_{\text{B}}\leq0.8$                                &             &                           &                           \\
                        &      &            & $E_{\bar{\nu}}^{\text{reco}}\geq5$ GeV      &             &                           &                           \\
    \noalign{\smallskip}\hline\noalign{\smallskip}
                        &      &            & $p^\mu \geq 3$ GeV$/c$                      &             &                           &                           \\
     BEBC (1983)        & 10   & [9, 121]   & $E^{\text{visible}} \geq 5$ GeV             & $3-5$\%     & \xmark                    & \cite{GRASSLER1983269}    \\
                        &      &            & $W^2 \geq 9$ GeV$^2/c^4$                    &             &                           &                           \\
    \noalign{\smallskip}\hline\noalign{\smallskip}
                        &      &            & $Q^2 \geq 0.1$ (GeV/$c$)$^2$                      &             &                           &                           \\
     BEBC (1990)        &  6   & [6, 144]   & $p^\mu \geq 3$ GeV$/c$                      & \emph{Statistical} & $W^2<10~\text{GeV}^2/c^4$ & \cite{Jones1990}          \\
                        &      &            & $W^2 \geq 4$ GeV$^2/c^4$                    &             &                           &                           \\
    \noalign{\smallskip}\hline\noalign{\smallskip}
     BEBC (1992)        &  5   & [12, 144]  & $p^\mu \geq 3$ GeV$/c$                      & \emph{Included}    & $W^2<60~\text{GeV}^2/c^4$ & \cite{BEBC1992}           \\
    \noalign{\smallskip}\hline\hline
    \end{tabular}
     \caption{Compilation of historical data from the \ac{BEBC} and FNAL 15 ft bubble chamber experiments on averaged charged hadron multiplicity in muon (anti)neutrino on hydrogen interactions. Information about the number of points in each dataset, $N_p$, the $W^2$ range covered and the cuts applied in each analysis is provided. Unless specified, the systematic errors were not included in the data release error bands and have been added in quadrature by the amount specified in this table, see details in Sec.~\ref{subsec:systerrors}. The sixth column specifies whether a dataset is included, discarded or partially included in the fit, see Sec.~\ref{subsec:req_data}. 
     The complete list of data points removed in this analysis is specified in the Sec.~\ref{sec:AGKYTune}.}
    \label{tab:summary_data_H}
\end{table*} 

Both experiments look for $\nu_\mu$ and $\bar{\nu}_\mu$ CC interactions on hydrogen to study the averaged charged multiplicity of the final state as a function of the event invariant mass.
The main requirement to select \ac{CC} events is to detect a muon track.
Muons are detected with a \ac{EMI}, and a minimum muon momentum, $p_\mu$, is usually required to guarantee good muon identification (ID). 
This is a consequence of the muon ID efficiency dependence on the muon momentum energy. 
For instance, in \ac{BEBC} experiment, the muon ID efficiency varies from $40\%$ to $100 \%$ in the muon momentum range of $3~\text{GeV}/c \leq p_\mu \leq 10~\text{GeV}/c$, with an average efficiency of 95\%. 
The FNAL 15 ft experiment also uses a kinematic technique to identify negative muons in neutrino interactions~\cite{PhysRevLett.45.1817}. 
Under this $\mu^-$-ID method, only events in which the $\mu^-$ candidate has transverse momentum, $p_{\bot}^{\mu}$, of at least $1$ GeV$/c$ are accepted. 

Selected events, which satisfy the conditions specified above, are analyzed to reconstruct the event topology and kinematics.
In particular, \ac{BEBC} uses the HYDRA program~\cite{GRASSLER1983269,Jones1990,BEBC1992} and FNAL 15 ft a modified version of the TVGP program~\cite{PhysRevD.25.1}.
Only a small fraction of the charged final state hadrons is identified by using energy loss, range in hydrogen, break point probability and kinematic fits~\cite{GRASSLER1983269}. 
If left unidentified, the remaining charged hadrons are assumed to be pions: this assumption can cause migration of particles from the backward to the forward going hemisphere.
For instance, the \ac{BEBC} experiment is able to identify about 30\% of the protons using the HYDRA algorithm, while the rest are classified as pions~\cite{Jones1990}. 

For $\nu_\mu$ CC interactions, because of charge conservation, the experiments scan for events with three or more charged particles in the final state.

The topology of neutrino and anti-neutrino events is expected to be different.
In anti-neutrino events, interactions with only one charged track can occur ($n_{\text{ch}}=0$).
Such events are not negligible at low $E_{\nu}$ and low-$W$.
However, these are removed due to low scanning efficiency and poor anti-neutrino energy reconstruction. 
Both \ac{BEBC} and FNAL 15 ft correct for the effect of removing one-prong contributions in anti-neutrino samples using \ac{MC} calculations~\cite{PhysRevD.25.624,GRASSLER1983269,Jones1990,BEBC1992}.
One-prong \ac{MC} events are weighted so that the fraction of one-prong events agree with the experimental estimate. 
The scanning efficiencies for three prong events are higher than 90\%, improving as the number of charged secondaries increases ($\geq 95\%$). 

In hydrogen and deuterium bubble chambers, the identification of neutral particles, such as $\pi^0$, is difficult due to the low $Z$ of the medium. 
As a consequence, the transverse momentum balance method is used to estimate the neutrino energy by assuming undetected neutral particles in the event~\cite{transversebalancemethod}, 
    \begin{equation}
        E_\nu^{\text{reco}} = p_L^\mu{c}+ p_L^{\text{ch}}{c} \left(1+\frac{|\boldsymbol{p}_\bot^\mu + \boldsymbol{p}_T^{\text{ch}}|}{\sum_{i=1}^{n_{\text{ch}}}|\boldsymbol{p}_{\bot i}|}\right).
        \label{eq:enureco}
    \end{equation}
The subscript $L$ and $\bot$ refer to longitudinal and transverse components of the momenta relative to the neutrino direction, whereas the $ch$ and $\mu$ labels denote the charged-hadron system and the muon respectively.
The index $i$ runs over the charged hadrons in the hadronic system. 
By using this method, there is a non-negligible bias for the neutrino energy reconstruction.
For instance, the \ac{BEBC} experiment estimated the reconstructed neutrino energy to differ from the true energy by $\sim10-15$\%~\cite{PhysRevD.27.47,GRASSLER1983269}. 
Both bubble chambers corrected for this effect, see Sec.~\ref{subsec:systerrors}.
In some analyses, cuts on the reconstructed neutrino energy, $E_\nu^{\text{reco}}$ are applied~\cite{PhysRevLett.36.124,GRASSLER1983269,PhysRevD.27.47}. 

Backgrounds from NC events, quasi-elastic (QEL) CC events or neutral particle induced events are removed from the final sample using kinematic cuts that depend on each analysis.
NC events can mimic CC events as a consequence of muon-hadron miss-ID.
On the one hand, for the FNAL 15 ft experiment, the muon-hadron miss-ID increases at high Bjorken inelasticity values ($y_{\text{B}}$) and a cut on $y_{\text{B}}$ is required to guarantee a good efficiency in selecting CC events~\cite{PhysRevD.25.624}.
In $\bar{\nu}_\mu$ events, backgrounds from low-energy neutrons as well as events caused by incoming hadron tracks that re-scatter within the chamber are controlled by requiring the total momentum in the forward hemisphere, $p_{\text{FW}}^{\text{tot}}$, to be greater than $2$~GeV$/c$~\cite{PhysRevD.25.624,PhysRevD.25.1}.
Moreover, FNAL 15 ft removes backgrounds from $K^0_L$ mesons by requiring the minimum total momentum from charged particles, $p_{\text{ch}}$, to be higher than $5$~GeV$/c$~\cite{PhysRevD.25.624}.
On the other hand, the \ac{BEBC} experiment applies kinematic cuts on either $W$ or/and $Q^2$ to remove QEL events~\cite{GRASSLER1983269,Jones1990,BEBC1992}. 
All cuts applied to the different analyses are shown in Tab.~\ref{tab:summary_data_H}.

\subsection{Deuterium data}
\label{sec:dataanalysesBBCH2H}
The analyses algorithm followed by the FNAL 15 ft and \ac{BEBC} bubble chamber experiments operating with deuterium aims to discriminate between interactions on proton and neutron. 
The data on deuterium considered in this work are those listed in Tab.~\ref{tab:summary_data_2H}.

\begin{table*}   
    \footnotesize
    \centering
    \begin{tabular}{@{\extracolsep\fill} c c c c c c c c} \hline\hline\noalign{\smallskip}
    \textbf{Experiment}& $\mathbf{N_{p}}\,\,\,\,$ & 
    \textbf{$W^2$ [GeV$^2/c^4$]} & \textbf{Cuts} & 
    \textbf{Syst.} & \textbf{In Fit} & \textbf{Ref.} \\    \noalign{\smallskip}\hline\hline\noalign{\smallskip}
    \multicolumn{7}{c}{$\nu_\mu + p \rightarrow \mu^- X^{++}$} \\ 
    \noalign{\smallskip}\hline\hline\noalign{\smallskip}
                       &    &             & $p_\mu \geq 5$ GeV$/c$            &               &                                    &                       \\
                       &    &             & $p^\bot_\mu \geq 1$ GeV$/c$       &               &                                    &                       \\
     FNAL 15 ft (1983) & 14 & $[1, 225]$  & $p_{\text{ch}}^L \geq 5$ GeV$/c$  & 10\%          & $W^2>4~\text{GeV}^2/c^4$ $^\dagger$ & \cite{PhysRevD.27.47} \\
                       &    &             & $p_p \leq 340$ MeV$/c$         &               &                                    &                       \\
                       &    &             & $p_p \geq 200$ MeV$/c$         &               &                                    &                       \\
                       &    &             & $W\geq 1.5$ GeV$/c^2$             &               &                                    &                       \\
                       &    &             & $E_\nu^{\text{reco}} \geq 10$ GeV &               &                                    &                       \\
    \noalign{\smallskip}\hline
                       &    &             & $\varepsilon_{\text{cut}}$        &               &                                    &                       \\
     BEBC (1989)       & 6  & $[4, 196]$  & $p^\mu \geq 4$ GeV$/c$            & \emph{Not Included}      & \xmark                             & \cite{Jongejans1989}  \\
                       &    &             & $p_p \leq 300$ MeV$/c$            &               &                                    &                       \\
    \noalign{\smallskip}\hline\hline\noalign{\smallskip}
    \multicolumn{7}{c}{$\nu_\mu + n \rightarrow \mu^- X^{+}$} \\ 
    \noalign{\smallskip}\hline\hline\noalign{\smallskip}
                       &    &             & $p^\bot_\mu \geq 1$ GeV$/c$       &               &                                    &                       \\
     FNAL 15 ft (1983) & 14 & $[1, 225]$  & $p^{\text{ch}}_L \geq 5$ GeV$/c$  & 10\%          & \checkmark                         & \cite{PhysRevD.27.47} \\
                       &    &             & $E_\nu^{\text{reco}} \geq 10$ GeV &               &                                    &                       \\
                       &    &             & $p_p \leq 340$ MeV$/c$          &               &                                    &                       \\
                       &    &             & $p_p \geq 200$ MeV$/c$         &               &                                    &                       \\
    \noalign{\smallskip}\hline
                       &    &             & $\varepsilon_{\text{cut}}$        &               &                                    &                       \\
                       &    &             & $p_\mu \geq 4$ GeV$/c$            &               &                                    &                       \\
     BEBC (1984)       & 8  & $[6, 112]$  & $Q^2 \geq 1$ (GeV$/c$)$^2$        & \emph{Statistical}   & \checkmark                         & \cite{Allasia1984}    \\
                       &    &             & $W^2 \geq 5$ GeV$^2/c^4$          &               &                                    &                       \\
                       &    &             & $p_p \leq 300$ MeV$/c$          &               &                                    &                       \\
    \noalign{\smallskip}\hline
                       &    &             & $\varepsilon_{\text{cut}}$        &               &                                    &                       \\
     BEBC (1989)       & 6  & $[4, 196]$  & $p_\mu \geq 4$ GeV$/c$            & \emph{Not Included}      & \xmark                             & \cite{Jongejans1989}  \\
                       &    &             & $p_p \leq 300$ MeV$/c$            &               &                                    &                       \\
                       &    &             & $W \geq 5$ GeV$/c^2$              &               &                                    &                       \\
    \noalign{\smallskip}\hline\hline\noalign{\smallskip}
    \multicolumn{7}{c}{$\bar{\nu}_\mu + p \rightarrow \mu^+ X^{0}$} \\ 
    \noalign{\smallskip}\hline\hline\noalign{\smallskip}
     BEBC (1982)       & 8  & $[5, 75]$   & $p_\mu \geq 4$ GeV$/c$            & \emph{Statistical} & \checkmark                         & \cite{Barlag1982}     \\
                       &    &             & $p_p \leq 300$ MeV$/c$            &               &                                    &                       \\
    \noalign{\smallskip}\hline
                       &    &             & $\varepsilon_{\text{cut}}$        &               &                                    &                       \\
     BEBC (1989)       & 6  & $[4, 196]$  & $p^\mu \geq 4$ GeV$/c$            & \emph{Not Included}      & \xmark                             & \cite{Jongejans1989}  \\
                       &    &             & $p_p \leq 300$ MeV$/c$            &               &                                    &                       \\
    \noalign{\smallskip}\hline\hline\noalign{\smallskip}
    \multicolumn{7}{c}{$\bar{\nu}_\mu + n \rightarrow \mu^+ X^{-}$} \\ 
    \noalign{\smallskip}\hline\hline\noalign{\smallskip}
     BEBC (1982)       & 8  & $[1.5, 56]$ & $p_\mu \geq 4$                    & \emph{Statistical}  & \checkmark                         & \cite{Barlag1982}     \\
                       &    &             & $p_p \leq 300$ MeV$/c$            &               &                                    &                       \\
    \noalign{\smallskip}\hline
                       &    &             & $\varepsilon_{\text{cut}}$        &               &                                    &                       \\
     BEBC (1989)       & 6  & $[4, 196]$  & $p_\mu \geq 4$ GeV/c    & \emph{Not Included}    & \xmark                             & \cite{Jongejans1989}  \\
                       &    &             & $p_p \leq 300$ MeV$/c$            &               &                                    &                       \\
    \noalign{\smallskip}\hline\hline
    \end{tabular}
    \caption{Compilation of historical data from the \ac{BEBC} and FNAL 15 ft bubble chamber experiments on averaged charged hadron multiplicity
             in muon (anti)neutrino on deuterium interactions.
             Information about the number of points in each dataset, $N_p$, the $W^2$ range covered and the cuts applied in each analysis is provided.
             Unless specified, the systematic errors were not included in the data release error bands and have been added in quadrature by the amount
             specified in this table, see details in Sec.~\ref{subsec:systerrors}. The sixth column specifies whether a dataset is included, discarded or partially included in the fit,
             see Sec.~\ref{subsec:req_data}.}
    \label{tab:summary_data_2H}
\end{table*}

Before classifying the event as a neutrino interaction on either proton or neutron, the analyses procedure is equivalent to the one described in Sec.~\ref{sec:dataanalysesBBCH}. 
Each event has to contain a muon, identified with the \ac{EMI}, that satisfies the cuts summarised in Tab.~\ref{tab:summary_data_2H}. 
The information about the event topology and kinematics is obtained using the TVGP-SQUAW or HYDRA algorithms for FNAL 15 ft~\cite{PhysRevD.27.47} and \ac{BEBC} respectively~\cite{Jongejans1989,Allasia1984,Barlag1982}.
Particles are classified as pions if the algorithm fails to identify them as any other particle.
The neutrino energy is reconstructed using the transverse momentum balance method.
Similar kinematic cuts to those specified for the hydrogen analyses are applied. 

\begin{figure*}
    \centering
    \begin{subfigure}{0.45\textwidth}
        \centering
        \includegraphics[width=0.9\columnwidth]{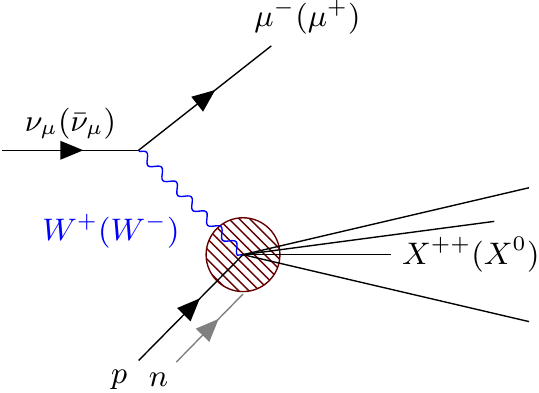}
        \caption{$\nu_\mu p$ topology.}
    \end{subfigure} 
    \begin{subfigure}{0.45\textwidth}
        \centering
        \includegraphics[width=0.9\columnwidth]{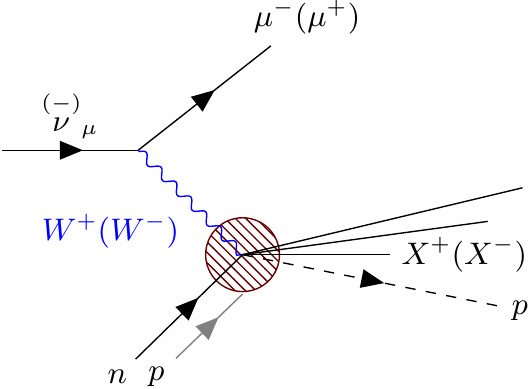}
    \caption{$\nu_\mu n$ topology.}
    \end{subfigure}
\caption{Bubble chamber analyses of $\nu_\mu$ and $\bar{\nu}_\mu$ on $^2$H data schematic procedure. The topology definition is based mainly on the number of prongs in each event. Possible visible proton spectators that satisfy the momentum requirements specified in Tab.~\ref{tab:summary_data_2H} are represented with dashed lines. $\bar{\nu}_\mu$ $^2$H one prong events are not considered.} 
\label{Fig:BBCHAnalaysis}
\end{figure*}

The main difference between both analyses is the particle identification of struck nucleons in the event.
A neutrino event is classified as a neutrino interaction on proton if the event topology has an odd number of prongs.
Alternatively, the event is classified as an interaction on neutron if the event has an even number of prongs with no visible spectator or an odd number of prongs that include a visible proton.
See a graphical interpretation in Fig.~\ref{Fig:BBCHAnalaysis}. 
The anti-neutrino case is similar except that the minimum prong multiplicity on proton is 1, instead of 3.
Because of the selection criteria explained in Sec.~\ref{sec:dataanalysesBBCH}, interactions with $n_{\text{ch}}=0$ are not considered, effectively making the selection criteria for anti-neutrinos the same as for neutrinos.

In the analyses, a prong is classified as a proton if it corresponds to  a particle moving backwards relatively to the beam direction ($\cos \theta_p < 0$) or a forward-going particle with low momentum. 
The maximum momentum cut is dataset dependent, see Tab.~\ref{tab:summary_data_2H}. 
If these conditions are not satisfied, the proton is not reconstructed and for the purpose of the analyses it is considered invisible.
In the FNAL 15 ft analyses, for a proton to be detected as a prong, its momentum has to be $p_p>200$~MeV$/c$. 

\begin{figure}
    \centering
    \begin{subfigure}{\columnwidth}
        \centering\includegraphics[width=6.5cm]{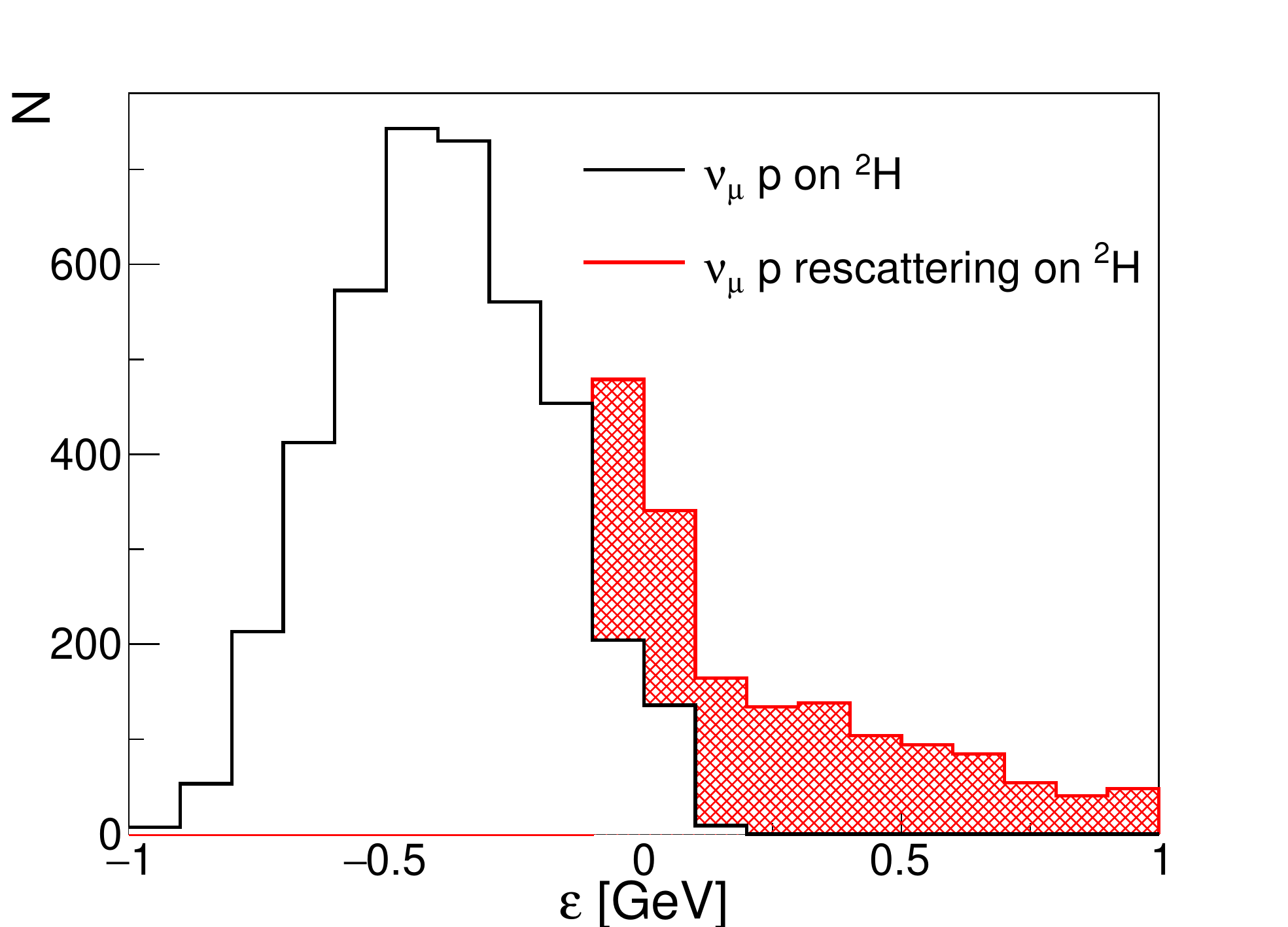}
        \caption{ $\nu_\mu p$ events under the odd prong topology assumption.   }   
    \end{subfigure}
    \begin{subfigure}{\columnwidth}
        \centering\includegraphics[width=6.5cm]{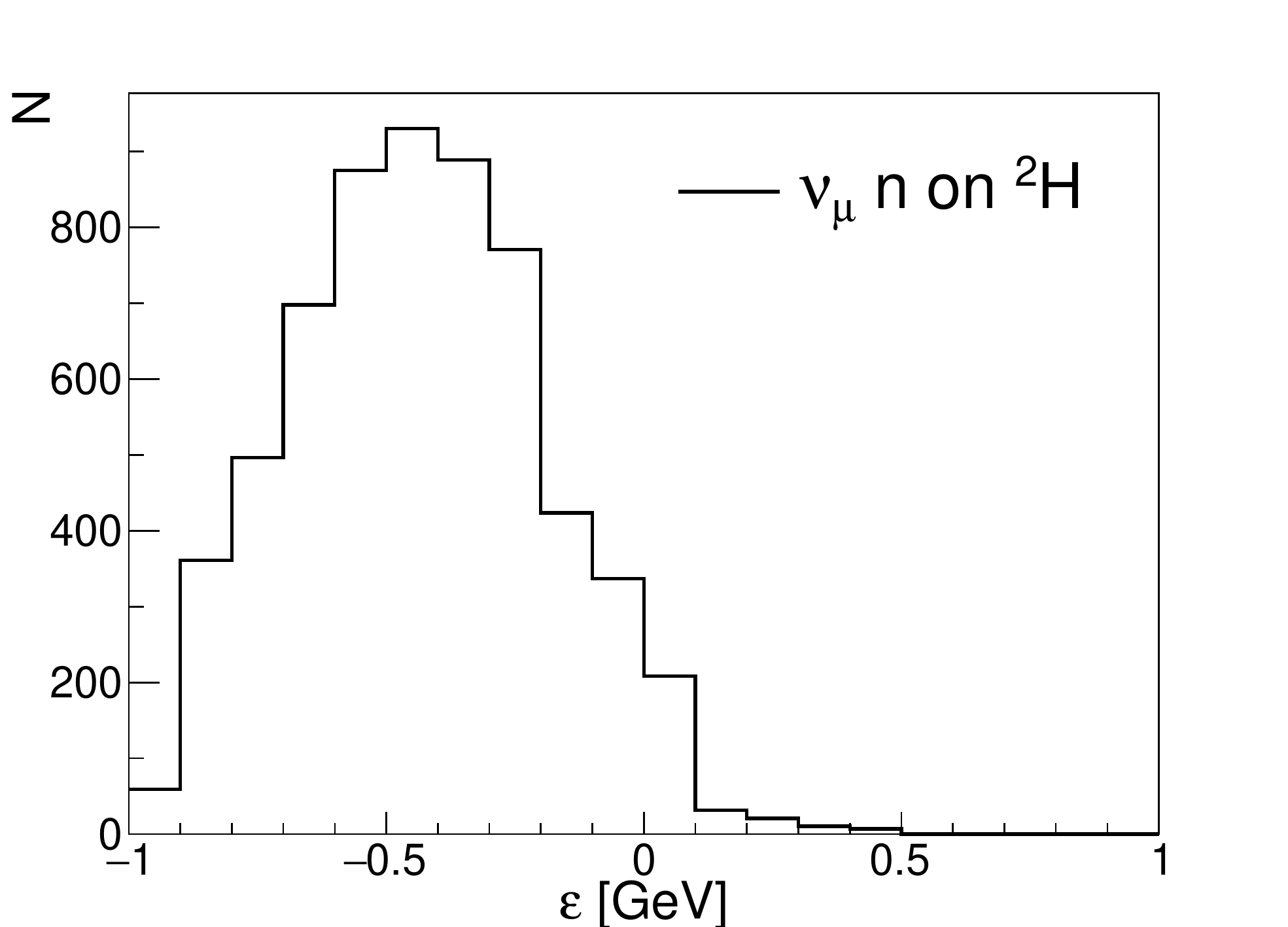}
        \caption{ $\nu_\mu n$ events under the even prong topology assumption. Neutron events with a spectator proton are not included. }   
    \end{subfigure}
\caption{ Energy balance distribution for $\nu_\mu$ events on proton and neutron candidates digitized from the \ac{BEBC} analyses paper~\cite{Allasia1984}. 
Events that do not satisfy the $\epsilon_{\text{reco}}$ correspond to rescattering events and are highlighted in red.
No rescattering contribution were observed in the $\nu_\mu n$ sample. }
\label{fig:nupscattering}
\end{figure}

The deuterium target can induce rescattering of the hit nucleon with the spectator: this can increase the number of hadrons in the final state~\cite{PhysRevLett.45.1817}. 
An odd number of prongs can occur in any possible neutrino interaction because of rescattering, independently of the hit nucleons, so the $\nu_\mu p$ sample will contain $\nu_\mu n$ events.
In contrast, the $\nu_\mu n$ sample can only contain $\nu_\mu p$ events because of detector inefficiencies. 
Rescattering events have an impact on the event kinematics, which can be quantified defining an energy balance as,
\begin{equation}
    \varepsilon \equiv \sum_i (E_i - {p_{L}}_ic) - Mc^2,
    \label{Eq:energybalance}
\end{equation}
where $E_i$ and ${p_{L}}_i$  are the $i$-th charged particle energy and longitudinal momentum component relative to the neutrino direction respectively while $M$ is the mass of the target nucleon assumed in the selection sample. 
Eq.~\ref{Eq:energybalance} assumes that the nucleon is at rest and that the neutrino direction is known.
In an ideal detector where all final state particles are identified, $\varepsilon = 0$~\cite{chang1983study}.
In a bubble chamber experiment, where only charged particles are detected, $\varepsilon<0$. 
Possible particle miss-ID reduces the $\varepsilon$ value further, as particles are assigned to be pions as a default, unless identified otherwise.
Rescattering events have a $\varepsilon>0$ with a maximum value of $M_{{}^2\text{H}}c^2-M_nc^2$, see Fig.~\ref{fig:nupscattering}.

The \ac{BEBC} experiment eliminates rescattering events from the sample by imposing a cut on the energy balance~\cite{Allasia1984,Jongejans1989,Barlag1982}. 
An event is rejected due to rescattering if the following conditions are satisfied:
\begin{itemize}
    \item[-] ${\varepsilon > 0.1}$~GeV,
    \item[-] ${\varepsilon > -0.1}$~GeV and the transverse missing momenta squared differs from zero, ${(p_\bot^{\text{miss}})^2 > 0.075~\text{(GeV$/c$)}^2}$. 
\end{itemize}
The FNAL 15 ft experiment did not correct for rescattering events.
In some of the analyses, additional cuts are considered for the deuterium analyses to remove backgrounds.
For instance, the FNAL 15 ft bubble chamber reduces the background from neutral hadron-induced events by applying a cut on the total charged-particle longitudinal momentum in the final state system ($p_{\text{ch}}^L$)~\cite{PhysRevD.27.47}. 

\subsection{Sources of systematic uncertainties in the FNAL 15 ft and BEBC experiments}
\label{subsec:systerrors}
\ac{MC} studies were preformed by the FNAL 15 ft and \ac{BEBC} bubble chamber experiments to correct for possible sources of errors. 
In particular, the different analyses correct for the following effects:
\begin{itemize}
    \item[-] \ac{EMI} geometrical inefficiency~\cite{Jones1990,BEBC1992,Allasia1984,Barlag1982}.
    \item[-] Efficiency losses due to possible hadron miss-ID and migration of particles from the forward to backward hemispheres~\cite{GRASSLER1983269,Jones1990,Allasia1984,BEBC1992}. 
    \item[-] $W^2$ smearing due to the uncertainty in the neutrino energy
             reconstruction~\cite{PhysRevD.27.47,GRASSLER1983269,Jones1990,BEBC1992,PhysRevLett.36.124,PhysRevD.25.624,Jongejans1989,Allasia1984,Barlag1982}.
    \item[-] Neutrino energy uncertainty associated the transverse balance
             method~\cite{PhysRevD.27.47,GRASSLER1983269,Jones1990,BEBC1992,PhysRevLett.36.124,PhysRevD.25.624,Jongejans1989,Allasia1984,Barlag1982}.
    \item[-] Neutral particle decays ($\gamma$, $K^0$ and $\Delta$) into charged particles that can lead to a higher charged multiplicity
             if the decay vertex is close to the primary one~\cite{PhysRevLett.36.124,PhysRevD.27.47,PhysRevD.25.624,GRASSLER1983269,Jones1990,BEBC1992,Barlag1982}. 
    \item[-] One prong event corrections~\cite{PhysRevD.25.624,GRASSLER1983269,Jones1990,Allasia1984,BEBC1992}.
             This kind of events occur for low-$W$ $\bar{\nu}p$ interactions in which only the $\mu^+$ is observed.
    \item[-] Efficiency to detect CC events~\cite{PhysRevD.27.47,PhysRevD.25.624}.
    \item[-] Corrections due to the Fermi motion in Deuterium~\cite{Allasia1984}.
    \item[-] Possible measurement errors~\cite{PhysRevD.27.47,GRASSLER1983269,Jones1990,Allasia1984,BEBC1992}.
\end{itemize} 

The information about the \ac{BEBC} systematic errors was obtained by using two \ac{MC} programs: the LUND MC and a \ac{LPS}~\cite{PhysRevD.19.1}.
Both MC were tuned to describe the \ac{BEBC} experiment.
From the MC generations, two samples are created: the initial, $d^{\text{MC}}_{\text{initial}}$, and modified, $d^{\text{MC}}_{\text{modified}}$, samples.
The initial sample contains the truth information of the event.
The modified sample includes modifications to mimic the analyses procedure. 
The ratio between the samples provides with a correction factor that it is applied to the data.

\ac{BEBC} systematic errors are obtained from the difference between both \ac{MC} calculations. 
The FNAL 15 ft corrected for some of the effects, but no clear information about the methodology followed to estimate the systematics is provided.
Some of these experiments provide error bars which already include an estimation of systematics, 
however, this is not the case for most of the data. 
In particular, there is three different ways in which the \ac{BEBC} and FNAL 15 ft experiments quote the systematic errors: (1) the systematic errors are already included in the total error, (2) the systematic uncertainty was quoted as a percentage with respect to the central value, (3) the systematic error is considered to be approximately of the same size of the statistical error or (4) no information is provided in the data release. For the cases (2) and (3), the systematic errors are added in quadrature to the statistical ones in this analysis.  
Particularly, for the datasets from Ref.~\cite{Jongejans1989}, information on systematic errors is not provided in the data release.
In Tab.~\ref{tab:summary_data_H} and Tab.~\ref{tab:summary_data_2H}, the information on the systematic error is provided. 
We label the different categories as (1) \emph{Included}, (2) with the percentage, (3) \emph{Statistical}, (4) \emph{Not Included} respectively.
No correlation matrices are provided by any of these experiments. 

\section{Review of previous tunes to hadronization data}
\label{sec:tune_review}

While summarising the experimental fits to averaged charged multiplicity data, this section also explains the origin of the \emph{2010} GENIE tune parameters. 
This is necessary to define proper selection criteria for a dataset to be included in a global fit.

\subsection{Fits to bubble chamber data}
\label{sec:data_review_lowW}
Both \ac{BEBC} and FNAL 15 ft experiments provided estimations of the $\alpha_{\text{ch}}$ and $\beta_{\text{ch}}$ parameters for every released dataset.
The individual fits were performed by fitting Eq.~\ref{Eq:Avnch} in each channel.
Fit results are summarised in Tab.~\ref{tab:summary_data}. 

\begin{table*}   
    \centering
    \begin{tabular}{@{\extracolsep\fill} c c c c c c} \hline\hline\noalign{\smallskip}
    \textbf{Experiment}        & 
    $\mathbf{W^2~[\textbf{GeV}^2/c^4]}$ & \textbf{Target} & $\boldsymbol\alpha_{\textbf{ch}}$ & $\boldsymbol\beta_{\textbf{ch}}$ & \textbf{Ref.} \\ 
    \noalign{\smallskip}\hline\hline\noalign{\smallskip}
    \multicolumn{6}{c}{$\nu_\mu + p \rightarrow \mu^- X^{++}$} \\ 
    \noalign{\smallskip}\hline\hline\noalign{\smallskip}
    FNAL 15 ft (1976)          & [1.5, 150]  &  H         & $1.09\pm0.38$    & $1.09\pm0.03$  & \cite{PhysRevLett.36.124} \\
    \noalign{\smallskip}
     BEBC (1983)               & [12, 112]   &  H         & $-0.05\pm0.11$   & $1.43\pm0.04$  & \cite{GRASSLER1983269}\\
    \noalign{\smallskip}
    \textbf{FNAL 15 ft (1983)} & [1.5, 160]  & $^2$H      & $0.05\pm0.07$    & $1.42\pm0.03$  & \cite{PhysRevD.27.47} \\
    \noalign{\smallskip}
     BEBC (1990)               & [6, 150]    &  H         & $0.911\pm0.224$  & $1.131\pm0.086$& \cite{Jones1990}\\
    \noalign{\smallskip}
     BEBC (1992)               & [12, 144]   &  H         & $0.40\pm0.13$    & $1.25\pm0.04$  & \cite{BEBC1992}\\
    \noalign{\smallskip}\hline\hline\noalign{\smallskip}
    \multicolumn{6}{c}{$\nu_\mu + n \rightarrow \mu^- X^{+}$} \\ 
    \noalign{\smallskip}\hline\hline\noalign{\smallskip}
    BEBC (1984)                & [6, 112]    & $^2$H      & $1.75\pm0.12$    & $1.31\pm0.04$  & \cite{Allasia1984}   \\
    \noalign{\smallskip}
    \textbf{FNAL 15 ft (1983)} & [1.5, 160]  & $^2$H      & $-0.20\pm0.07$   & $1.42\pm0.03$  & \cite{PhysRevD.27.47} \\
    \noalign{\smallskip}\hline\hline\noalign{\smallskip}
    \multicolumn{6}{c}{$\bar{\nu}_\mu + p \rightarrow \mu^+ X^{0}$} \\ 
    \noalign{\smallskip}\hline\hline\noalign{\smallskip}
     FNAL 15 ft (1982)         & [1.7, 74]   &  H         & $-0.44\pm0.13$   & $1.48\pm0.06$  & \cite{PhysRevD.25.624}\\
    \noalign{\smallskip}
    \textbf{BEBC (1982)}       & [5, 75]     & $^2$H      & $0.02\pm0.20$    & $1.28\pm0.08$  & \cite{Barlag1982}\\
    \noalign{\smallskip}
     BEBC (1983)               & [12, 96]    &  H         & $-0.56\pm0.25$   & $1.42\pm0.08$  & \cite{GRASSLER1983269}\\
    \noalign{\smallskip}
    BEBC (1990)                & [6, 144]    &  H         & $0.222\pm0.362$& $1.117\pm0.100$& \cite{Jones1990}\\
    \noalign{\smallskip}
    BEBC (1992)                & [12, 144]   &  H         & $-0.44\pm0.20$   & $1.30\pm0.06$  & \cite{BEBC1992}\\
    \noalign{\smallskip}\hline\hline\noalign{\smallskip}
    \multicolumn{6}{c}{$\bar{\nu}_\mu + n \rightarrow \mu^+ X^{-}$} \\ 
    \noalign{\smallskip}\hline\hline\noalign{\smallskip}
     \textbf{BEBC (1982)}      & [1.5, 56]   & $^2$H      & $0.80\pm0.09$    & $0.95\pm0.04$  & \cite{Barlag1982}\\
    \noalign{\smallskip}\hline\hline\noalign{\smallskip}
    \multicolumn{6}{c}{$\nu_\mu + A $} \\ 
    \noalign{\smallskip}\hline\hline\noalign{\smallskip}
    OPERA (2018)               & [1.6, 54.6] & Pb         & $-0.19\pm0.18$   & $0.76\pm0.07$  & \cite{collaboration2017study} \\
    CHORUS (2007)              & [1, 148]    & Fuji ET-B7 & $1.07\pm0.05$    & $1.32\pm0.11$  & \cite{Kayis_Topaksu_2007}  \\
    \noalign{\smallskip}\hline\hline
    \end{tabular}
    \caption{Compilation of best fit values for the intercept $\alpha_{\text{ch}}$ and           slope $\beta_{\text{ch}}$ obtained from individual fits
             to Eq.(\ref{Eq:Avnch}) against mean charged hadron multiplicity data as a function of $W^2$. The parameters for charged-current
             $\nu_\mu$ and $\bar{\nu}_\mu$ scattering data on hydrogen, deuterium, $^{207}$Pb and the Fuji ET-B7 emulsion are shown in the table.
             \emph{2010} GENIE parameters are extracted from the analyses highlighted in bold.}
    \label{tab:summary_data}
\end{table*} 

There are 6 channels in total: $\nu_\mu$ or $\bar{\nu}_\mu$ on proton or neutron, while the information on interactions on proton can be from data with hydrogen or deuterium targets.
Information about neutrino interaction on neutron can only be extracted from deuterium samples. 
The  \ac{BEBC} and/or FNAL 15 ft experiments performed individual fits to each of the available channels.

From the best-fit values extracted for each dataset, we observe clear discrepancies for the $\alpha_{\text{ch}}$ and $\beta_{\text{ch}}$ values between data releases and between the \ac{BEBC} and FNAL 15 ft data (e.g.\ for $\nu_\mu p$ interactions on hydrogen).
Discrepancies between hydrogen and deuterium samples are also present.
This target-related discrepancy can also be observed in fits to OPERA and CHORUS data~\cite{collaboration2017study,Kayis_Topaksu_2007}.
These discrepancies could have different origins: the $W^2$ range, the beam energy or the kinematic cuts applied in the analyses. 

The \emph{2010} GENIE AGKY parameter values presented in Tab.~\ref{tab:agkyParameters} correspond to the analyses on deuterium targets highlighted in Tab.~\ref{tab:summary_data}. 
Notice that the parameters used in the \emph{2010} GENIE prediction come from fits to Eq.~\ref{Eq:Avnch} over the whole $W^2$ range.
This procedure is not adequate as the $\alpha_\text{ch}$ and $\beta_\text{ch}$ should be extracted from a fit to data over the low-$W$ validity range given that the \ac{AGKY} model differs from the simplified linear behaviour.

The description of the shower particle content is linked to several observables whose correlation is still unknown. 
For instance, the averaged charged multiplicity and dispersion observables can be correlated.
The full list of available hadronization data is shown in Ref.~\cite{Yang_2009}.
Ideally, the AGKY tune should improve the agreement with all hadronization related observables. 
The extraction of the averaged charged multiplicity parameters, such as $\alpha_\text{ch}$ and $\beta_\text{ch}$, strongly relays on the precise understanding of the datasets described in Sec.~\ref{sec:dataanalysesBBCH} and Sec.~\ref{sec:dataanalysesBBCH2H}.
However, the analyses of historical averaged charged multiplicity datasets already show clear disagreements between each of the different data releases, as summarised in Tab.~\ref{tab:summary_data}. 
For these reasons, on this work, we focus on the description and tune of averaged charged multiplicity data on hydrogen and deuterium samples.

\subsection{The HERMES tune}
\label{subsec:HERMESTUNE}
\begin{figure}
    \centering
    \begin{subfigure}{\columnwidth}
         \centering\includegraphics[width=0.955\columnwidth]{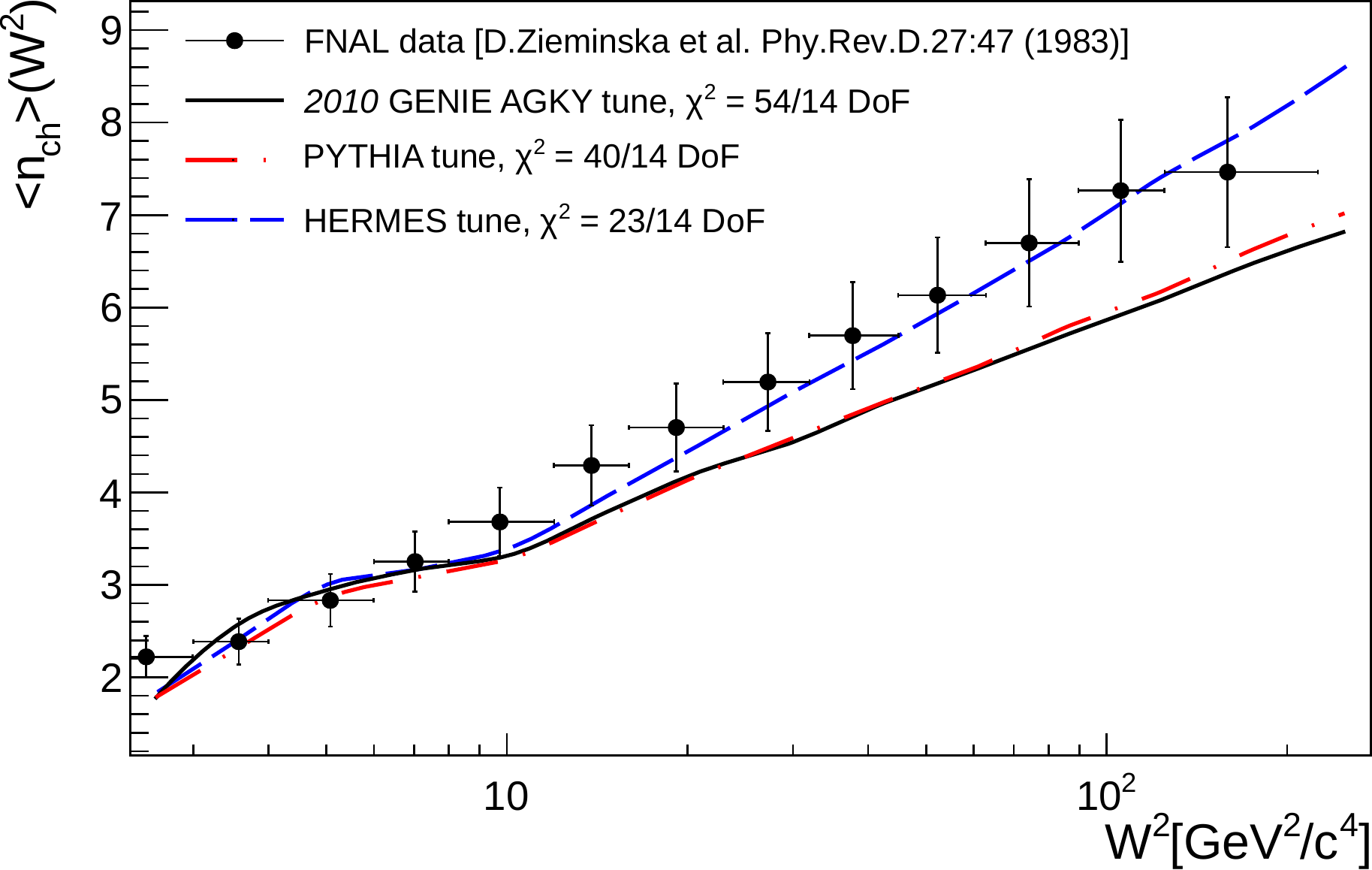}
         \caption{$\nu_{\mu}p\rightarrow\mu^-X^{++}$ on $^2$H.}
    \end{subfigure}
    \begin{subfigure}{\columnwidth}
         \centering\includegraphics[width=\columnwidth]{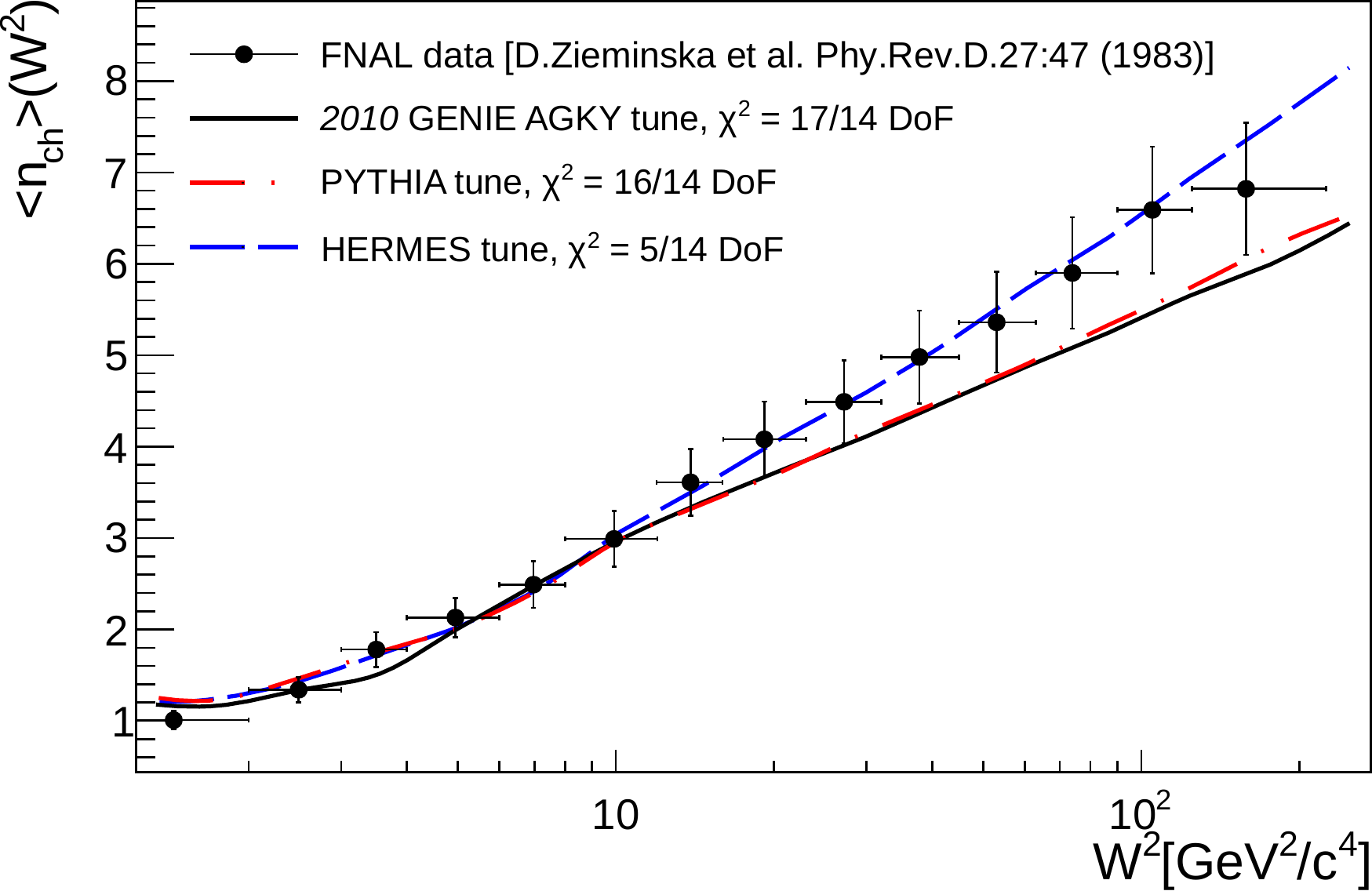}
          \caption{$\nu_{\mu}n\rightarrow\mu^-X^{+}$ on $^2$H.}
    \end{subfigure}
\caption{Comparison of FNAL average charged multiplicity deuterium data against GENIE predictions obtained with the parameterisations specified in Tab.~\ref{tab:PYTHIAParameters_default}.}
\label{fig:OldTunes}   
\end{figure}

The PYTHIA parameters are extracted from high energy $e^-e^+$ experiments, see Sec.~\ref{subsec:PYTHIA}.
From Fig.~\ref{fig:OldTunes} we see that PYTHIA underestimates the averaged charged multiplicity. 
However, PYTHIA has not been tuned using data from neutrino experiments.

The \emph{2010} GENIE tune, summarised in Sec.~\ref{subsec:PYTHIA}, aimed to improve the agreement with different hadronization observables by incorporating the results from a NUX PYTHIA tune~\cite{NUX}.
However, this was not sufficient to improve the agreement of PYTHIA with average charged multiplicity data from bubble chambers experiments.
Moreover, the tune lacked of information about the uncertainties of the fit parameters.

Information on PYTHIA parameters at lower energy was provided by the HERMES experiment, which tuned PYTHIA using $e^\pm p$ data at $27$~GeV~\cite{HERMES}.
Ref.~\cite{TeppeiTUNE} suggests that the HERMES tune improves the agreement with neutrino data, as summarized in Fig.~\ref{fig:OldTunes}. 
The main differences between the HERMES tune and the GENIE \emph{2010} re-tune are the modification of the Lund $a$ and Lund $b$ parameters, suggesting higher (lower) values of Lund $a$ (Lund $b$).

The PYTHIA parameters with most impact on the average charged multiplicity for the \emph{2010} GENIE AGKY and HERMES tunes are summarised in Tab.~\ref{tab:PYTHIAParameters_default}. 

\subsection{Requirements for including a dataset in the AGKY multiplicity tune}
 \label{subsec:req_data}
Only the averaged charged multiplicity data on hydrogen and deuterium is taken into account in this AGKY fit.
Tables~\ref{tab:summary_data_H}--\ref{tab:summary_data_2H} summarize the information about which datasets are included in the tune.
If possible, only the latest analysis of each experiment is included. 
Previous analyses are only considered if:
\begin{enumerate}
    \item Its reanalyses did not cover all the original $W^2$ range, 
    \item The prediction interpolation by Professor fails to describe the GENIE prediction (see Sec.~\ref{sec:professorcondition}),
    \item The data release lack of sufficient information about systematic errors.   
\end{enumerate}
In requirement (1), previous analyses are used to complement the covered $W^2$ range as those points were not documented in the revisited ones. 
If the datasets are  only included partially, the approximate $W^2$ range used is provided.
An example is the \ac{BEBC} $\nu_\mu$ on H data, in particular datasets [BEBC,1] and [BEBC,2].
In this case, the data point at $W^2 < 10 $~GeV$^2/c^4$ from the earlier release ([BEBC,1]) is included in the fit, while the others are not because the later [BEBC,2] is covering the same $W$ region.
This approach has already been implemented in other studies~\cite{Kuzmin_2013}.
The exact $W^2$ range after requirement (2) is given in Sec.~\ref{sec:AGKYTune}.

Global fits can be used to expose datasets that pull the results in different directions. 
This is the case of the most recent $\bar{\nu}_\mu$ measurement by \ac{BEBC} experiment~\cite{Jongejans1989}, which did not provide information on systematic errors and, consequently, the total error on this data tends to be much smaller than the rest, see Sec.~\ref{sec:AGKYTune}.
Such small errors give a strong preference to this dataset and, as a consequence, this measurement is in tension with other data, including older $\bar{\nu}_\mu$ \ac{BEBC} measurements~\cite{Barlag1982} which information on the systematic uncertainty was provided, see Tab.~\ref{tab:summary_data_2H}. 
Given that the \ac{BEBC}~\cite{Jongejans1989} analyses did not provide enough information on the systematic errors and they are in clear disagreement with the other ones, these are not considered in the tune and are shown for comparison only.

\section{Parameterisation of model uncertainties}
\label{sec:parametrization}

This section discusses the impact on AGKY parameters on the predictions. 
The predictions are generated with the \texttt{G18\_02a\_02\_11a} tune of GENIE version 3.0.6. 
This tune was previously obtained to improve the agreement with pion production data on free nucleon~\cite{mypaper_1}.
The complete model list for this tune is summarised in Tab.~\ref{tab:G18_02a_02_11a_details}.
As introduced in Sec.~\ref{sec:AGKY}, hadronization is modelled with the \ac{AGKY} model~\cite{Yang_2009}. 
Interactions with nuclei are calculated within the relativistic Fermi Gas framework, using the Bodek-Ritchie model~\cite{PhysRevD.23.1070}, and hadronic re-interactions are simulated using INTRANUKE hA. The main contributions to the averaged charged multiplicity predictions come from CC DIS and non-resonance SIS~\cite{0709.4378}.
As the DIS and models are common for all GENIE v3 tunes, the choice of the base configuration does not affect the hadronization predictions.
An updated version of the \texttt{G18\_02a\_02\_11a} tune, named \texttt{G18\_02a\_02\_11b}, has been recently released in Ref.~\cite{mypaper_1}.
In terms of the hadronization predictions, these \ac{CMC}'s are interchangeable and the results of this work are valid within the updated version.

\begin{table}
    \centering
    \begin{tabular}{c c}
    Simulation domain & Model \\ \noalign{\smallskip} \hline \hline \noalign{\smallskip}
      Nuclear model & Fermi Gas~\cite{PhysRevD.23.1070} \\ \noalign{\smallskip}
     \hline\noalign{\smallskip}
      QEL   &  Llewellyn Smith~\cite{LLEWELLYNSMITH1972261}  \\
      QEL Charm &  Kovalenko~\cite{Kovalenko:1990zi} \\
      QEL $\Delta S=1$ & Pais~\cite{PAIS1971361} \\
      RES   &  Rein-Sehgal~\cite{REIN198179}  \\
      SIS/DIS   &  Bodek-Yang~\cite{0709.4378} \\
      DIS $\Delta S=1$ & Aivazis-Olness-Tung~\cite{aivazis1993nexttoleading}\\\noalign{\smallskip}
      Coherent $\pi$ production &  Rein-Sehgal~\cite{REIN198179}  \\ \hline\noalign{\smallskip}
      Hadronization & \ac{AGKY}~\cite{Yang_2009} \\\hline\noalign{\smallskip}
      FSI & INTRANUKE hA~\cite{Andreopoulos:2015wxa} \\ \noalign{\smallskip}  \hline \hline
     \end{tabular}
    \caption{Complete list of models used for the \texttt{G18\_02a\_02\_11a/b} CMC in GENIE v3. }
    \label{tab:G18_02a_02_11a_details}
\end{table}

The subset of parameters controlling the averaged charged hadron multiplicity is the target of our tune.
The list contains the parameters $\alpha_\text{ch}$ and $\beta_\text{ch}$ defined in Eq.~\ref{Eq:Avnch} and the five PYTHIA parameters discussed before in Sec.~\ref{subsec:PYTHIA}. 
The ranges for $\alpha_{\text{ch}}$ and $\beta_{\text{ch}}$ parameters are chosen in a such a way that they cover the values reported by experimental fits, see Tab.~\ref{tab:summary_data}.
The same approach is followed to define the PYTHIA parameters range from the HERMES tune, see Tab.~\ref{tab:PYTHIAParameters_default}.

\begin{figure*}
    \centering
    \includegraphics[width=\textwidth]{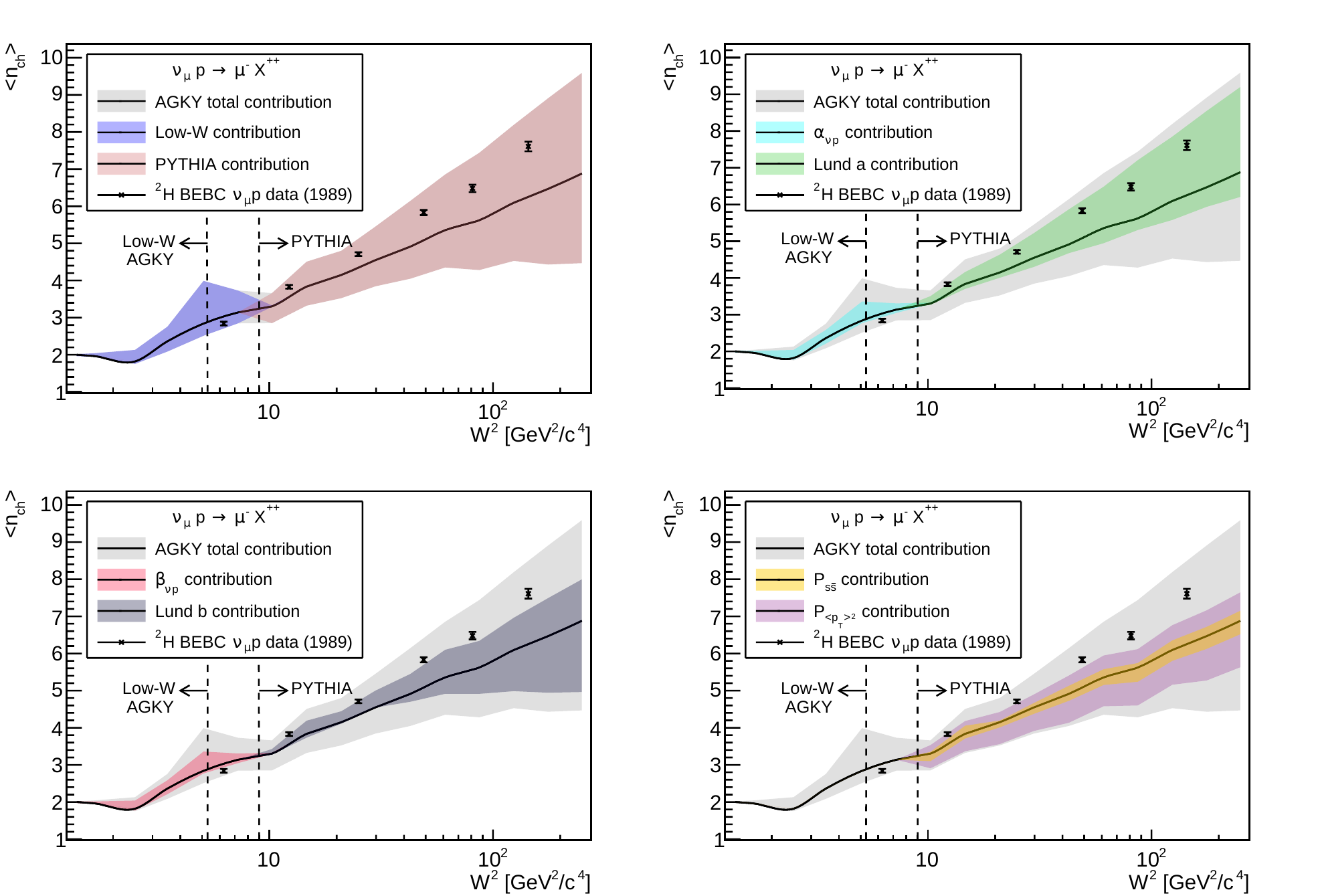}
    \caption{Impact of fit parameters on the prediction of the averaged charged multiplicity, as a function of W, for $\nu_\mu p\rightarrow\mu^-X^{++}$ interaction on a deuterium target. 
    Each parameter has been varied within the range of study specified in Tab.~\ref{tab:fitResultParameters}. 
    The top left plot shows the total contributions from low-$W$ and PYTHIA parameters. 
    All the other plots specify the contribution from specific parameters compared to the total, which is always rendered with the grey area. 
    Dashed lines correspond to $W_{\min}^{\text{tr}}$ and $W_{\max}^{\text{tr}}$, defining the transition region. 
    \ac{BEBC} data~\cite{Jongejans1989} are shown for reference. }
    \label{fig:parameter_range}
\end{figure*}

The impact of each parameter range on the predictions of averaged charge multiplicity for $\nu_\mu$ CC interactions on proton is shown in Fig.~\ref{fig:parameter_range}. 
As expected both $\alpha_\text{ch}$ and $\beta_\text{ch}$ act on low-$W$ and their effect vanishes gradually over the transition region. 
In the PYTHIA region, the largest contribution comes from the Lund $a$ and Lund $b$ parameters.
In the transition region, the prediction will be determined by both sets of parameters: as a consequence we anticipate a correlation between PYTHIA and the low-$W$ AGKY parameters after the fit. 
The parameter ranges that defines the parameter space are defined in Tab.~\ref{tab:fitParameters}.

\section{Construction of the GENIE predictions and evaluation of the likelihood}
\label{sec:Likelihood}
In order to build the hadronization prediction for the data described in Sec.~\ref{sec:dataanalysesBBCH}, $\nu_\mu$ and $\bar{\nu}_\mu$  charged current (CC) events on H and $^2$H are simulated. 
Events are generated using a "$1/E$"-like flux, with a $0.1-200$~GeV energy range.
This is sufficient as the observables are given in terms of $W$, hence, the neutrino flux is factorized out.

In order to compute the prediction associated to the $i$-th dataset from Tab.~\ref{tab:summary_data_H} and Tab.~\ref{tab:summary_data_2H}, we select events simulated with the neutrino flux and target of the corresponding experiment and processed using the same experimental cuts. 
For each selected event we reconstruct $E_\nu$ and $W$ following the recipes described in Sec.~\ref{sec:dataanalysesBBCH}.
The events are classified in bins according to the reconstructed $W$ and for each bin we evaluate the average charged multiplicity $\langle n_{\text{ch}}\rangle_i( W )$.
This operation is repeated for a number of points in the parameter space $\boldsymbol{\theta}$ defined in Tab.~\ref{tab:fitParameters}. 
Each experiments has a different binning system and therefore we identify the $W$ bins using two indices: one for the dataset ($i$) and the other one for the bin index inside the dataset ($j$).
Labeling with $\boldsymbol{\theta}$ the vector of coordinates of a point belonging to the  parameters space, we can define our predictions associated to the $i$-th dataset and a given $j$-th $W$ bin as $\langle n_{\text{ch}}\rangle_i \left( W_{ij}|\boldsymbol{\theta}\right)$.
The statistical error due to the MC sample size is also evaluated and this is refereed to as $\sigma_{ij} \left( \boldsymbol{\theta}\right)$.

We use Professor~\cite{Professor} to generate a parameterisation denoted as $\widetilde{n}_{ij}(\boldsymbol{\theta})$ and  $\widetilde{\sigma}_{ij} \left( \boldsymbol{\theta}\right)$ interpolating the values of $\langle n_{\text{ch}} \rangle_{i}(W_{ij}|\boldsymbol{\theta})$ and $\sigma_{ij} \left( \boldsymbol{\theta}\right)$ as a function of~$\boldsymbol{\theta}$. 
The parameterisation is a generic polynomial of order $M$ in the P-dimensional space~\cite{Professor}, whose analytical form is: 
\begin{eqnarray}
  \widetilde{n}_{ij}(\boldsymbol{\theta}) & = & \alpha_{0}^{ijk} + \sum_{n=1}^P \beta_{n}^{ijk} \theta_{n} + \sum_{n \le m} \gamma_{nm}^{ijk} \theta_{n} \theta_m \nonumber \\
    & + & \ldots + \sum_{n_1 \le \ldots \le n_M } \! \! \! \! \! \! \xi ^{ijk} _{n_1 \ldots n_M } \prod _{\ell=1} ^M \theta_{n_\ell}, \label{eqn:parameterisation}
\end{eqnarray}
where $\theta_{n}$ is the coordinate of the $n$-th parameter and $M$ the polynomial order, set to 4$^{th}$ order in this work.
The coefficients $\alpha_{0}^{ijk}$, $\beta_{n}^{ijk}, \gamma_{(nm)}^{ijk}, \ldots,\xi ^{ijk} _{(n_1 \ldots n_M)} $ are determined by Professor fitting the parameterisation against the computed  $\langle n_{\text{ch}}\rangle_i \left( W_{ij}|\boldsymbol{\theta}\right)$ obtained by generating O(10$^{4}$) points uniformly spread within parameter space defined in Tab.~\ref{tab:fitParameters}. 
Non-physical regions in the sampled parameter space are avoided applying a veto function.
In particular, every combination of $\boldsymbol{\theta}$ has to verify that $\langle n_{\text{ch}} \rangle \geq  0 $ at the pion production threshold.
The parameterisation $\widetilde{n}_{ij}(\boldsymbol{\theta})$ is used instead of the exact predictions in order to to estimate the best-fit parameters by minimising the $\chi^2$.
The main advantage of this method is the reduction of the brute-force scans computational complexity while allowing for massive parallelisation.

\subsection{Professor interpolation Cut-off condition}
\label{sec:professorcondition}

\begin{figure}
    \centering
    \includegraphics[width=0.8\columnwidth]{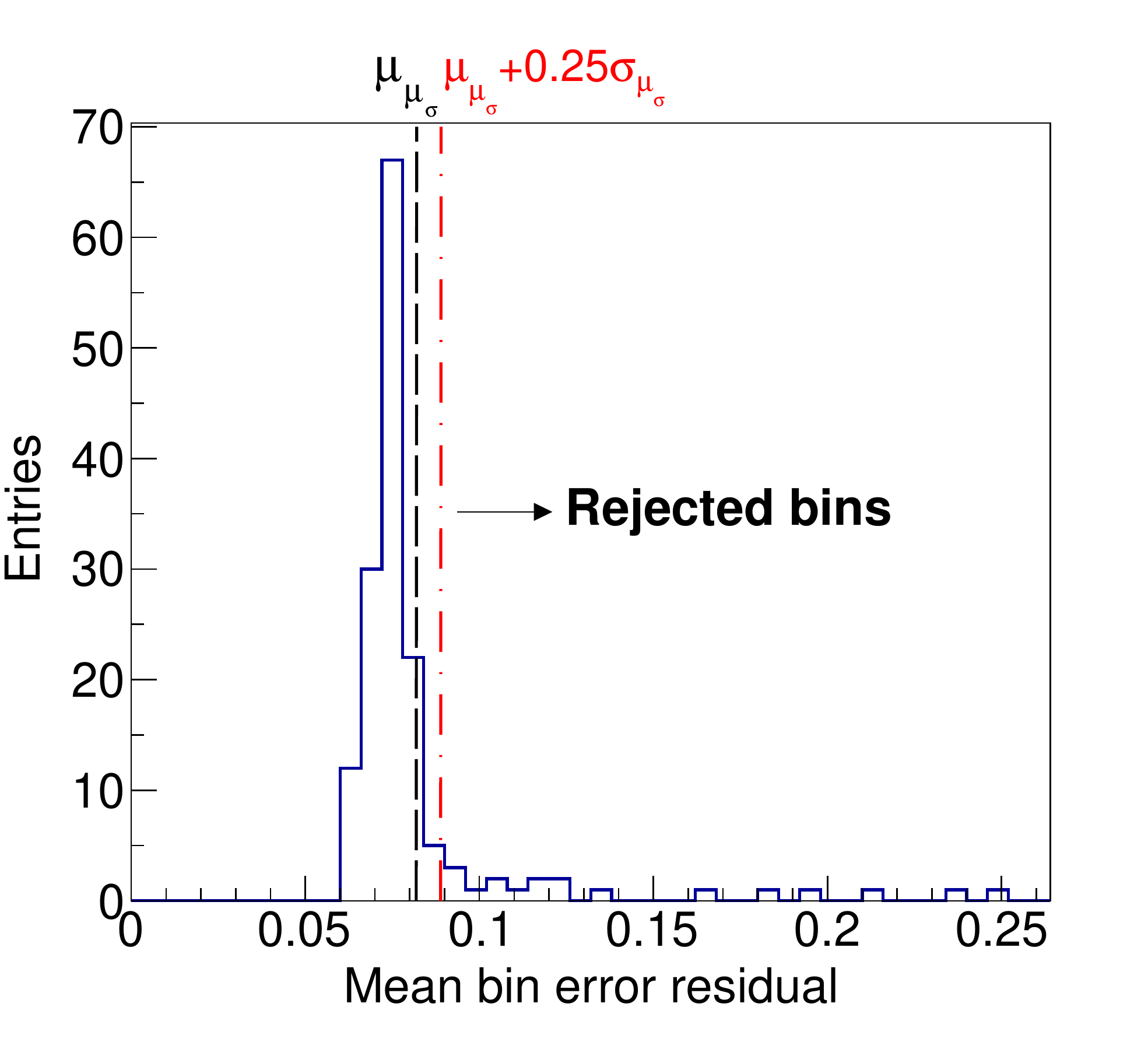}
    \caption{Distribution of mean bin error residual for all data points. 
    The distribution mean value ($\mu_{\mu_\sigma}$) is shown with a dashed black line.
    Data points with a mean value higher than 0.25 of the mean bin error variance ($\sigma_{\mu_\sigma}$) are rejected.
    This cut-off value is shown with the dashed red line.
    }
    \label{fig:Mean_error_distribution}
\end{figure}

As mentioned in Sec.~\ref{subsec:req_data}, in view of the fact that the Professor interpolation is just an approximation, it can fail to describe the actual prediction.
When this happens, we remove from the analysis data points whose Professor interpolation, of the predicted mean value or predicted error, disagree too much with the GENIE prediction corresponding to that data point. 
The relative difference between the interpolation and the GENIE prediction is known as residual.
For each data point, we calculate the bin central value and bin error residuals for all the points in our parameter space.
The distributions of the residual for central values and errors are monitored and whenever the means or the variances of a bin are too far from the average values among all bins, the corresponding data point is removed. 
In this analysis, the cut-off condition requires that any data points with a mean central value or error that exceeds the average values among all bins by 0.25$\sigma_{\mu_\mu}$ or 0.25$\sigma_{\mu_\sigma}$ respectively is removed from the analysis.
An example of the cut-off condition on the error residual distribution for all data points is shown in Fig.~\ref{fig:Mean_error_distribution}. 

\begin{figure*}
    
    \begin{subfigure}{0.35\columnwidth}
        \includegraphics[width=0.9\columnwidth]{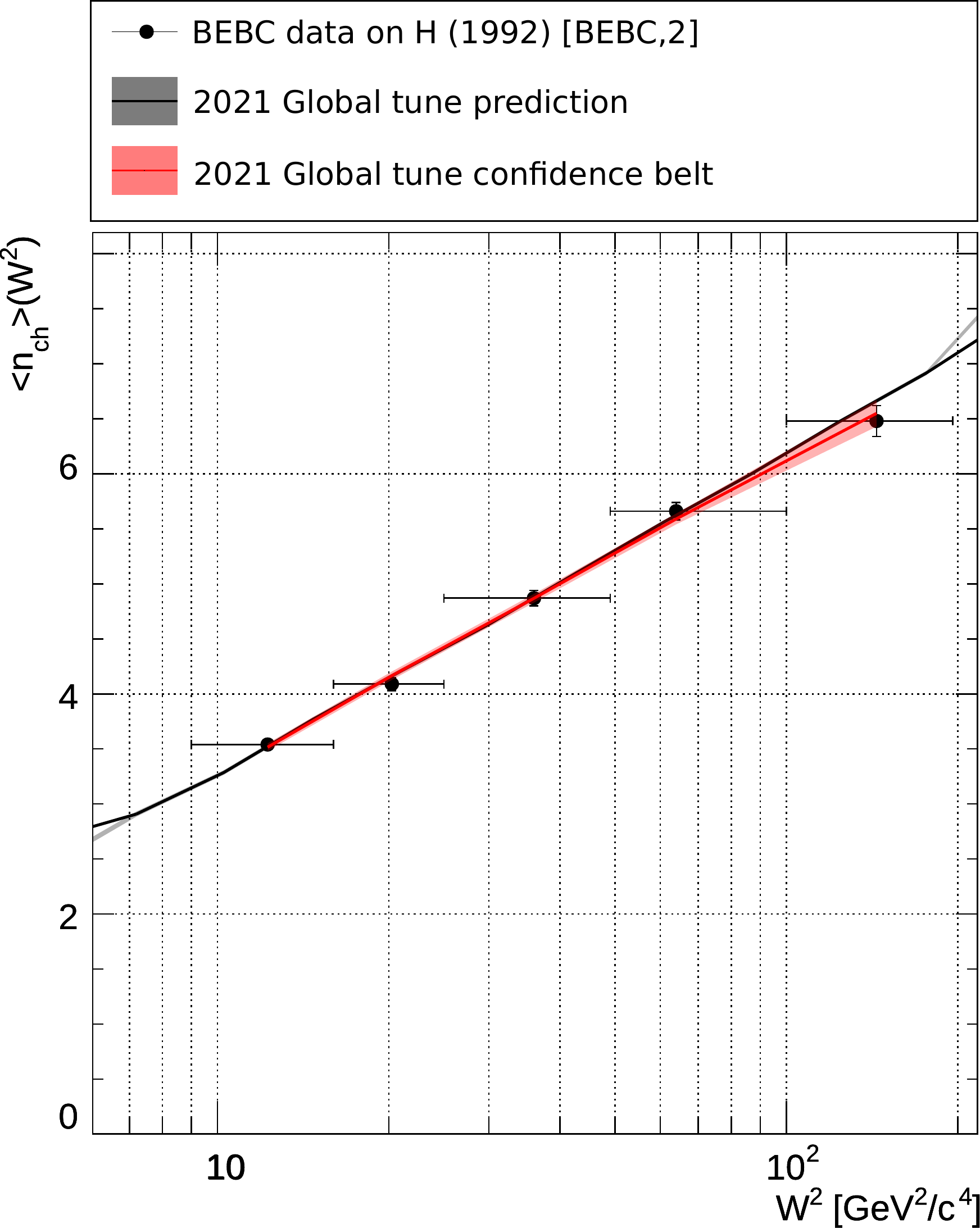}
        \caption{Dataset with good interpolations.}
    \end{subfigure}
    \begin{subfigure}{0.35\columnwidth}
        \begin{subfigure}{\columnwidth}
            \includegraphics[width=0.9\columnwidth]{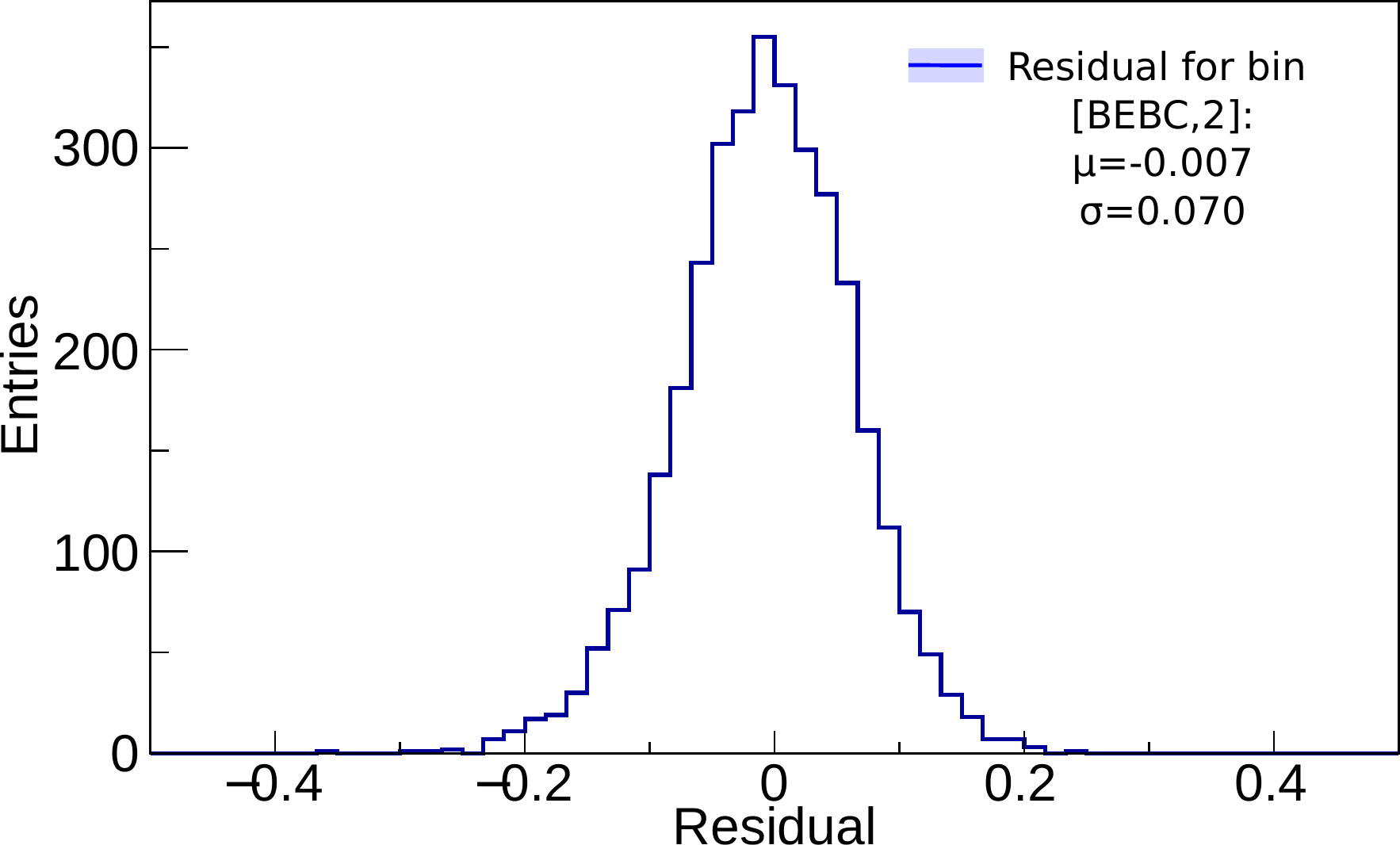}
            \caption{Accepted central value residual distribution.}
        \end{subfigure}
        
        \begin{subfigure}{\columnwidth}
            \includegraphics[width=0.8\columnwidth]{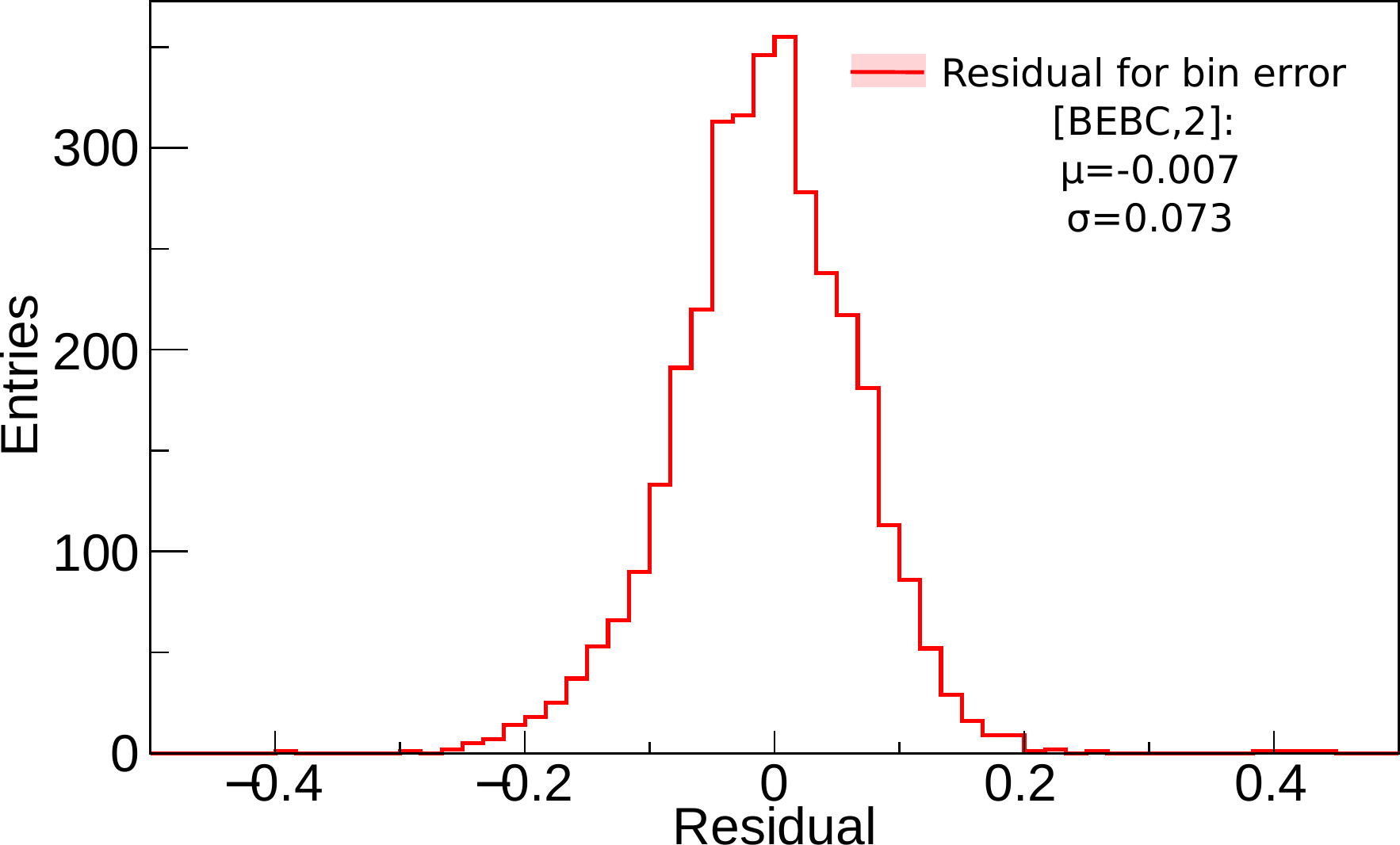}
            \caption{Accepted error residual distribution.}
        \end{subfigure}
    \end{subfigure}
    
    \bigskip
    
    \begin{subfigure}{0.35\columnwidth}
        \includegraphics[width=0.9\columnwidth]{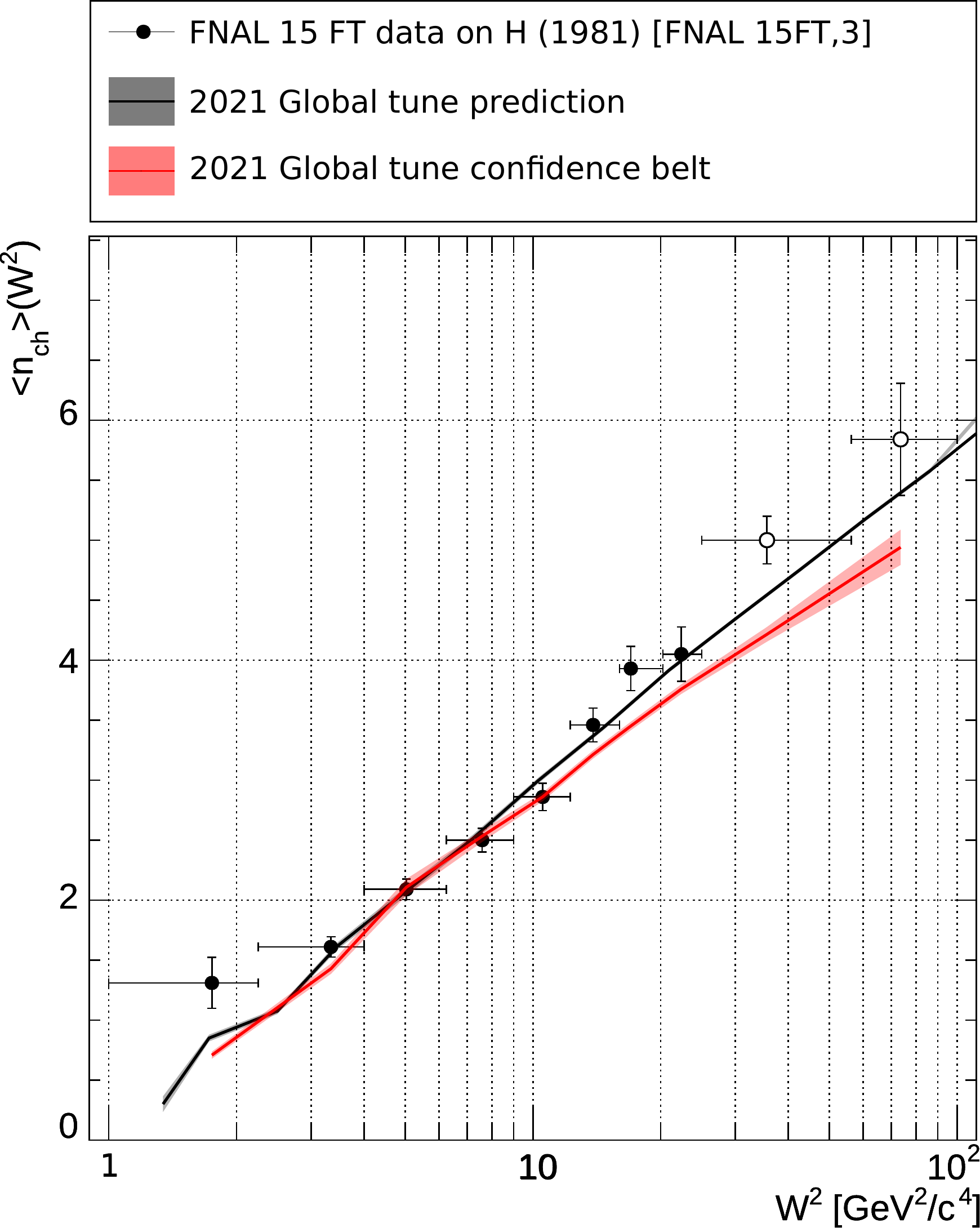}
        \caption{Dataset with some rejected interpolations.}
    \end{subfigure}
    \begin{subfigure}{0.35\columnwidth}
        \begin{subfigure}{\columnwidth}
            \includegraphics[width=0.9\columnwidth]{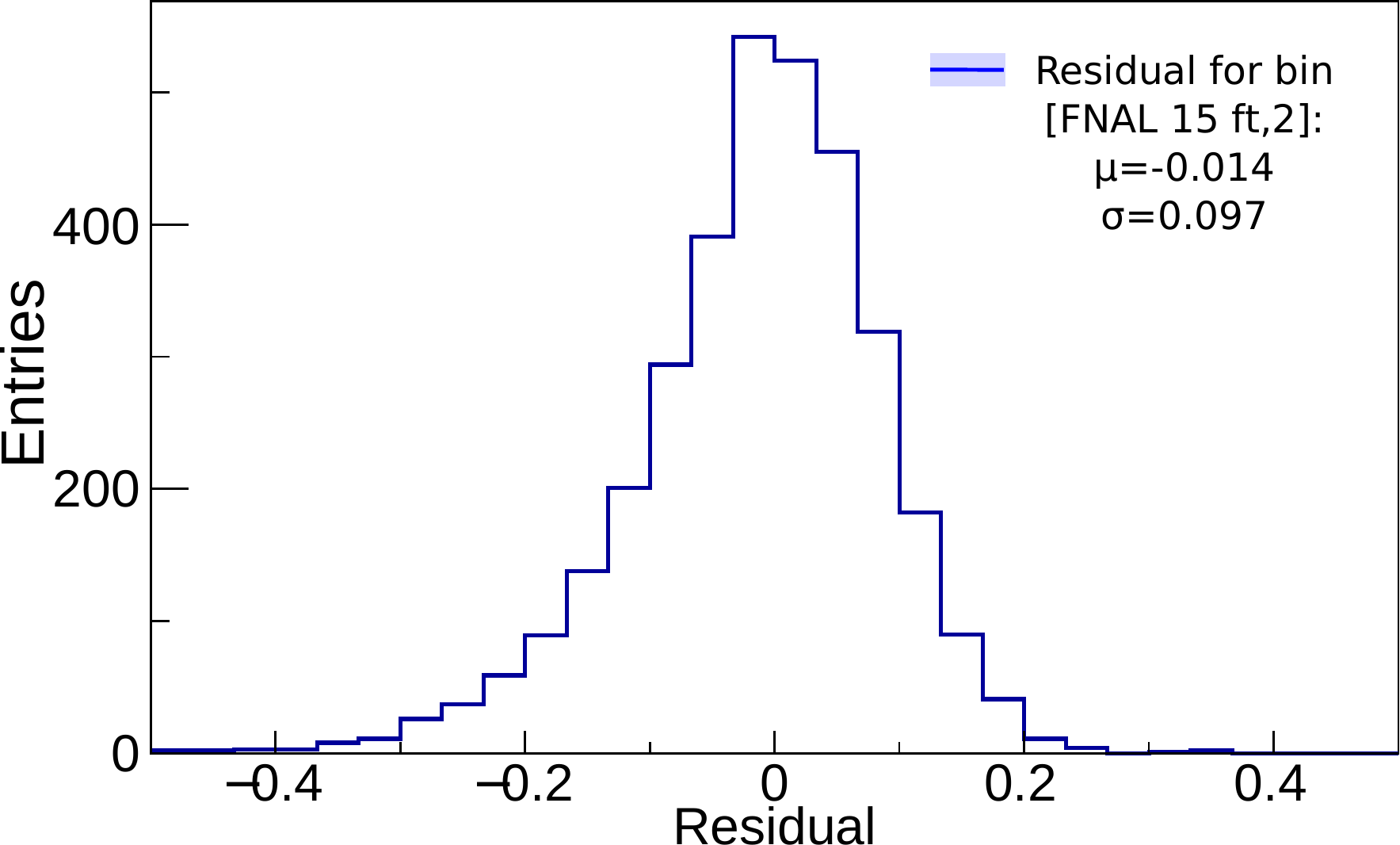}
            \caption{Rejected central value residual distribution.}
        \end{subfigure}
        
        \begin{subfigure}{\columnwidth}
            \includegraphics[width=0.8\columnwidth]{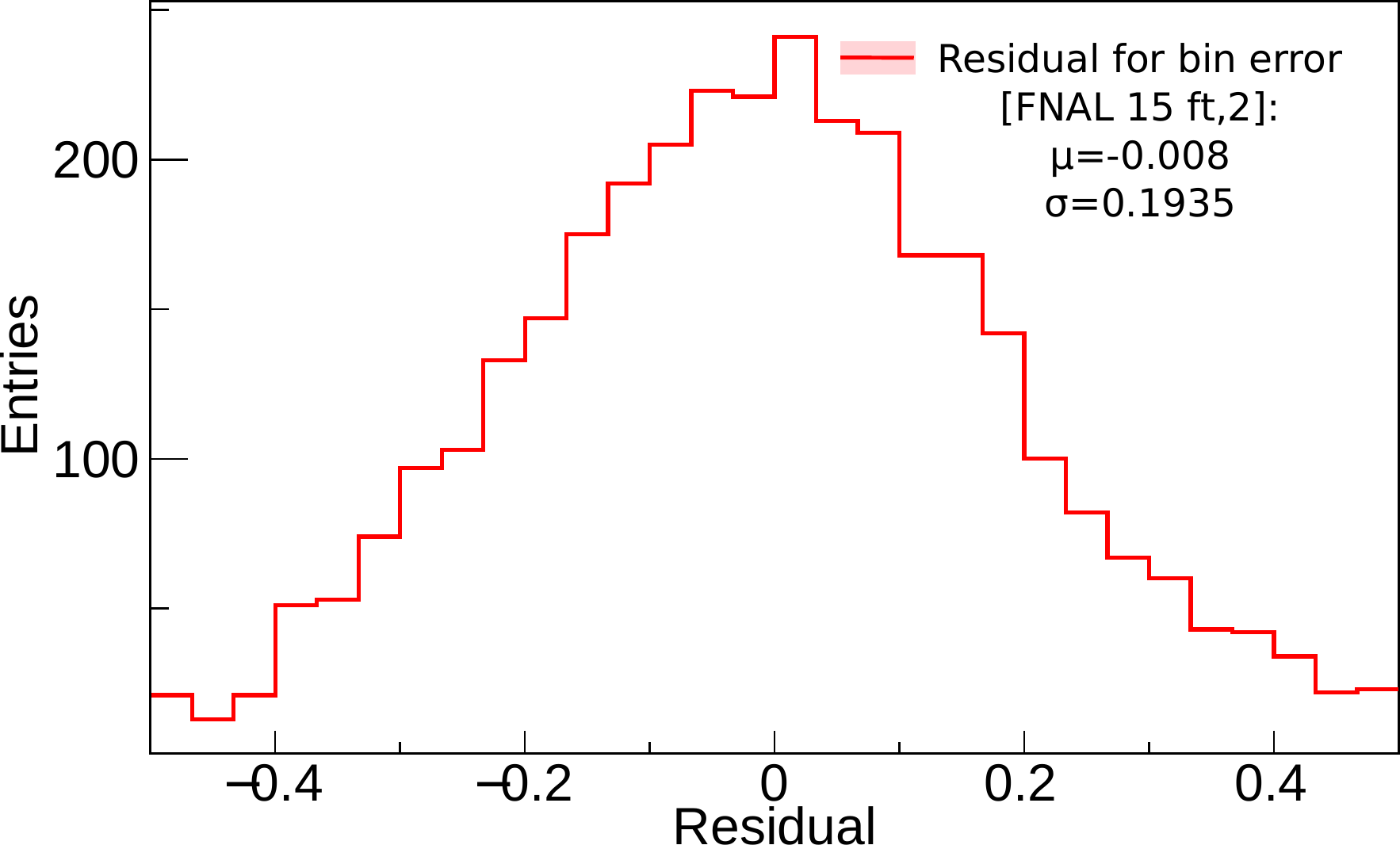}
            \caption{Rejected error residual distribution.}
        \end{subfigure}
    \end{subfigure}
    \caption{Comparisons of $\nu_\mu$ data on H against predictions obtained from the Professor parametrisation (red confidence belt) at the best-fit value for the AGKY global tune and the actual GENIE prediction (black line).
    Datapoints that do not satisfy our selection criteria are shown in empty markers.
    For the Professor parameterisation, the uncertainties of the tuned parameters are propagated to the prediction considering the full covariance matrix.
    For three selected bins (the bin with highest $W$ for [BEBC,2] (b-c) and the two higher $W$ bins for the [FNAL 15FT,3] plot (e-f) on the left column), the central values and error residual distributions are shown, blue and red respectively: accepted parameterisations at the top, rejected parameterisation at the bottom. 
    It can be seen that the residual distribution of the rejected bins is wider than its accepted counterpart.
    In this particular case, the two data points with higher $W$ are neglected as the parameterisation of the bin value and error do not satisfy the required criteria.
 \label{fig:ProfessorPred} }
\end{figure*}

The corresponding distribution associated to the bin central value and bin error residual for the last bin of the BEBC~\cite{BEBC1992} and the two lasts bins of the FNAL 15 ft~\cite{PhysRevD.27.47} datasets are shown in Fig.~\ref{fig:ProfessorPred} (b-c) and (e-f) respectively.
Two examples are given in Fig.~\ref{fig:ProfessorPred}: a dataset in which the interpolation is accurate for all the $W^2$ range and a dataset in which the interpolation fails for some of the dataset points, highlighted with empty markers.
This criteria allows us to ensure that the Professor parameterisation does not fail for the data considered in the tune.
A total of $\sim18\%$ of the data points have been removed due to this requirement.
In this work, it has been observed that the residual variance increases with $W^2$, with few exceptions.
The complete list of removed datapoints is specified in Sec.~\ref{sec:AGKYTune}.

The variance of the residual distribution for a given data point can be improved by increasing the order of the polynomial used for the Professor interpolation.
In this case, a polynomial of order four is used.
However, specifically in this particular tune where 13 parameters are tuned, an increase of the order requires the generation of a much higher number of MC samples, which can be very computationally demanding.

\subsection{Parameter priors}
\label{sec:tunepriors}
Our parameters of interest affect other hadronization observables and not only the averaged multiplicity.
This is taken into account by using Gaussian priors. 

The $s\bar{s}$ suppression factor not only impacts the averaged multiplicity data but also the $\eta$ multiplicity production, see Fig.~\ref{fig:ssProd}. 
A prior of $0.30\pm0.05$ is considered in order to preserve a good agreement with the SKAT data~\cite{Agababyan:2008gg}.

\begin{figure}
    \centering
    \begin{subfigure}[t]{\columnwidth}
        \centering
        \includegraphics[width=0.8\columnwidth]{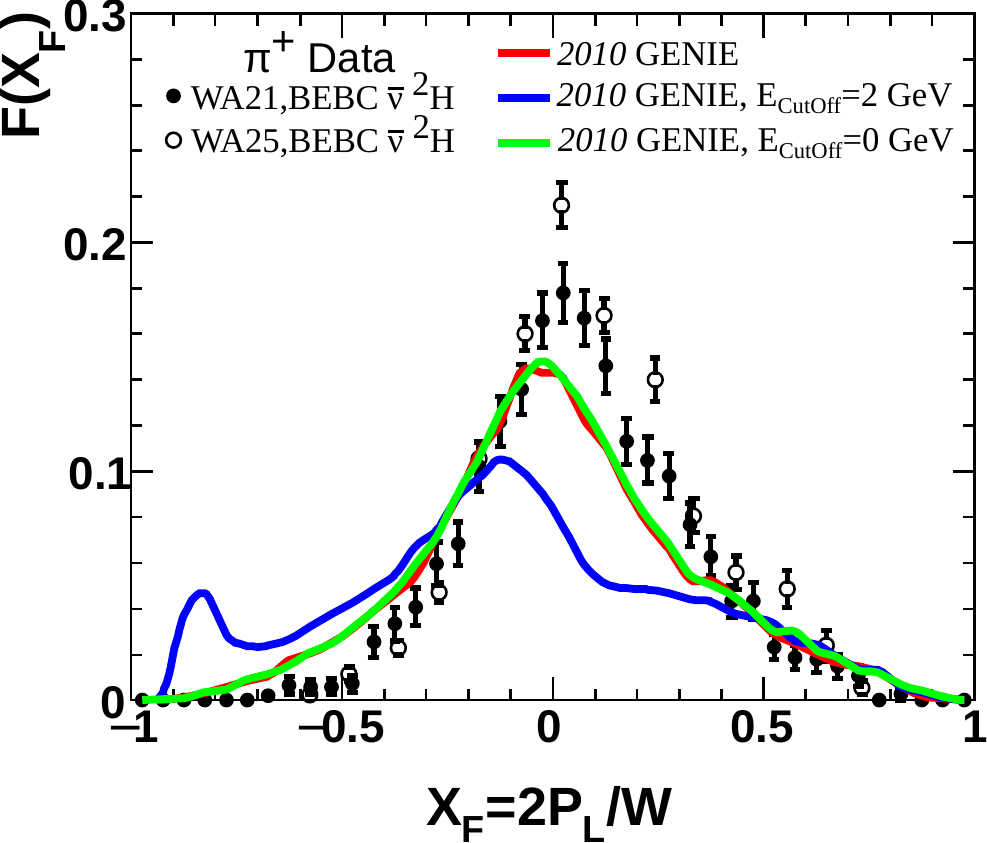}
    \caption{}\label{fig:xf_positive}
    \end{subfigure}
    \\
    \begin{subfigure}[t]{\columnwidth}
        \centering
        \includegraphics[width=0.8\columnwidth]{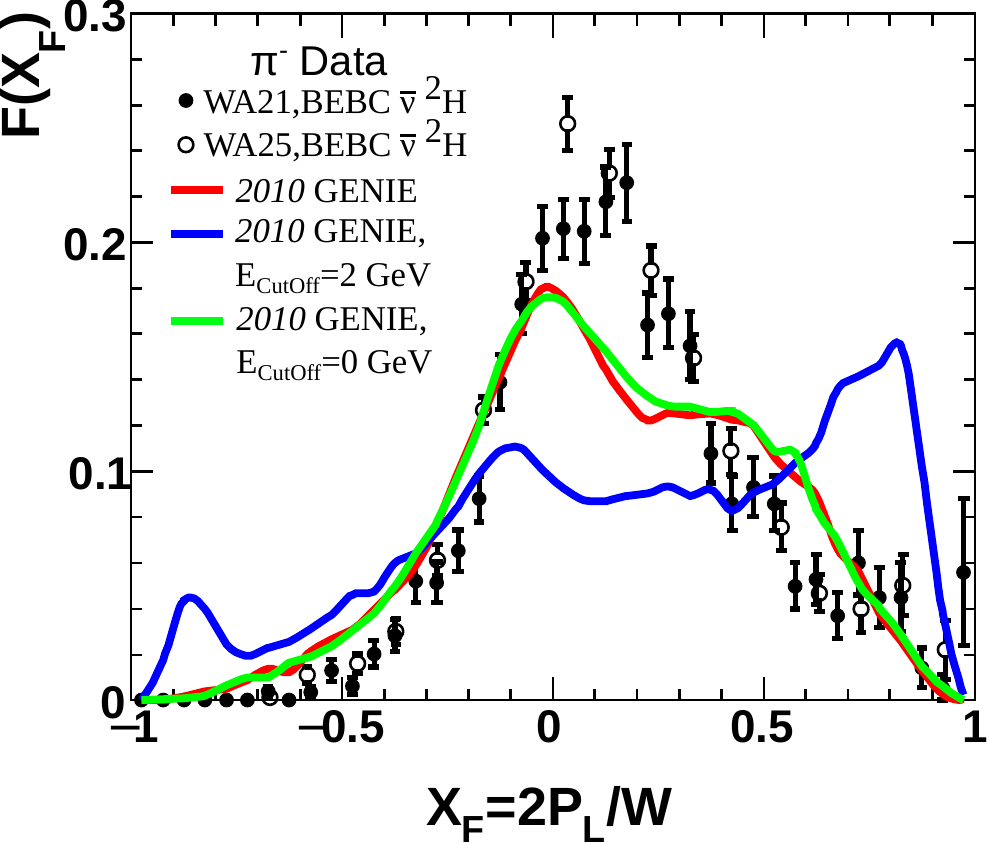}
        \caption{}
        \label{fig:xf_negative}
    \end{subfigure} 
    \caption{Effect of $E_{\text{CutOff}}$ on the $x_{\text{F}}$ invariant distributions for $\pi^+$ (a) and $\pi^-$ (b) in $\bar{\nu}_\mu$ data on $^2$H from the \ac{BEBC} experiment~\cite{Allasia1984}. The \emph{2010} GENIE tune value for the energy cut-off is $E_{\text{CutOff}}=0.2$ GeV.}
    \label{fig:xf_distribution}
\end{figure}

Variations of $E_{\text{CutOff}}$ affect the shape of $F(x_{\text{F}})$ invariant distribution, defined as:
\begin{equation}
    F(x_{\text{F}}) = \frac{1}{N_{\text{ev}}}\cdot\frac{1}{\pi}\cdot\frac{E}{p^{L\,\max}c}\cdot\frac{dN}{dx_{\text{F}}},
\end{equation}
where $x_{\text{F}}$ is the Feynman variable, $N_{\text{ev}}$ the total number of events, and $E$ and $p^{L\,\max}$ the energy and maximum longitudinal momentum of the final state hadron in the hadronic center of mass respectively. 
The $F(x_{\text{F}})$ invariant distribution describes the fragmentation process for the forward and backward hemispheres and it allows to study the symmetry between this two fragmentation regions.
In Fig.~\ref{fig:xf_distribution}, the $F(x_{\text{F}})$ invariant distribution for $\bar{\nu}_\mu$ data on $^2$H~\cite{Allasia1984} is compared against GENIE predictions obtained by varying the $E_{\text{CutOff}}$ within a $[0,2]$~GeV range.
The main conclusion is that small values of this parameter preserve the agreement with data.
In order to avoid an increase of $F(x_{\text{F}})$ at $|x_{\text{F}}|\sim 1$, we apply a prior on $E_{\text{CutOff}}$ of $0.25\pm0.05$~GeV. 

Another parameter that has a strong impact on other observables is  $\langle p_\bot^2 \rangle$. 
As demonstrated in Fig.~\ref{fig:pt2validation}, low values of $\langle p_\bot^2 \rangle$ are not in agreement with data for $\langle p_{T}^2\rangle$ distributions.
Thus, we also apply a prior on the parameter to guarantee the agreement with this data of $0.44\pm0.05~(\text{GeV}/c)^2$. 
No priors are applied to the remaining parameters.

\subsection{Final form of the $\chi^2$}

Using  the  parameterisation  and  the  corresponding  set  of points belonging to the $i$-th dataset, ${\mathcal{D}_{ij}\pm\delta\mathcal{D}_{ij}}$, we seek to estimate the best-fit parameters $\hat{\boldsymbol{\theta}}$ by minimising the quantity:
\begin{equation}
\chi^2 (\boldsymbol{\theta}) = 
  \sum_{i,j} w_{ij} \frac{ \left( \widetilde{n}_{ij}(\boldsymbol{\theta}) -\mathcal{D}_{ij} \right)^2}{ \widetilde{\sigma}^2_{ij}(\boldsymbol{\theta}) + \delta\mathcal{D}_{ij}^2   } 
   + \sum_{l} \frac{(\theta_{l}-\mu_l)^2}{\sigma^2_l}.
  \label{eq:chi2}
\end{equation}
The first term allows the minimization between data and prediction while applying weights, $w_{ij}$, that allow to consider only specific data points in the fit. 
The second term adds uncorrelated Gaussian priors for a given parameter; the vectors of central values and variances are denoted $\mu_l$ and $\sigma_l$ respectively. 

\section{AGKY tune results}
\label{sec:AGKYTune}    

Starting from $\nu_\mu$ and $\bar{\nu}_\mu$ hadronization data, two tunes were considered: a global tune (\emph{2021} GENIE global) and a deuterium only tune (\emph{2021} $^{2}$H).
The reason for a deuterium only fit is because other studies showed tensions between data on H and $^2$H targets on bubble chamber experiments~\cite{Kuzmin_2013}. 
The goal of the global tune is to improve the agreement with hydrogen and deuterium targets, regardless of these tensions, while the deuterium only tune was performed to quantify the tensions within the same framework. 
An hydrogen only fit was not considered because it could not constrain the neutron related parameters of the low-$W$ empirical model. 

The analyses procedure outlined in the previous sections was applied to both tunes.
The likelihood function, Eq.~\ref{eq:chi2}, was minimized against the averaged charged multiplicity data that satisfies our selection criteria, see Sec.~\ref{sec:dataanalysesBBCH}. 
The best-fit parameter set for both tunes and the $\chi^2$ values obtained using the Professor parameterisations and Eq.~\ref{eq:chi2} are summarised in Tab.~\ref{tab:fitResultParameters}.

\begin{table*}   
    \centering
    \begin{tabular}{@{\extracolsep\fill} c c c c c c} \hline\hline\noalign{\smallskip}
    Parameter & GENIE parameter name & \emph{2010} GENIE\,\, & Allowed range\,\, & \emph{2021} Global Fit\,\, & \emph{2021} $^2$H Fit \\
    \noalign{\smallskip}\hline\hline\noalign{\smallskip}
    \multicolumn{6}{c}{ Low-$W$ empirical model } \\ 
    \noalign{\smallskip}\hline\hline\noalign{\smallskip} 
    $\alpha_{\nu p}$                             & \tt{KNO-Alpha-vp}                  &  0.40 & $[-1.0, 2.0]$ & $1.1\pm0.3$            & $1.2\pm0.4$          \\ \noalign{\smallskip}
    $\alpha_{\nu n}$                             & \tt{KNO-Alpha-vn}                  & -0.20 & $[-1.0, 2.0]$ & $1.75^{+0.14}_{-0.11}$ & $-0.58\pm0.07$       \\ \noalign{\smallskip}
    $\alpha_{\bar{\nu}p}$                        & \tt{KNO-Alpha-vbp}                 &  0.02 & $[-1.0, 2.0]$ & $1.32^{+0.16}_{-0.14}$ & $1.9\pm0.08$         \\ \noalign{\smallskip}
    $\alpha_{\bar{\nu}n}$                        & \tt{KNO-Alpha-vbn}                 &  0.80 & $[-1.0, 2.0]$ & $1.11\pm0.09$          & $1.07\pm0.3$         \\ \noalign{\smallskip}
    $\beta_{\nu p}$                              & \tt{KNO-Beta-vp}                   &  1.42 & $[ 0.0, 2.5]$ & $0.79\pm0.15$          & $0.9\pm0.3$          \\ \noalign{\smallskip}
    $\beta_{\nu n}$                              & \tt{KNO-Beta-vn}                   &  1.42 & $[ 0.0, 2.5]$ & $0.5\pm0.1$            & $1.9\pm0.3$          \\ \noalign{\smallskip}
    $\beta_{\bar{\nu}p}$                         & \tt{KNO-Beta-vbp}                  &  1.28 & $[ 0.0, 2.5]$ & $0.8\pm0.1$            & $0.3\pm0.1$          \\ \noalign{\smallskip}
    $\beta_{\bar{\nu}n}$                         & \tt{KNO-Beta-vbn}                  &  0.95 & $[ 0.0, 2.5]$ & $0.88^{+0.09}_{-0.08}$ & $0.9\pm0.2$          \\ \noalign{\smallskip}
    \noalign{\smallskip}\hline\hline\noalign{\smallskip}
    \multicolumn{6}{c}{ PYTHIA } \\ 
    \noalign{\smallskip}\hline\hline\noalign{\smallskip}
    $P_{s\bar{s}}$                               & \tt{PYTHIA-SSBarSuppression}       &  0.30 & $[ 0.0, 1.0]$ & $0.27\pm0.04$          & $0.29\pm0.05$        \\ \noalign{\smallskip}
    $\langle p_\bot^2 \rangle$ [GeV$^2/c^2$] & \tt{PYTHIA-GaussianPt2}            &  0.44 & $[ 0.1, 0.7]$ & $0.46\pm0.05$          & $0.43\pm0.04$        \\ \noalign{\smallskip}
    $E_{\text{CutOff}}$ [GeV]                  & \tt{PYTHIA-RemainingEnergyCutoff}  &  0.20 & $[ 0.0, 1.0]$ & $0.30\pm0.04$          & $0.24\pm0.05$        \\ \noalign{\smallskip}
    Lund $a$                                  & \tt{PYTHIA-Lunda}                  & 0.30  & $[ 0.0, 2.0]$ & $1.53\pm0.13$          & $1.85\pm0.15$        \\ \noalign{\smallskip}
    Lund $b$ [c$^4$/GeV$^2$]                      & \tt{PYTHIA-Lundb}                  & 0.58  & $[ 0.0, 1.5]$ & $1.16\pm0.09$          & $1.0\pm0.2$          \\ \noalign{\smallskip}
    \noalign{\smallskip}\hline\hline \noalign{\smallskip}
                                                 &                                    &       & $\chi^2 = $   & $87.9/62~\text{DoF}$   & $29.5/32~\text{DoF}$ \\
    \noalign{\smallskip}\hline\hline
    \end{tabular}
     \caption{Best fit result parameters for the AGKY \emph{2021} global tune and $^{2}$H only tune.
              The range of study and priors used in the tune are specified in the table. 
              The \emph{2010} GENIE AGKY parameter values are also specified for reference, as well as the parameter name used in the GENIE software.
              See Sec.~\ref{subsec:AGKYGlobalTune} and Sec.~\ref{subsec:AGKY2HTune} for the details on the error estimation of each tune.
              Posterior distributions are not always symmetric: in that case the interval is reported accordingly.
              The total $\chi^2$ obtained from each fit is obtained from the minimization of Eq.~\ref{eq:chi2}.
             }
    \label{tab:fitParameters}
    \label{tab:fitResultParameters}
\end{table*}

\begin{figure*}
    \centering
    \begin{subfigure}{0.4\textwidth}
        \centering\includegraphics[width=0.9\columnwidth]{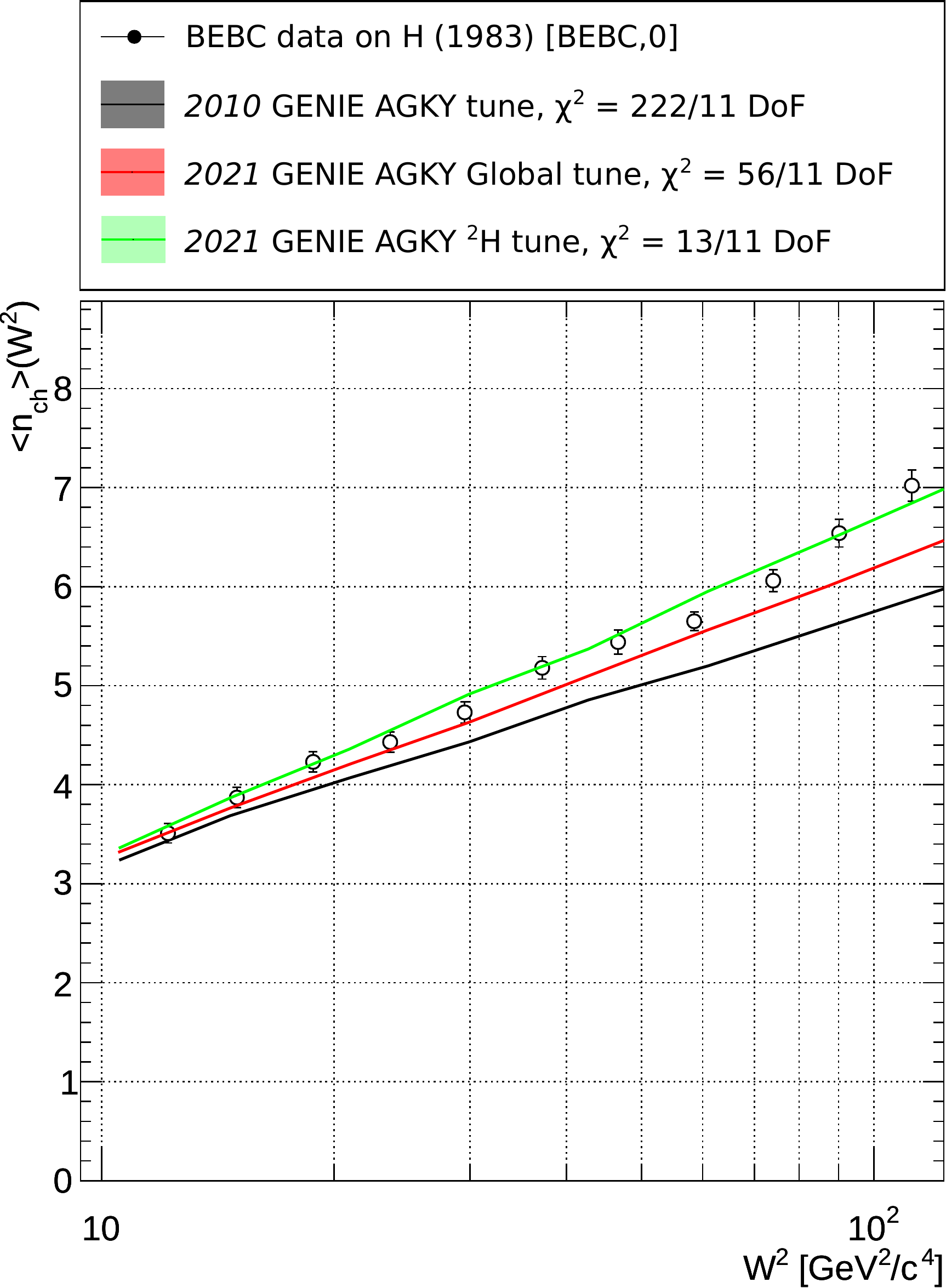}
        \caption{}
    \end{subfigure} 
    \begin{subfigure}{0.4\textwidth}
        \centering\includegraphics[width=0.9\columnwidth]{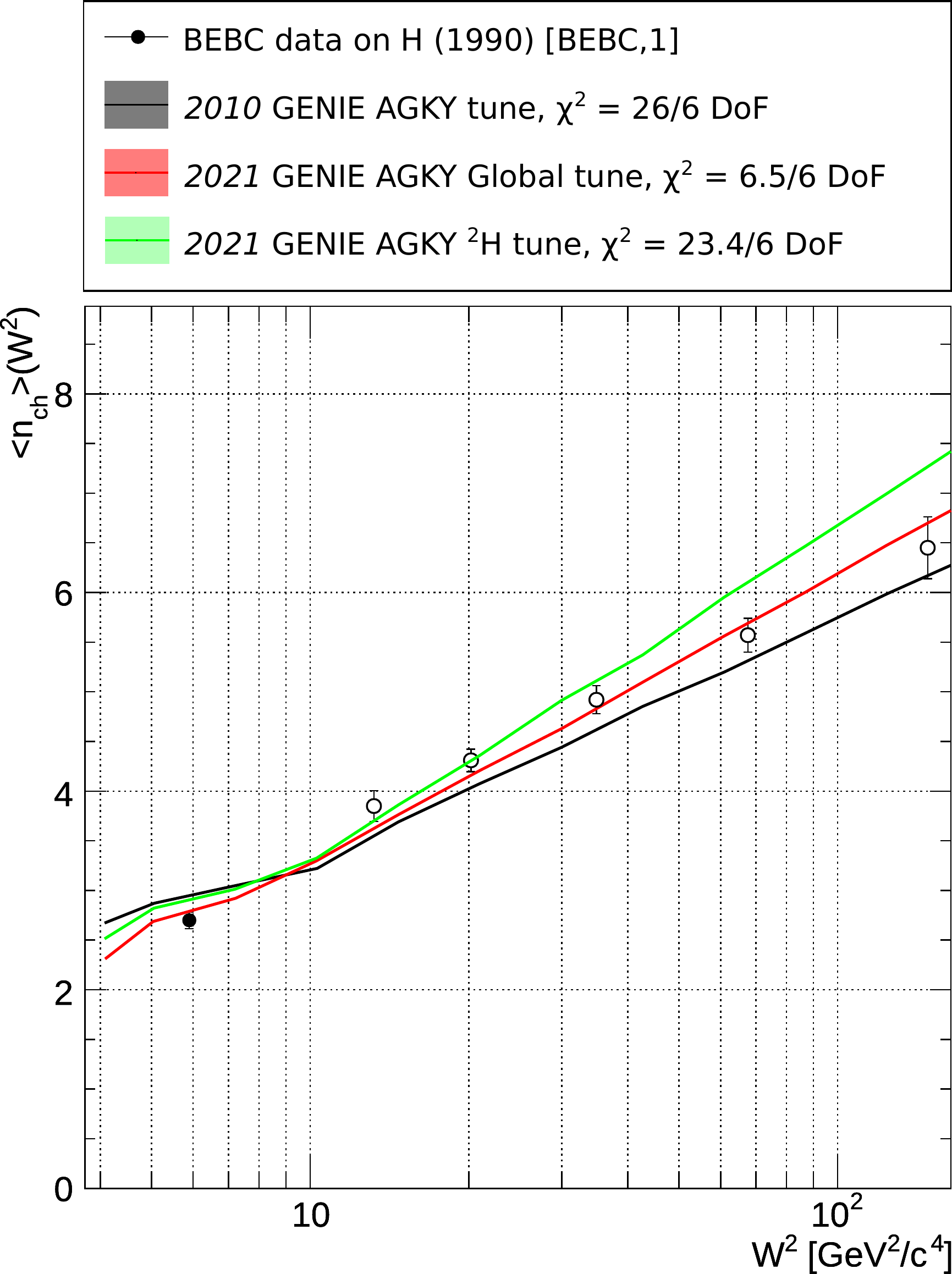}
         \caption{}
    \end{subfigure} 
        \\
    \begin{subfigure}{0.4\textwidth}
        \centering\includegraphics[width=0.9\columnwidth]{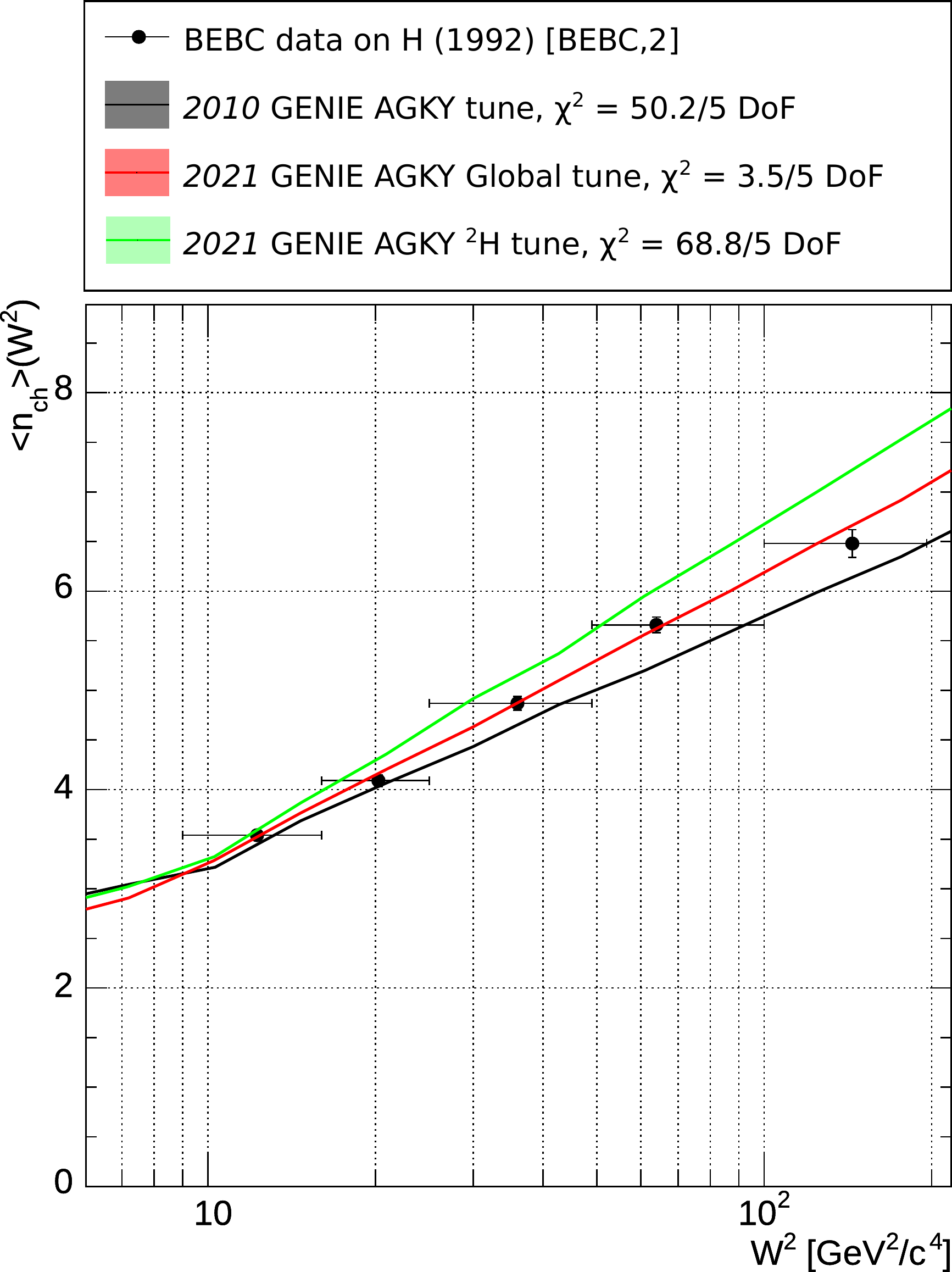}
        \caption{}
    \end{subfigure} 
    \begin{subfigure}{0.4\textwidth}
        \centering\includegraphics[width=0.9\columnwidth]{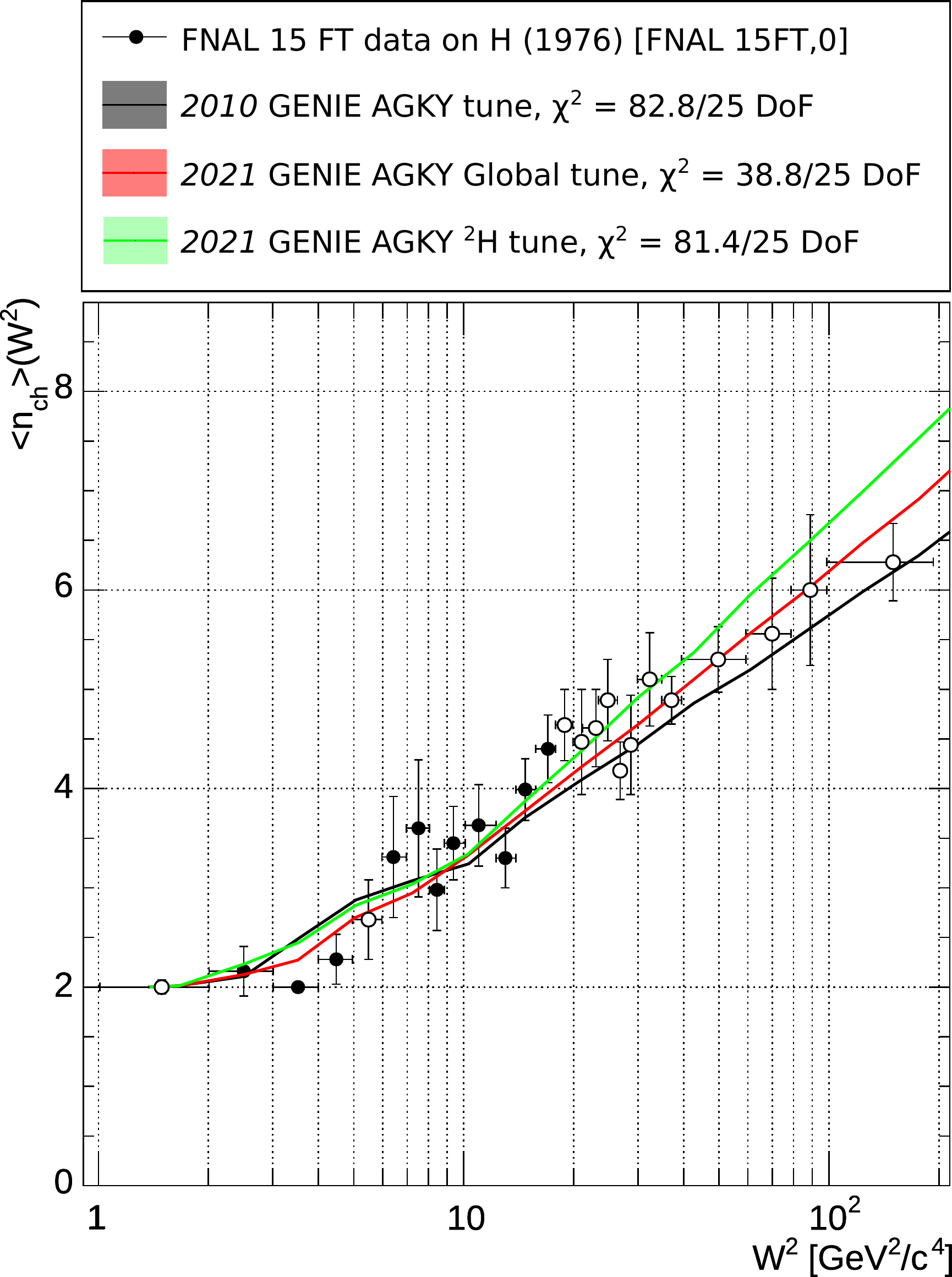}
        \caption{}
    \end{subfigure}
    \caption{Comparison of $\langle n_{\text{ch}}\rangle$ against neutrino-induced hadronization data on $\nu_\mu+p$ interactions on H from \ac{BEBC}~\cite{GRASSLER1983269,Jones1990,BEBC1992} and FNAL~\cite{PhysRevLett.36.124} bubble chamber experiments filled with H. 
    Datapoints used in the AGKY \emph{2021} global tune analysis are shown as filled black markers. 
    Discarded datapoints are represented using empty markers.
    The $^{2}$H tune prediction is shown for comparison only. The predictions are computed using the parameters specified in Tab.~\ref{tab:fitResultParameters}. The $\chi^2$ values are calculated against all the data from each experiment.  See definition of \emph{Tags} in Tab.~\ref{tab:references}. }
    \label{fig:pPredictions}
\end{figure*}

\begin{figure*}
    \centering
    \begin{subfigure}{0.3\textwidth}
        \centering\includegraphics[width=\columnwidth]{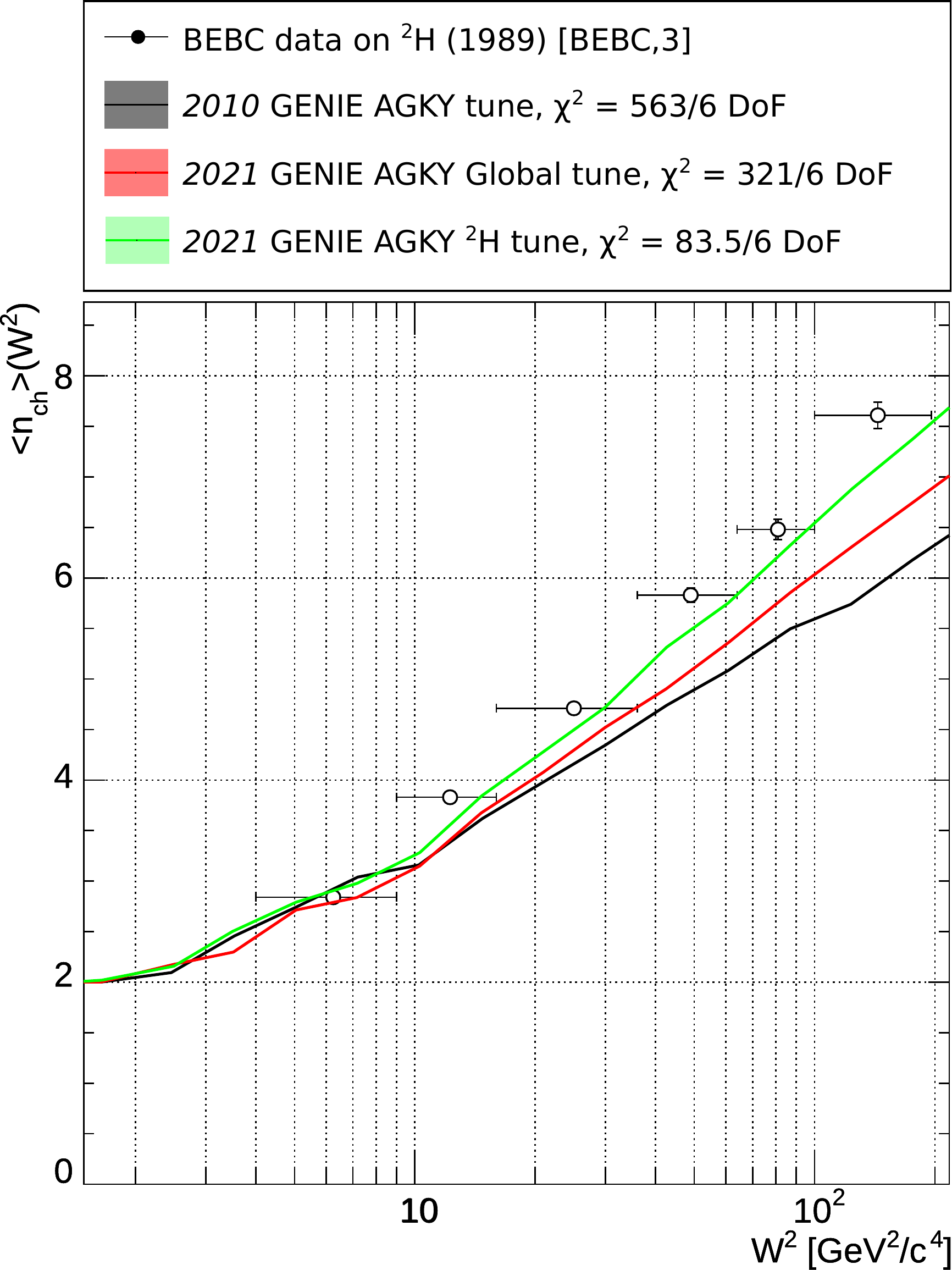}
        \caption{$\nu_\mu+p\rightarrow\mu^-X^{++}$ }   
    \end{subfigure} \,\,\,
    \begin{subfigure}{0.3\textwidth}
        \centering\includegraphics[width=\columnwidth]{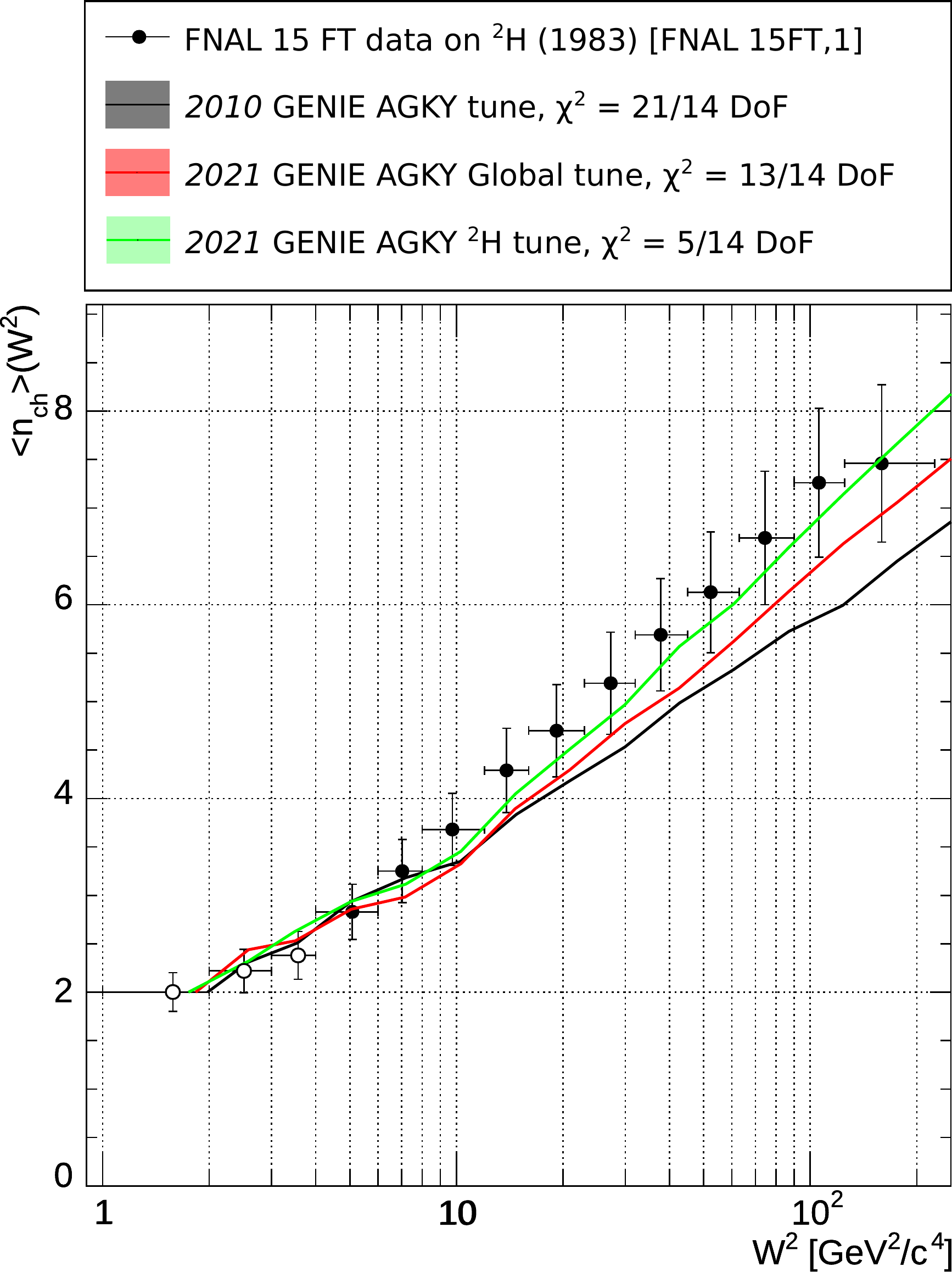}
        \caption{$\nu_\mu+p\rightarrow\mu^-X^{++}$}   
    \end{subfigure} \,\,\,
    \begin{subfigure}{0.3\textwidth}
        \centering\includegraphics[width=\columnwidth]{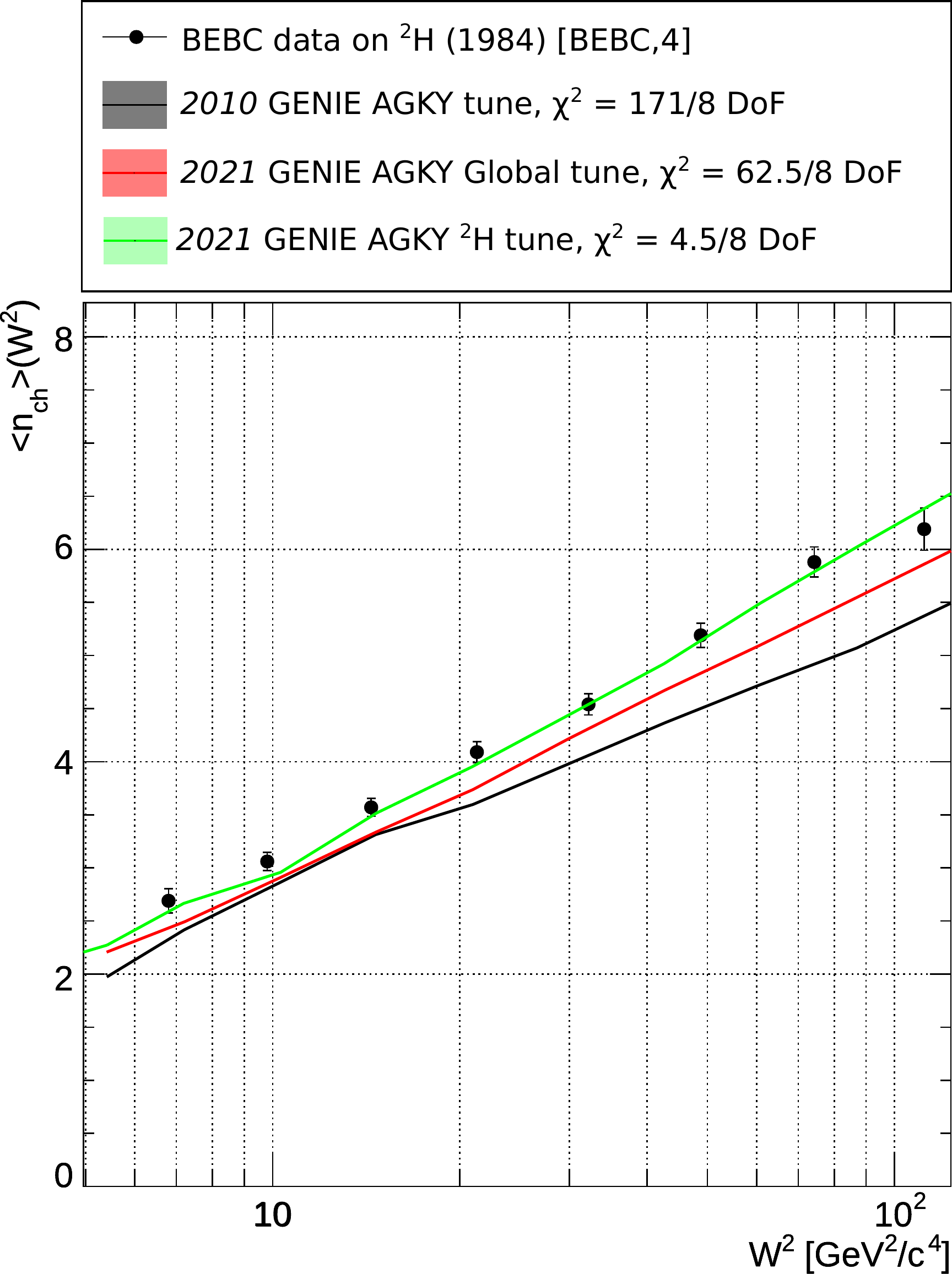}
        \caption{$\nu_\mu+n\rightarrow\mu^-X^+$ }   
    \end{subfigure} 
    
    \begin{subfigure}{0.3\textwidth}
        \centering\includegraphics[width=\columnwidth]{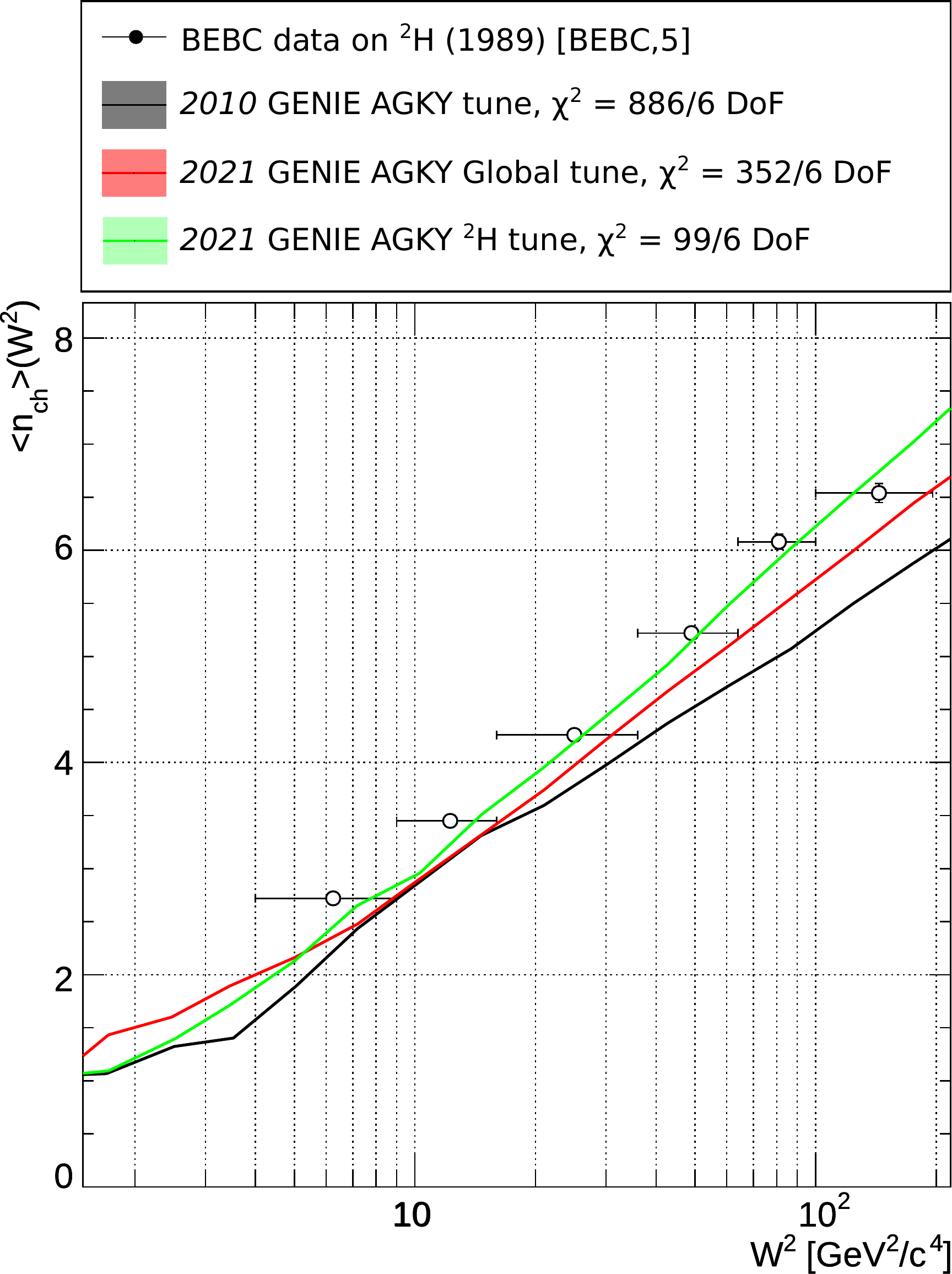}
        \caption{$\nu_\mu+n\rightarrow\mu^-X^+$ }   
    \end{subfigure} \,\,\,
    \begin{subfigure}{0.3\textwidth}
        \centering\includegraphics[width=\columnwidth]{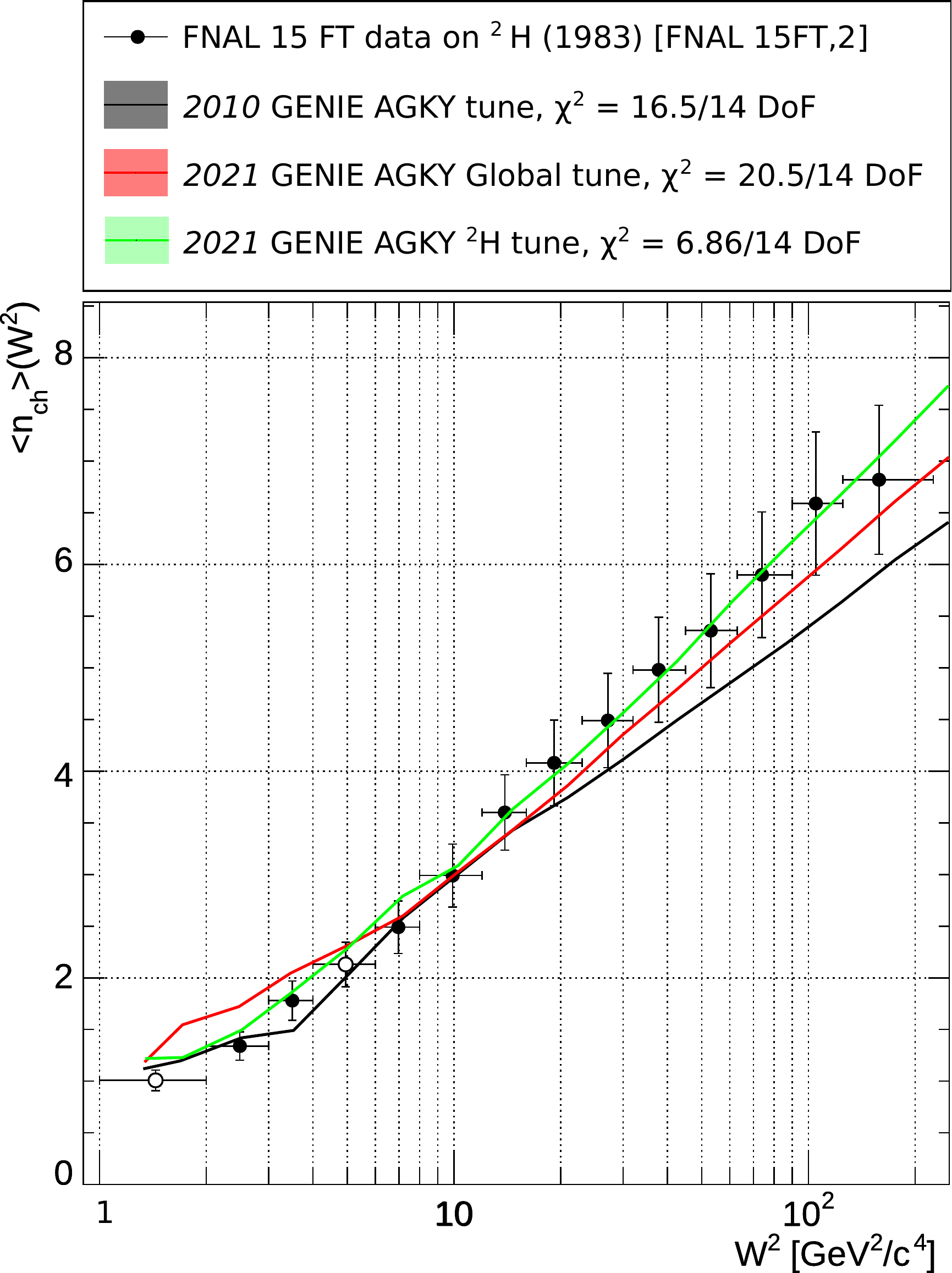}
        \caption{$\nu_\mu+n\rightarrow\mu^-X^+$ }   
    \end{subfigure} 
    \caption{Comparison of $\langle n_{\text{ch}}\rangle$ against neutrino-induced hadronization data on $\nu_\mu$ interactions on p and n from the \ac{BEBC} bubble chamber experiment filled with $^2$H~\cite{Barlag1982,Jongejans1989}.
    Datapoints used in the AGKY \emph{2021} global tune analysis are shown as filled black markers. 
    Discarded datapoints are represented using empty markers.
    The predictions are computed using the parameters specified in Tab.~\ref{tab:fitResultParameters}. The $\chi^2$ values are calculated against all the data from each experiment.  See definition of \emph{Tags} in Tab.~\ref{tab:references}.}
    \label{fig:2HPredictions}
\end{figure*}

\begin{figure*}
    \centering
    \begin{subfigure}{0.4\textwidth}
        \centering\includegraphics[width=0.85\columnwidth]{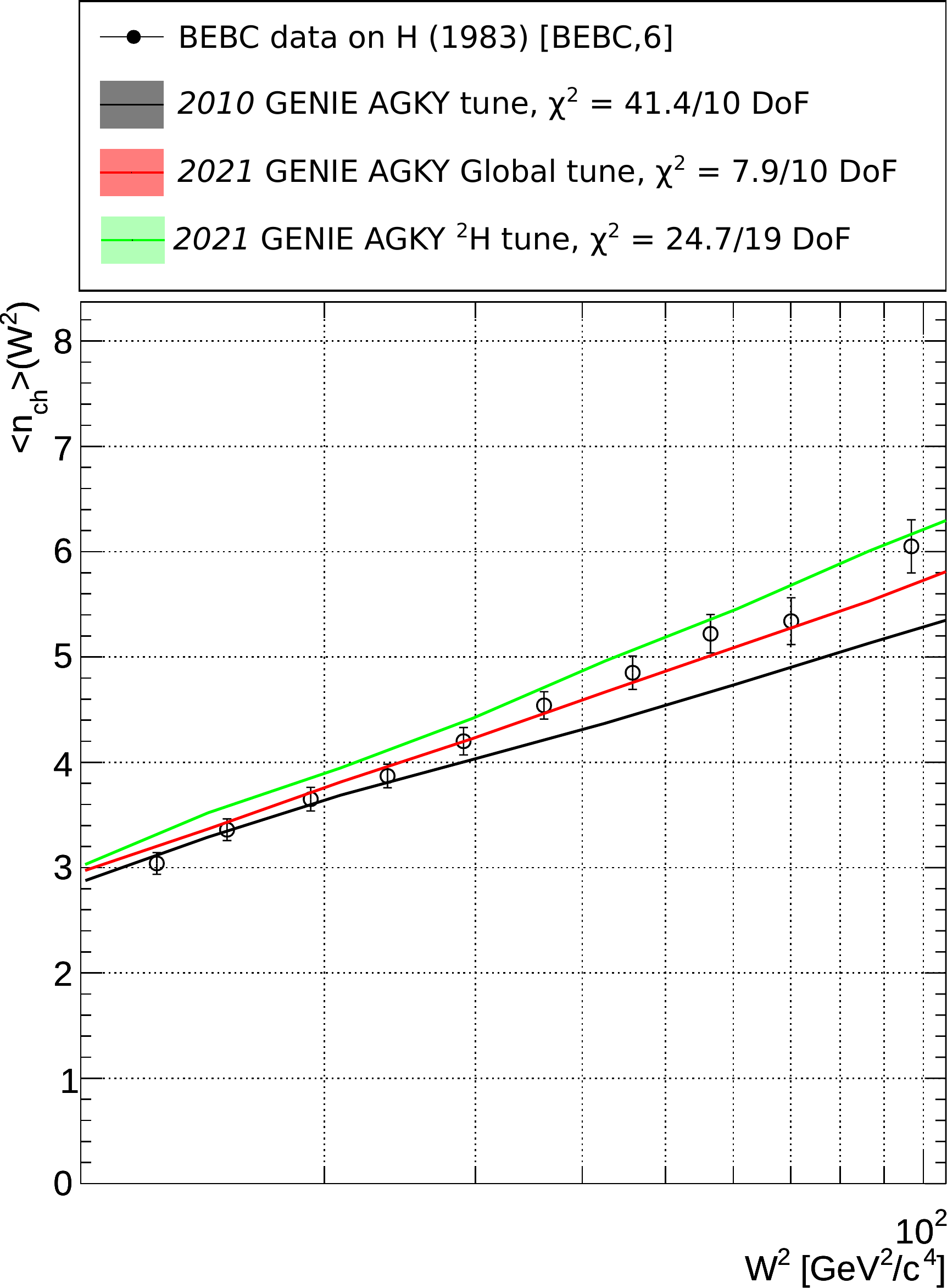}
        \caption{}
    \end{subfigure}
    \begin{subfigure}{0.4\textwidth}
        \centering\includegraphics[width=0.85\columnwidth]{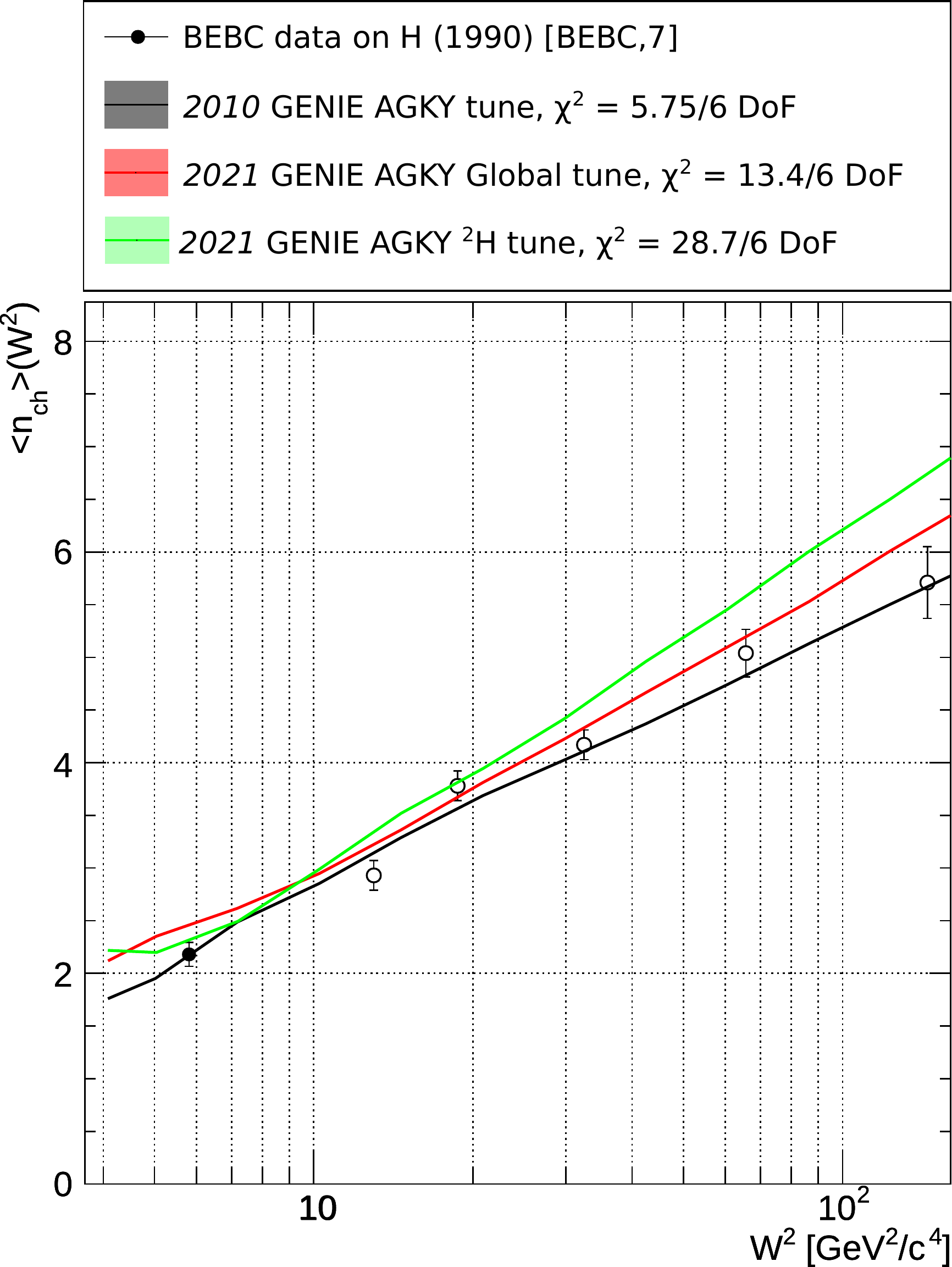}
        \caption{}
    \end{subfigure} 
        
    \begin{subfigure}{0.4\textwidth}
        \centering\includegraphics[width=0.85\columnwidth]{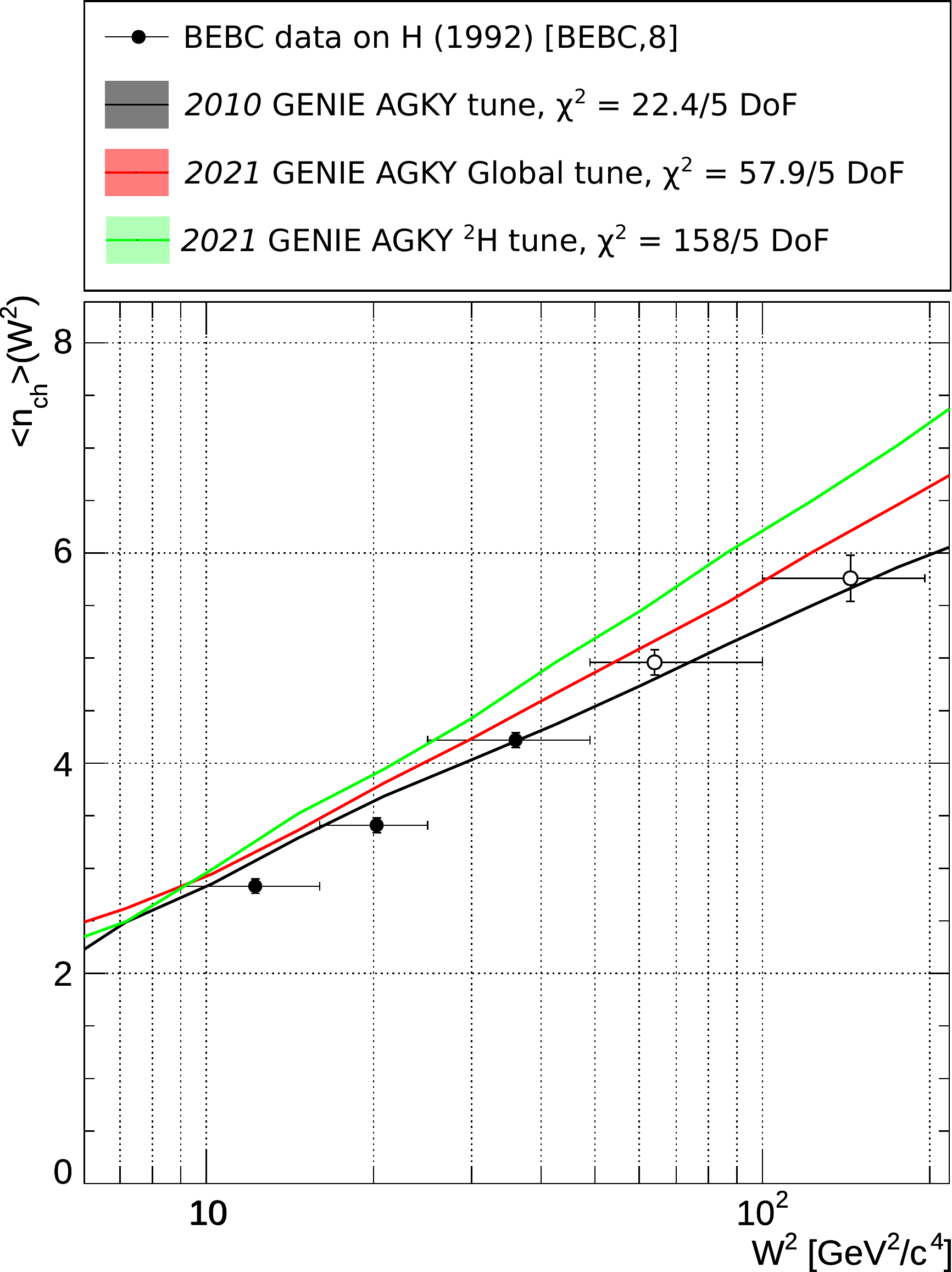}
        \caption{}
    \end{subfigure} 
    \begin{subfigure}{0.4\textwidth} \centering\includegraphics[width=0.85\columnwidth]{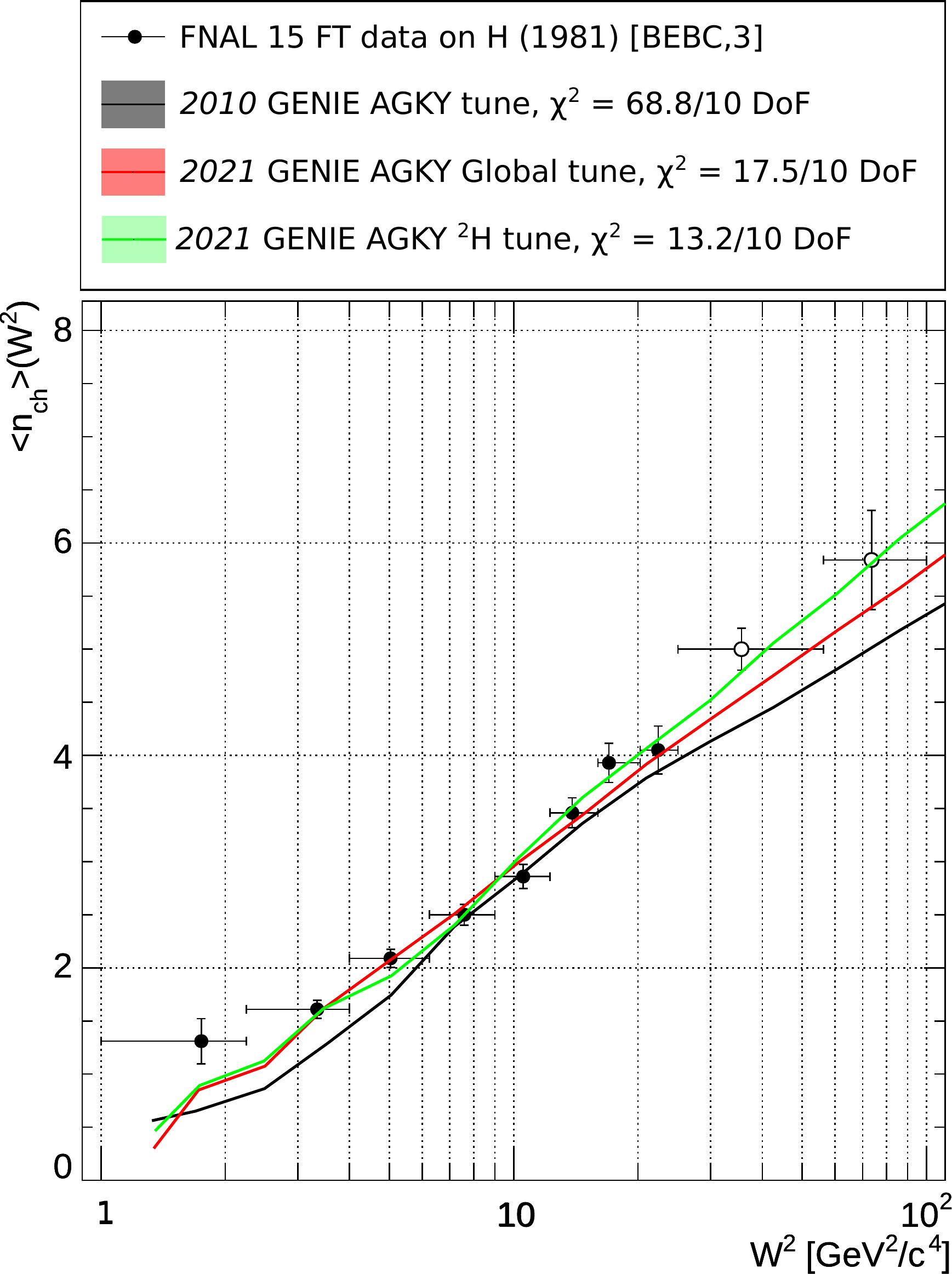}
        \caption{}
    \end{subfigure} 
    \caption{Comparison of $\langle n_{\text{ch}}\rangle$ against neutrino-induced hadronization data on $\bar{\nu}_\mu+p$ interactions on H from the \ac{BEBC}~\cite{GRASSLER1983269,Jones1990,BEBC1992} and FNAL~\cite{PhysRevD.25.624} bubble chamber experiment filled with H. 
    Datapoints used in the AGKY \emph{2021} global tune analysis are shown as filled black markers. 
    Discarded datapoints are represented using empty markers.
    The predictions are computed using the parameters specified in Tab.~\ref{tab:fitResultParameters}. The $\chi^2$ values are calculated against all the data from each experiment. See definition of \emph{Tags} in Tab.~\ref{tab:references}.}
    \label{fig:barpPredictions}
\end{figure*}

\begin{figure*}
    \centering
    \begin{subfigure}{0.4\textwidth}
        \centering\includegraphics[width=0.85\columnwidth]{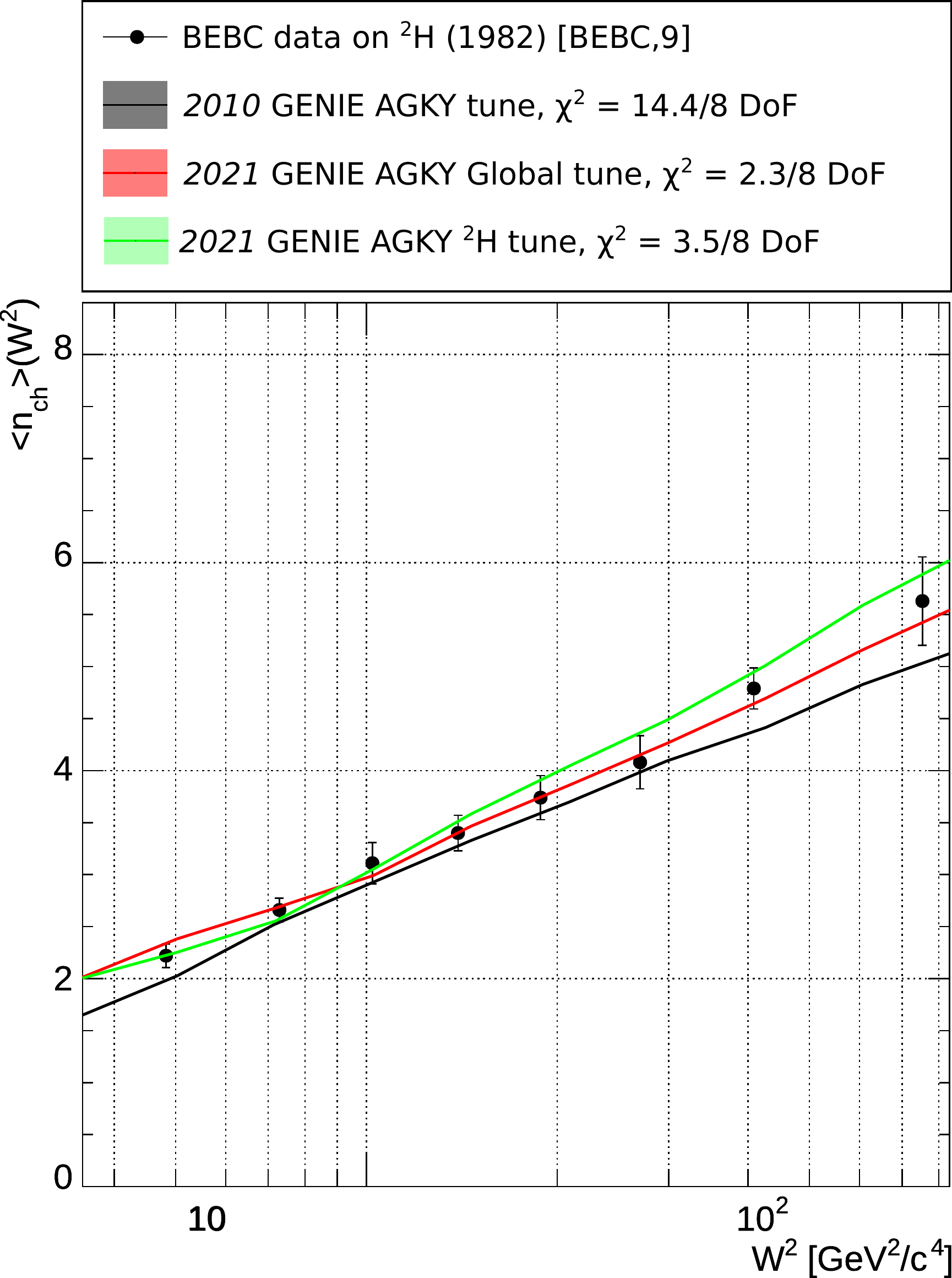}
        \caption{$\bar{\nu}_{\mu}+p\rightarrow\mu^+X^0$}   
    \end{subfigure} 
    \begin{subfigure}{0.4\textwidth}
        \centering\includegraphics[width=0.85\columnwidth]{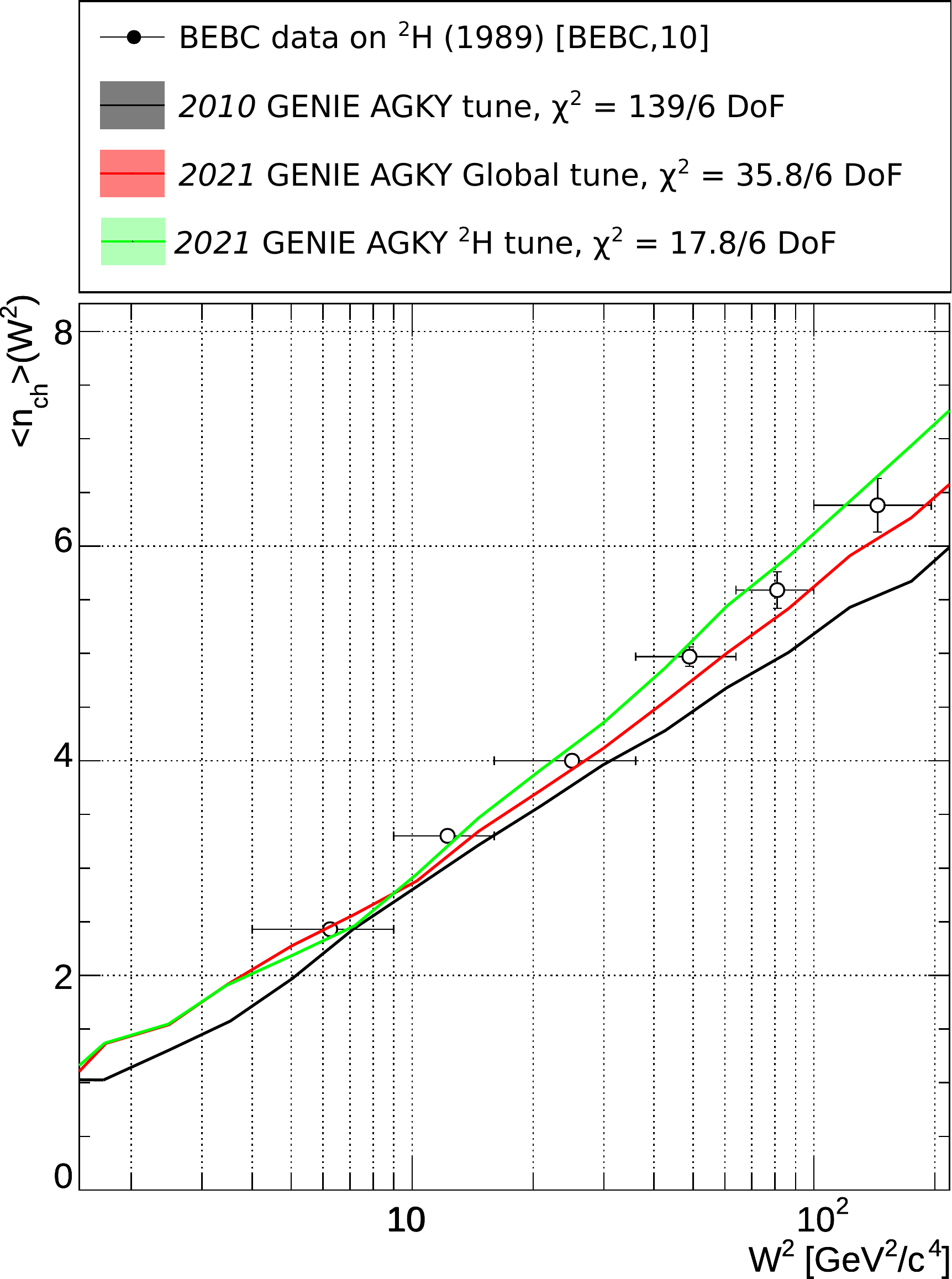}
        \caption{$\bar{\nu}_{\mu}+p\rightarrow\mu^+X^0$}   
    \end{subfigure} 
     
    \centering
    \begin{subfigure}{0.4\textwidth}
        \centering\includegraphics[width=0.85\columnwidth]{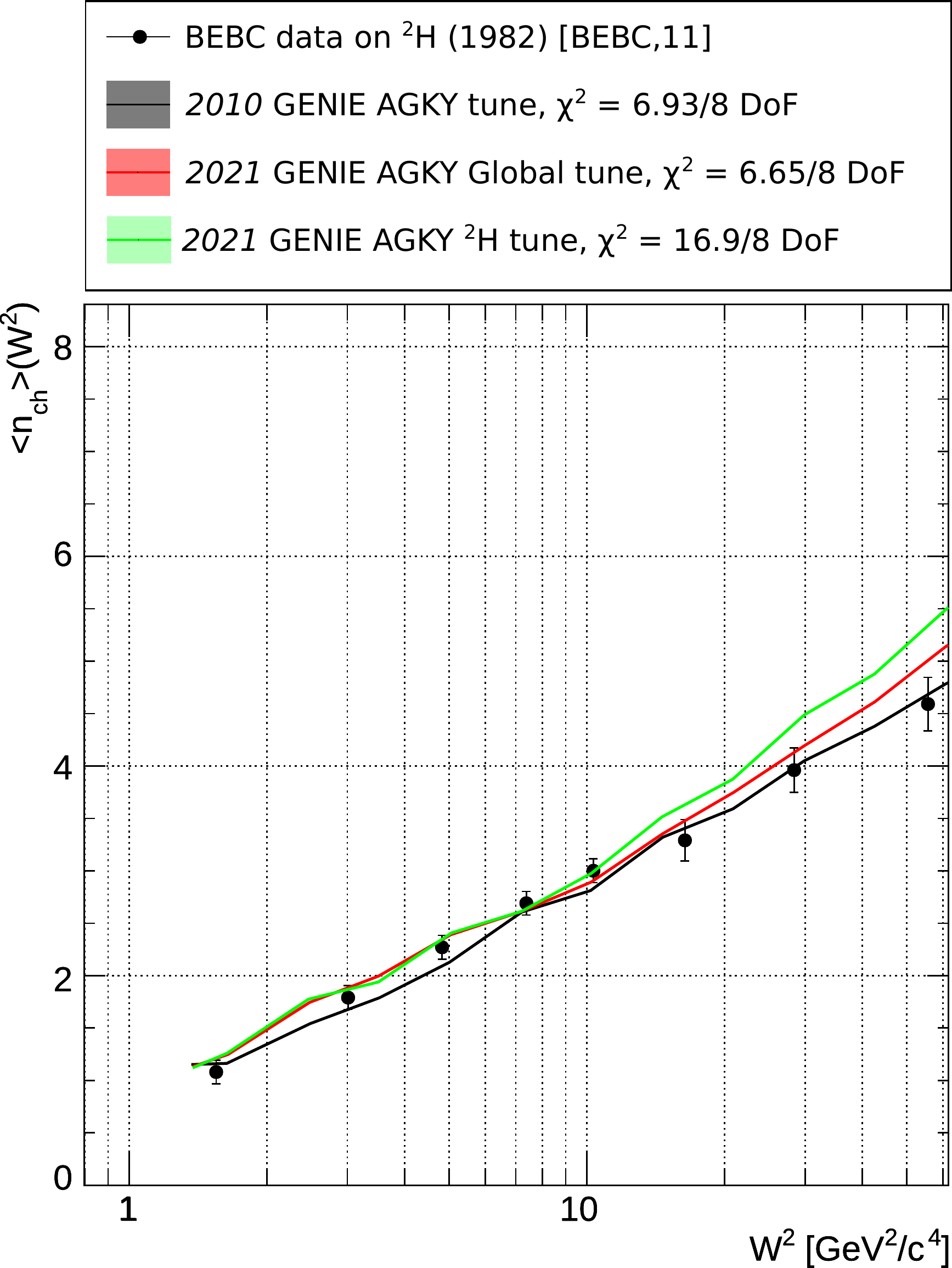}
        \caption{$\bar{\nu}_{\mu}+n\rightarrow\mu^+X^-$ }   
    \end{subfigure} 
    \begin{subfigure}{0.4\textwidth}
        \centering\includegraphics[width=0.85\columnwidth]{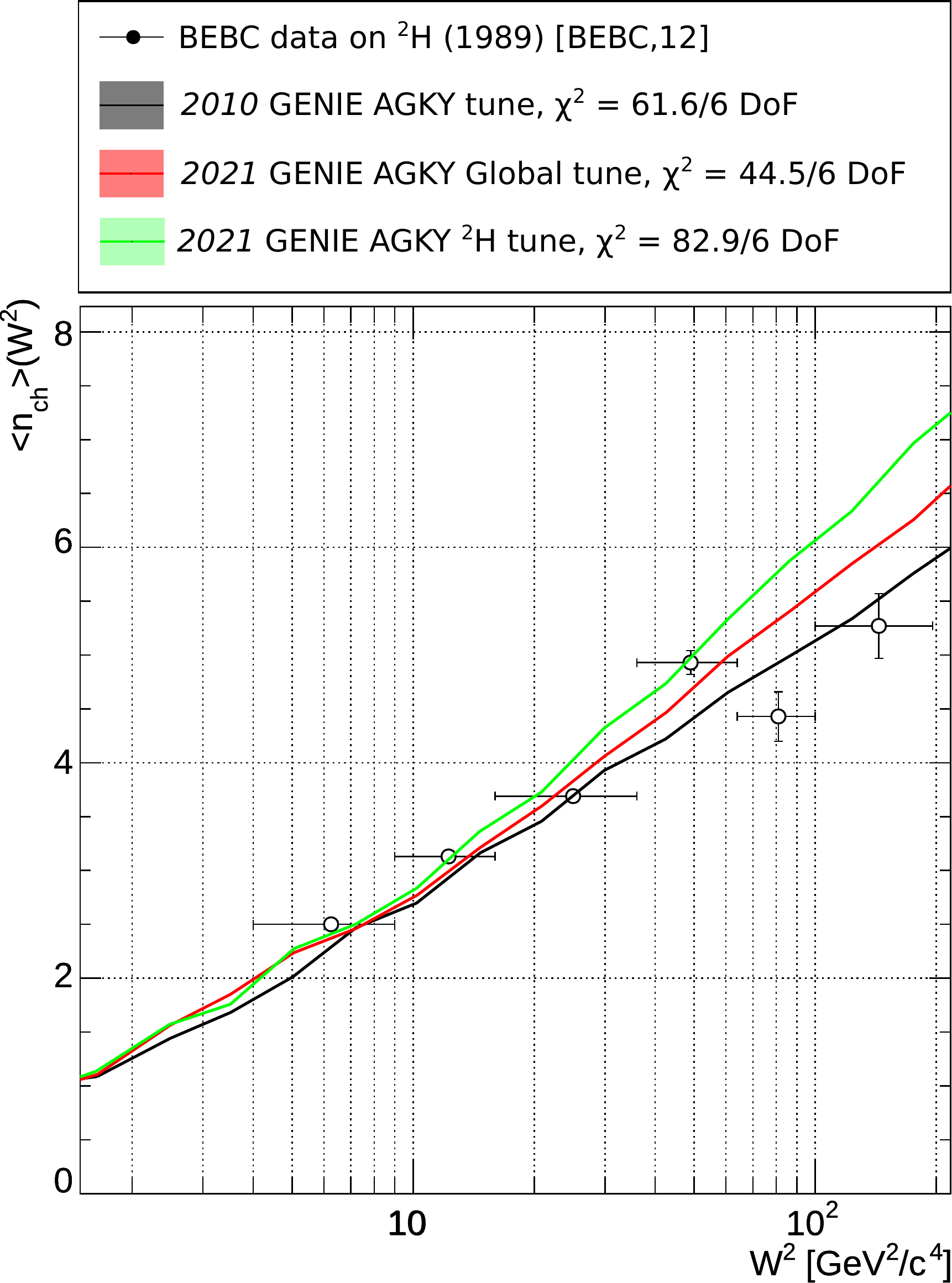}
        \caption{$\bar{\nu}_{\mu}+n\rightarrow\mu^+X^-$ }   
    \end{subfigure} 
    \caption{Comparison of $\langle n_{\text{ch}}\rangle$ against neutrino-induced hadronization data on $\bar{\nu}_\mu$ interactions on p and n from the \ac{BEBC} bubble chamber experiment filled with $^2$H~\cite{Barlag1982,Jongejans1989}. 
    Datapoints used in the AGKY \emph{2021} global tune analysis are shown as filled black markers. 
    Discarded datapoints are represented using empty markers.
    The predictions are computed using the parameters specified in Tab.~\ref{tab:fitResultParameters}. The $\chi^2$ values are calculated against all the data from each experiment. See definition of \emph{Tags} in Tab.~\ref{tab:references}.}
    \label{fig:bar2HPredictions}
\end{figure*}

GENIE predictions for all the averaged charged multiplicity data available are shown in Figs.~\ref{fig:pPredictions}--\ref{fig:bar2HPredictions} before and after the AGKY tunes. 
The results show the prediction for the \emph{2021} GENIE global tune in red and \emph{2021} GENIE $^{2}$H tune in green.
To distinguish data points used in the analyses from those that were not, the used one points have completely black markers, the others are empty circles. 
Each dataset is associated to a Tag, defined in Tab.~\ref{tab:references}. 
Vertical error bars include statistical and systematic uncertainties following our data review. 
Horizontal bars correspond to the bin width used in the data release, and are only shown if those are available in the original paper.

\begin{table}
    \footnotesize
    \centering
    \begin{tabular}{@{\extracolsep\fill} c c c c} \hline\hline\noalign{\smallskip}
    \textbf{Experiment}& Target & Tag & \textbf{Ref.} \\  
    \noalign{\smallskip}\hline\hline\noalign{\smallskip}
    \multicolumn{4}{c}{$\nu_\mu + p \rightarrow \mu^- X^{++}$} \\ 
    \noalign{\smallskip}\hline\hline\noalign{\smallskip}
    FNAL 15 ft (1976)  & H & FNAL 15 ft,0 &\cite{PhysRevLett.36.124} \\ \noalign{\smallskip}
    BEBC (1983)  & H & BEBC,0 &\cite{GRASSLER1983269}\\ \noalign{\smallskip}
    BEBC (1990)  & H & BEBC,1 &\cite{Jones1990}\\ \noalign{\smallskip}
    BEBC (1992)  & H & BEBC,2 &\cite{BEBC1992}\\ \noalign{\smallskip}
    FNAL 15 ft (1983)  & $^2$H & FNAL 15 ft,1 &\cite{PhysRevD.27.47} \\ \noalign{\smallskip}
    BEBC (1989)  & $^2$H & BEBC,3 &\cite{Jongejans1989} \\\noalign{\smallskip}
    \noalign{\smallskip}\hline\hline\noalign{\smallskip}
    \multicolumn{4}{c}{$\nu_\mu + n \rightarrow \mu^- X^{+}$} \\ \noalign{\smallskip}
    \noalign{\smallskip}\hline\hline\noalign{\smallskip}
    FNAL 15 ft (1983)  & $^2$H & FNAL 15 ft,2 &\cite{PhysRevD.27.47} \\\noalign{\smallskip}
    BEBC (1984)  & $^2$H & BEBC,4 &\cite{Allasia1984}   \\\noalign{\smallskip}
    BEBC (1989)  & $^2$H & BEBC,5 &\cite{Jongejans1989} \\\noalign{\smallskip}
    \noalign{\smallskip}\hline\hline\noalign{\smallskip}
    \multicolumn{4}{c}{$\bar{\nu}_\mu + p \rightarrow \mu^+ X^{0}$} \\ \noalign{\smallskip}
    \noalign{\smallskip}\hline\hline\noalign{\smallskip}
    FNAL 15 ft (1981)  & H & FNAL 15 ft,3 &\cite{PhysRevD.25.624}\\\noalign{\smallskip}
    BEBC (1983)  & H & BEBC,6 &\cite{GRASSLER1983269}\\\noalign{\smallskip}
    BEBC (1990)  & H & BEBC,7 &\cite{Jones1990}\\\noalign{\smallskip}
    BEBC (1992)  & H & BEBC,8 & \cite{BEBC1992}\\  \noalign{\smallskip}
    BEBC (1982)  & $^2$H & BEBC,9 &\cite{Barlag1982}\\\noalign{\smallskip}
    BEBC (1989)  &  $^2$H & BEBC,10 &\cite{Jongejans1989} \\\noalign{\smallskip}
    \noalign{\smallskip}\hline\hline\noalign{\smallskip}
    \multicolumn{4}{c}{$\bar{\nu}_\mu + n \rightarrow \mu^+ X^{-}$} \\ \noalign{\smallskip}
    \noalign{\smallskip}\hline\hline\noalign{\smallskip}
    BEBC (1982)  &  $^2$H & BEBC,11&\cite{Barlag1982}\\\noalign{\smallskip}
    BEBC (1989)  &  $^2$H & BEBC,12 &\cite{Jongejans1989} \\\noalign{\smallskip}
    \hline \hline
    \end{tabular}
    \caption{Summary of data used for comparisons in Figs.~\ref{fig:nupscattering}, \ref{fig:barpPredictions}, \ref{fig:2HPredictions}, and \ref{fig:bar2HPredictions}. This table links the experiment and the tag used for the legend in each plot to the corresponding reference.}
    \label{tab:references}
\end{table} 

In terms of prediction differences, the \emph{2021} GENIE global tune tends to underpredict deuterium data whereas the \emph{2021} GENIE $^{2}$H tune overpredicts the hydrogen data.
This especially true for the PYTHIA region, at high-$W$. 
This is translated in the parameters with an increase (decrease) of Lund $a$ (Lund $b$) for the deuterium tune with respect to the global tune.

The summary of the $\chi^2$ values per dataset as well as the total contributions are shown in Tab.~\ref{tab:summarychi2}.
Three different $\chi^2$ values are presented: $\chi^2_{\text{2010}}$, $\chi^2_{\text{2021(Global)}}$ and $\chi^2_{2021(^2\text{H})}$ using, respectively, the \emph{2010} GENIE, \emph{2021} GENIE global and \emph{2021} GENIE $^{2}$H tune parameters. 
The $\chi^2$ values per dataset are computed by comparing the GENIE predictions against all the data points in each dataset, regardless of the point being used in the fit or not. 
Differences between the $\chi^2$ obtained with Eq.~\ref{eq:chi2} and the one calculated using the GENIE predictions directly are expected. One of the reasons is that Eq.~\ref{eq:chi2} only considers the datapoints included in the tune.
Moreover, further differences arise form the fact that the Professor parameterisation $\widetilde{n}_{ij}$ is not exact, as explained in Sec.~\ref{sec:Likelihood}.

\begin{table}
    \footnotesize
    \centering
    \begin{tabular}{@{\extracolsep\fill} c c c c c c} \hline\hline\noalign{\smallskip}
    \textbf{Experiment}& $\chi_{\text{2010}}^2$ & $\chi_{\text{2021(Global)}}^2$ &  $\chi_{2021({^2\text{H}})}$ & DoF & In tune\\
    \noalign{\smallskip}\hline\hline\noalign{\smallskip}
    \multicolumn{6}{c}{$\nu_\mu + p \rightarrow \mu^- X^{++}$} \\ 
    \noalign{\smallskip}\hline\hline\noalign{\smallskip}
    \multicolumn{6}{c}{Data on hydrogen} \\ \noalign{\smallskip}
    \hline \noalign{\smallskip}
    FNAL 15 ft,0  & 83 & 39 & 81 & $25$  & Partially \\ \noalign{\smallskip}
    BEBC,0  & 222 & 56 & 13 & $11$  & \xmark \\ \noalign{\smallskip}
    BEBC,1  & 26 & 7 & 23 & $6$ & Partially\\ \noalign{\smallskip}
    BEBC,2  & 50.2 & 3.5 & 68.8 & $5$ & \checkmark \\ \noalign{\smallskip}
     \hline \noalign{\smallskip}
    \multicolumn{6}{c}{Data on deuterium} \\ \noalign{\smallskip}
    \hline \noalign{\smallskip}
    FNAL 15 ft,1  & 21 & 13 & 5 &  $14$ & Partially \\ \noalign{\smallskip}
    BEBC,3  & 563 & 321 & 84 & $6$ & \xmark  \\\noalign{\smallskip}
    \hline \noalign{\smallskip}
    \textbf{Total for} $\nu_\mu p$ & \textbf{965} & \textbf{447} & \textbf{275} & \textbf{67} &  \\
    \noalign{\smallskip}\hline\hline\noalign{\smallskip}
    \multicolumn{6}{c}{$\nu_\mu + n \rightarrow \mu^- X^{+}$} \\ \noalign{\smallskip}
    \noalign{\smallskip}\hline\hline\noalign{\smallskip}
    FNAL 15 ft,2  & 17 & 21 & 7 &  $14$ & Partially \\\noalign{\smallskip}
    BEBC,4  & 171 & 6 & 5 & $8$ & \checkmark \\\noalign{\smallskip}
    BEBC,5  & 886 & 352 & 99 &  $6$ & \xmark \\\noalign{\smallskip}
    \hline \noalign{\smallskip}
    \textbf{Total for} $\nu_\mu n$ & \textbf{1,074}& \textbf{435} & \textbf{111} & \textbf{28} & \\
    \noalign{\smallskip}\hline\hline\noalign{\smallskip}
    \multicolumn{6}{c}{$\bar{\nu}_\mu + p \rightarrow \mu^+ X^{0}$} \\ \noalign{\smallskip}
    \noalign{\smallskip}\hline\hline\noalign{\smallskip}
    \multicolumn{6}{c}{Data on hydrogen} \\ \noalign{\smallskip}
    \hline \noalign{\smallskip}
    FNAL 15 ft,3 & 69 & 18 & 13 & $10$ & Partially \\\noalign{\smallskip}
    BEBC,6  & 41 & 8 & 25 & $10$ & \xmark \\\noalign{\smallskip}
    BEBC,7  & 5.8 & 13.4 & 28.7 & $6$ &  Partially \\\noalign{\smallskip}
    BEBC,8  & 22.4 & 57.9 & 158.0 & $5$ & \checkmark \\  \noalign{\smallskip}
    \hline\noalign{\smallskip}
    \multicolumn{6}{c}{Data on deuterium} \\ \noalign{\smallskip}
    \hline\noalign{\smallskip}
    BEBC,9  & 14 & 2 & 4 & $8$ & \checkmark\\\noalign{\smallskip}
    BEBC,10  & 139 & 36 & 18 & $6$ & \xmark \\\noalign{\smallskip}
    \hline \noalign{\smallskip}
    \textbf{Total for} $\bar{\nu}_\mu p$ & \textbf{292} & \textbf{135} & \textbf{246} & \textbf{45} & \\
    \noalign{\smallskip}\hline\hline\noalign{\smallskip}
    \multicolumn{6}{c}{$\bar{\nu}_\mu + n \rightarrow \mu^+ X^{-}$} \\ \noalign{\smallskip}
    \noalign{\smallskip}\hline\hline\noalign{\smallskip}
    BEBC,11  & 6.9 & 6.7 & 16.9 & $8$ & \checkmark\\\noalign{\smallskip}
    BEBC,12  & 61.6 & 44.5 & 82.9 &  $6$  & \xmark \\\noalign{\smallskip}
    \hline \noalign{\smallskip}
    \textbf{Total for} $\bar{\nu}_\mu n$ & \textbf{69} &\textbf{51} & \textbf{100} & \textbf{14} &  \\
        \noalign{\smallskip}\hline\hline\noalign{\smallskip}
    \multicolumn{6}{c}{$\chi^2$ Summary} \\ \noalign{\smallskip}
    \noalign{\smallskip}\hline\hline\noalign{\smallskip}
    \textbf{All data} & \textbf{2,398} & \textbf{1,068} & \textbf{731}& \textbf{154} &  \\\noalign{\smallskip}
    \textbf{All $^2$H data} & \textbf{1,879} & \textbf{858} & \textbf{320}& \textbf{76} &  \\ \noalign{\smallskip}
    \textbf{All H data} & \textbf{519} & \textbf{202} & \textbf{411}& \textbf{78} &  \\\noalign{\smallskip}
    \noalign{\smallskip}\hline \hline \noalign{\smallskip}
    \end{tabular}
    \caption{Summary of $\chi^2$ values for the datasets shown in Figs.~\ref{fig:2HPredictions}, \ref{fig:bar2HPredictions}, \ref{fig:pPredictions}, and \ref{fig:barpPredictions}. The table shows the $\chi^2$ per dataset and interaction channel as well as the total and per channel $\chi^2$. The $\chi^2$ values are calculated using the GENIE predictions for each tune: \emph{2010} GENIE, $\chi_{\text{2010}}^2$, \emph{2021} GENIE, $\chi_{\text{2021(Global)}}^2$, and \emph{2021} GENIE, $\chi_{\text{2021}(^{2}\text{H})}^2$.  } 
    \label{tab:summarychi2}
\end{table} 

It is important to stress that the total $\chi^2$ from Tab.~\ref{tab:summarychi2} are not providing any information related to goodness of fit, but it simply shows the general agreement with respect to available datasets.
A sense of the goodness of fit can be obtained looking at the total $\chi^2$ calculated with the datasets included in the fit only, see Tab.~\ref{tab:summarychi2_fit_only}. 

\begin{table}
    \footnotesize
    \centering
    \begin{tabular}{@{\extracolsep\fill} c c c c c} \hline\hline\noalign{\smallskip}
    \textbf{Datasets}& $\chi_{\text{2010}}^2$ & $\chi_{\text{2021(Global)}}^2$ &  $\chi_{\text{2021}(^{2}\text{H})}^2$ & DoF \\
    \noalign{\smallskip}\hline\hline\noalign{\smallskip}
    {All Data in tune} & {486} & {242} & {410}&  {109} \\\noalign{\smallskip}
    {$^2$H Data in tune} & {230} & {105} & {37}&  {52} \\\noalign{\smallskip}
    {H Data in tune} & {256} & {138} & {374}&  {57} \\
    \hline \hline \noalign{\smallskip}
    \end{tabular}
    \caption{Total $\chi^2$ calculated with the datasets included in each fit: \emph{2010} GENIE, $\chi_{\text{2010}}^2$, \emph{2021} GENIE, $\chi_{\text{2021(Global)}}^2$, and \emph{2021} GENIE, $\chi_{\text{2021}(^{2}H)}^2$. }
    \label{tab:summarychi2_fit_only}
\end{table} 

\begin{figure}
    \centering
    \begin{subfigure}{\columnwidth}
        \includegraphics[width=\columnwidth]{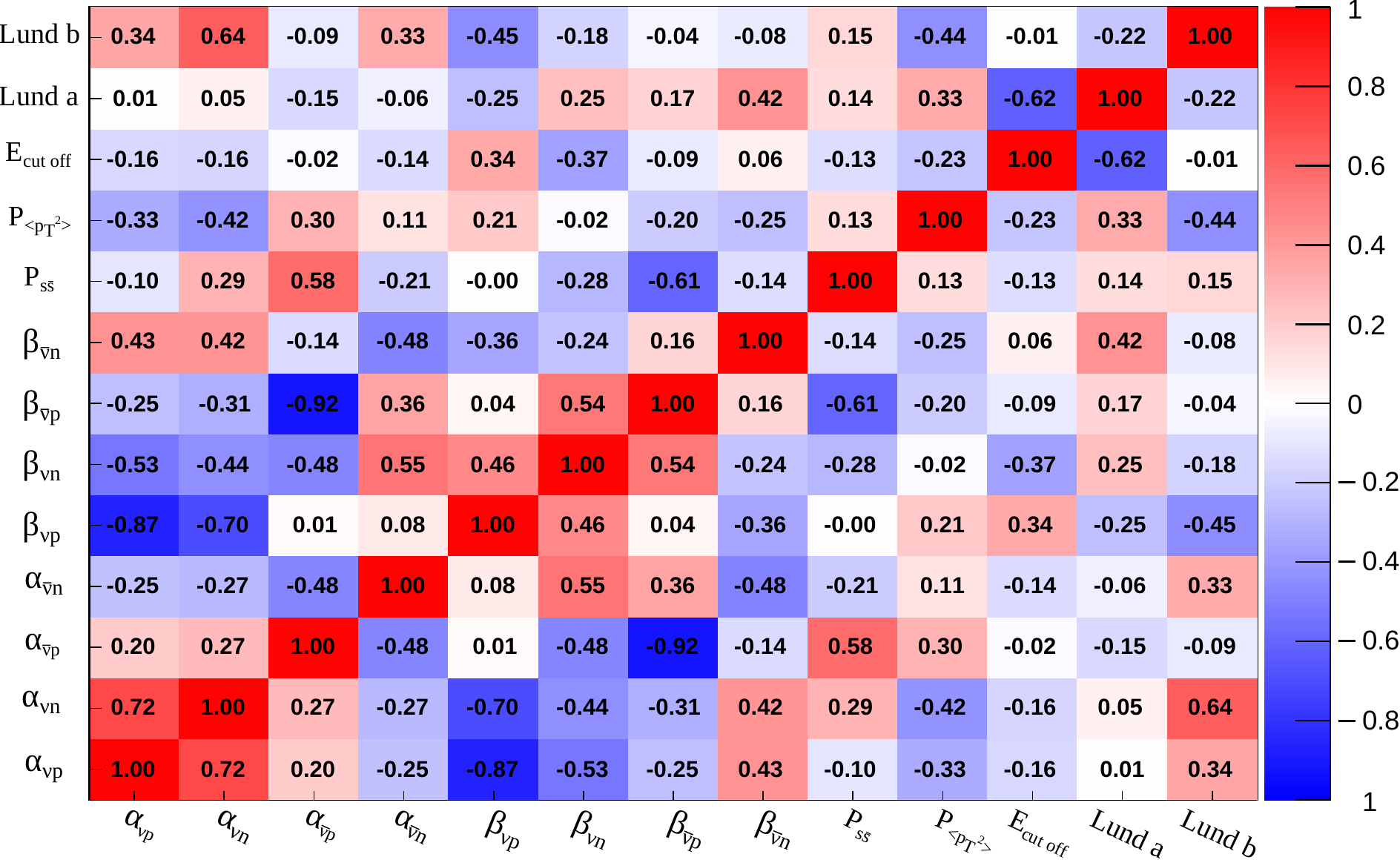}
        \caption{Global tune correlation matrix.}
    \end{subfigure}
    
        \begin{subfigure}{\columnwidth}
        \includegraphics[width=\columnwidth]{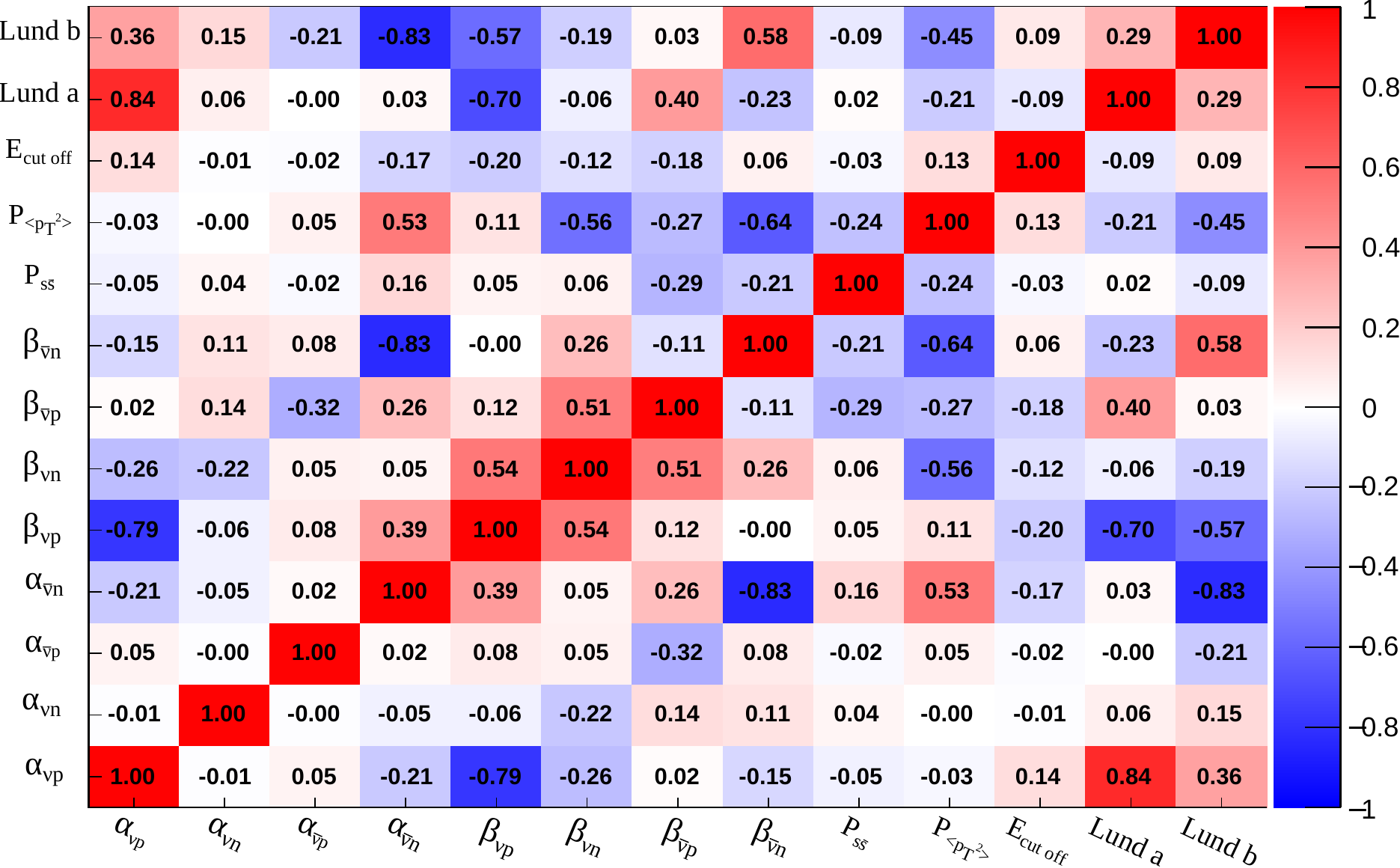}
        \caption{Deuterium only tune correlation matrix.}
    \end{subfigure}
    \caption{Parameter correlation matrix for the \emph{2021} GENIE AGKY tunes against averaged charged multiplicity data. }
    \label{fig:corrGlobal}
\end{figure}

The parameters covariance matrices for both tunes are obtained by inverting the Hessian of the log-likelihood function at the best fit point, see Tab.~\ref{tab:covGlobal} and Tab.~\ref{tab:cov2H}.
As expected, the low-$W$ AGKY and PYTHIA parameters are now correlated in both tunes because of the interplay of the models in the transition region, with a number of parameters showing a correlation above $50\%$.
See a graphical representation of the correlation matrix in Fig.~\ref{fig:corrGlobal}. 

The results form the \emph{2021} GENIE AGKY tunes will available in GENIE v3.2.0.
Users can run the \emph{2021} GENIE tunes global and $^2$H tunes out of the box using the \texttt{G18\_02a\_03\_330} and \texttt{G18\_02a\_03\_320} comprehensive configurations respectively.

\subsection{The \emph{2021} GENIE AGKY global tune}
\label{subsec:AGKYGlobalTune}    

After the AGKY global tune, the GENIE predictions show a better agreement to the data.
In particular, for the datasets included in the \emph{2021} GENIE global tune, the $\chi^2$ associated to the prediction is $\chi^2_{\text{2010}}=486/109$~DoF.
After the tune, the $\chi^2_{\text{2021(Global)}}$ is $242/109$~DoF. 
This is clearly an improvement although the agreement is not completely satisfactory since the p-value is $4\times 10^{-12}$.

The improvement in the data description is general and both deuterium and hydrogen samples have a better agreement.
Moreover, after the tune both hydrogen and deuterium samples have similar goodness of fit, and, in general, the level of agreement is the same. 
This can be noted from the $\chi^2$ contributions from Tab.~\ref{tab:summarychi2_fit_only}.

The agreement with the datasets not included in the tune has also improved, as shown in Tab.~\ref{tab:summarychi2}. 
The total $\chi^2$ computed using all available data is reduced significantly for both H and $^2$H datasets.
Particularly, the global tune shows a better agreement against all hydrogen data. 
As expected from Sec.~\ref{subsec:req_data}, the datasets with highest contribution to the total $\chi^2$ after the global tune are [BEBC,3] and [BEBC,5].

The main effect of the tune is observed in the PYTHIA region, at $W>3$~GeV$/c^2$, where the prediction of $\langle n_{\text{ch}} \rangle$ increased. 
This is a direct consequence of the increase on Lund $a$ and Lund $b$. 
This behaviour is consistent with the HERMES tune, summarised in Sec.~\ref{subsec:PYTHIA}. 

\begin{figure*}
    \centering
    \begin{subfigure}{0.4\textwidth}
        \includegraphics[width=\columnwidth]{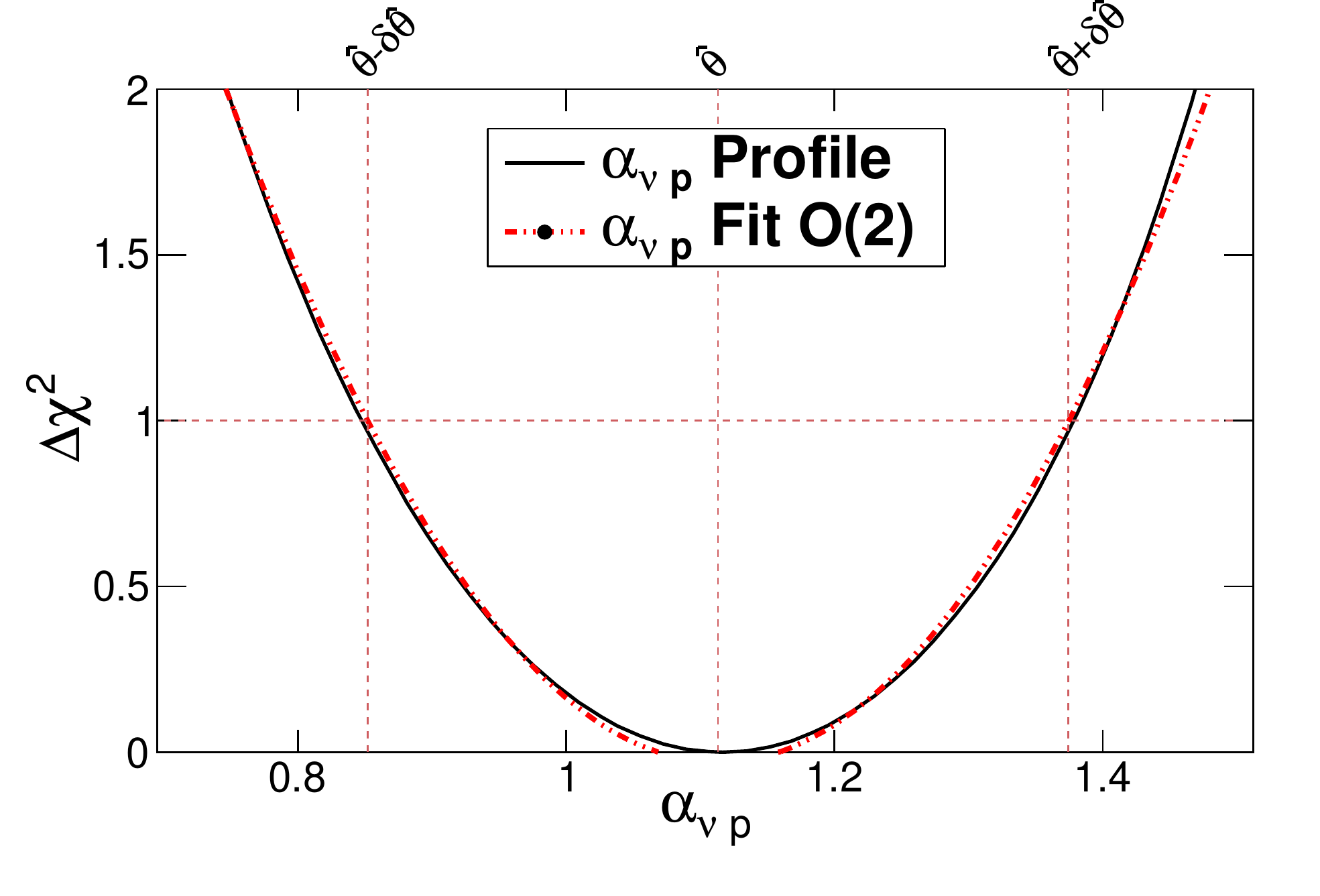}
        \caption{Symmetric parameter profile.}
    \end{subfigure}
    \begin{subfigure}{0.55\textwidth}
        \begin{subfigure}{\columnwidth}
        \includegraphics[width=0.49\columnwidth]{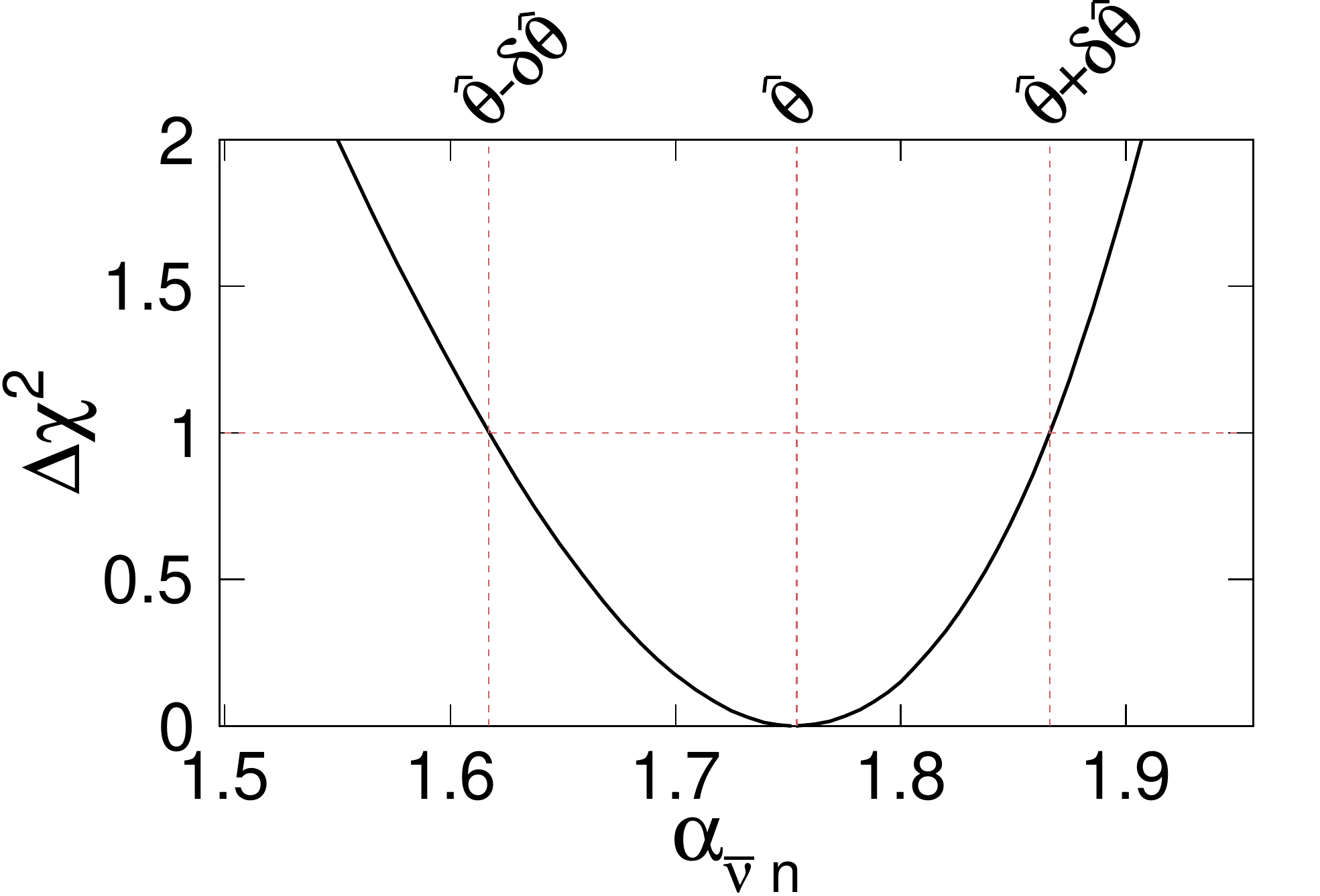}
        \includegraphics[width=0.49\columnwidth]{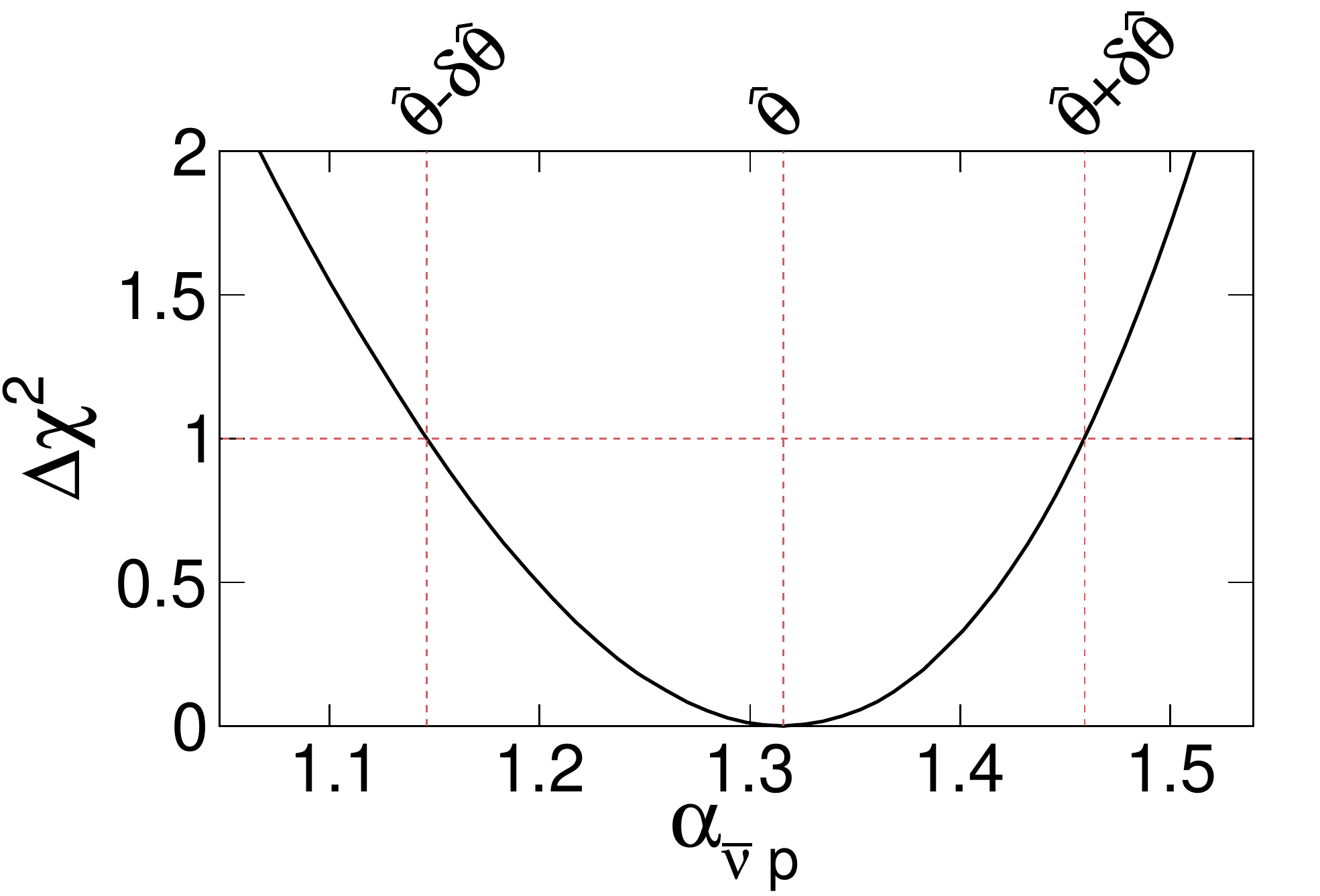}
        
        \includegraphics[width=0.45\columnwidth]{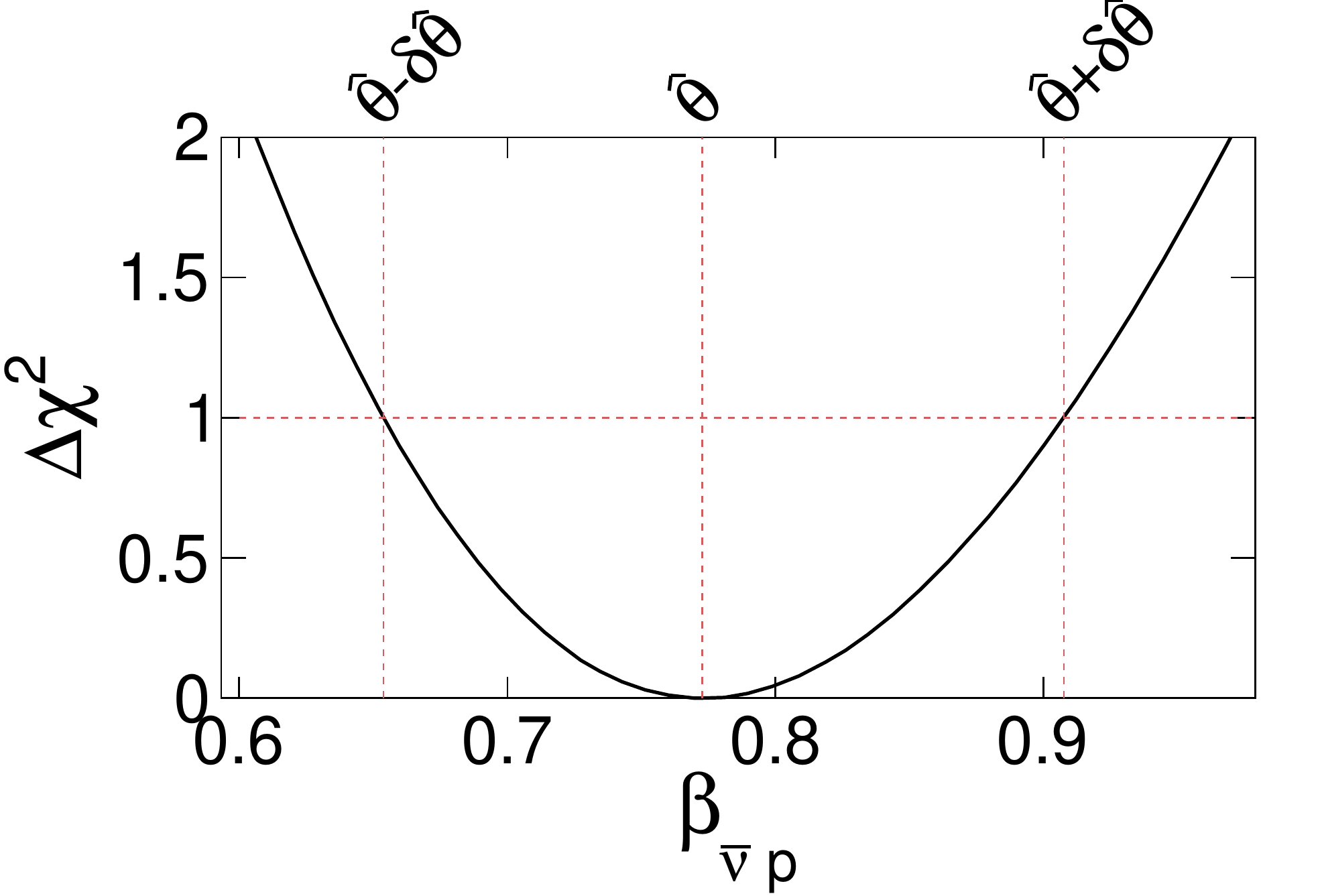}
        \end{subfigure}
        \caption{Asymmetric parameter profiles.}
    \end{subfigure}
    \caption{Joint function obtained fixing the two parameters under study and minimizing $\Delta\chi^2_{\text{profile}}(\boldsymbol{\theta})$ respect the other parameters in the \emph{2021} GENIE global tune.
    The dashed lines represent the parameter range that satisfies the condition  $\Delta\chi^2_{\text{profile}}(\theta_i)<1$.
    This is also denoted as $\hat{\theta}\pm\delta\hat{\theta}$.
    }
    \label{fig:GlobalProfiles}
\end{figure*}

\begin{figure*}
    \centering
    \begin{subfigure}{0.45\columnwidth}
        \includegraphics[width=0.9\columnwidth]{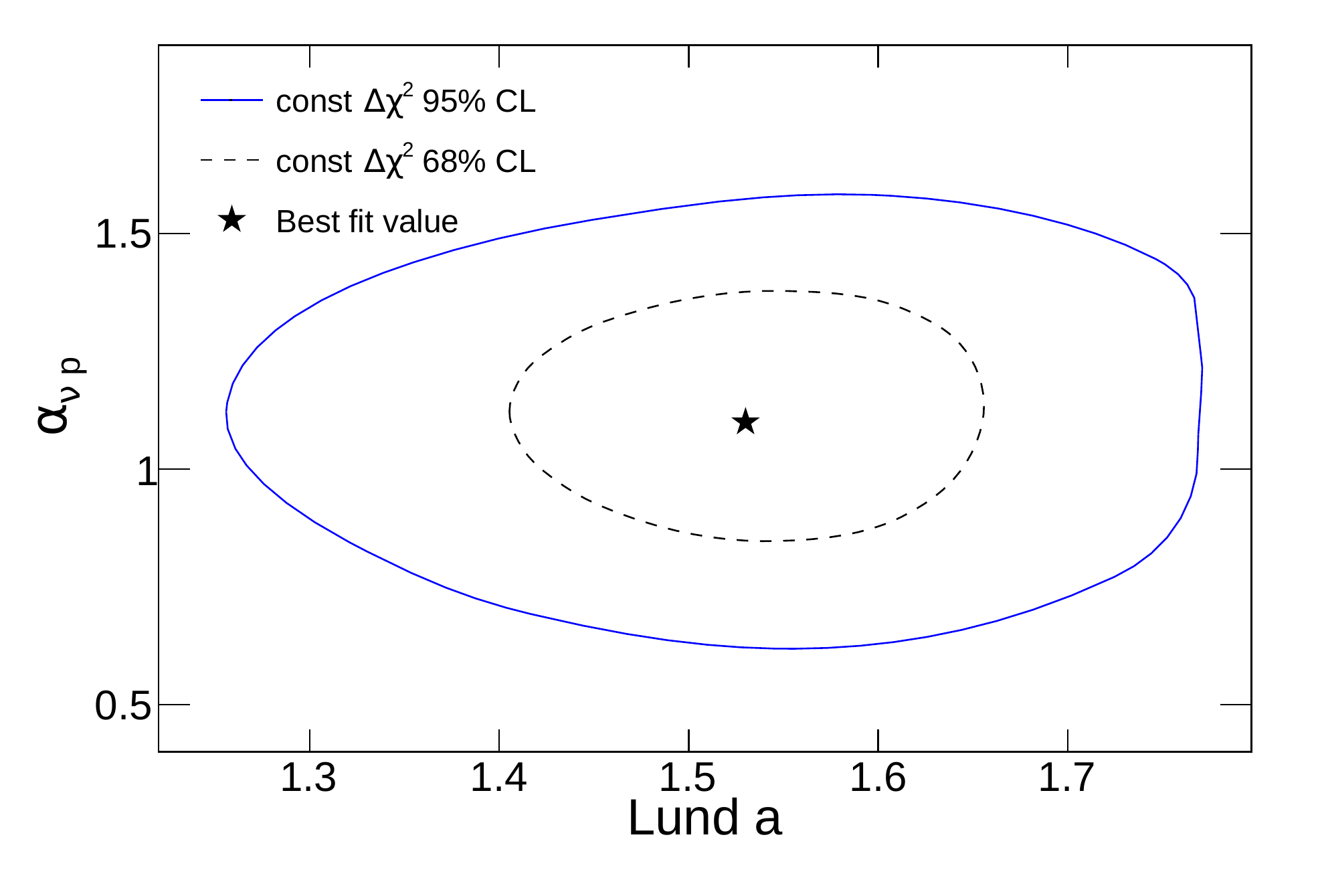}
    \end{subfigure}
        \begin{subfigure}{0.45\columnwidth}
            \includegraphics[width=0.9\columnwidth]{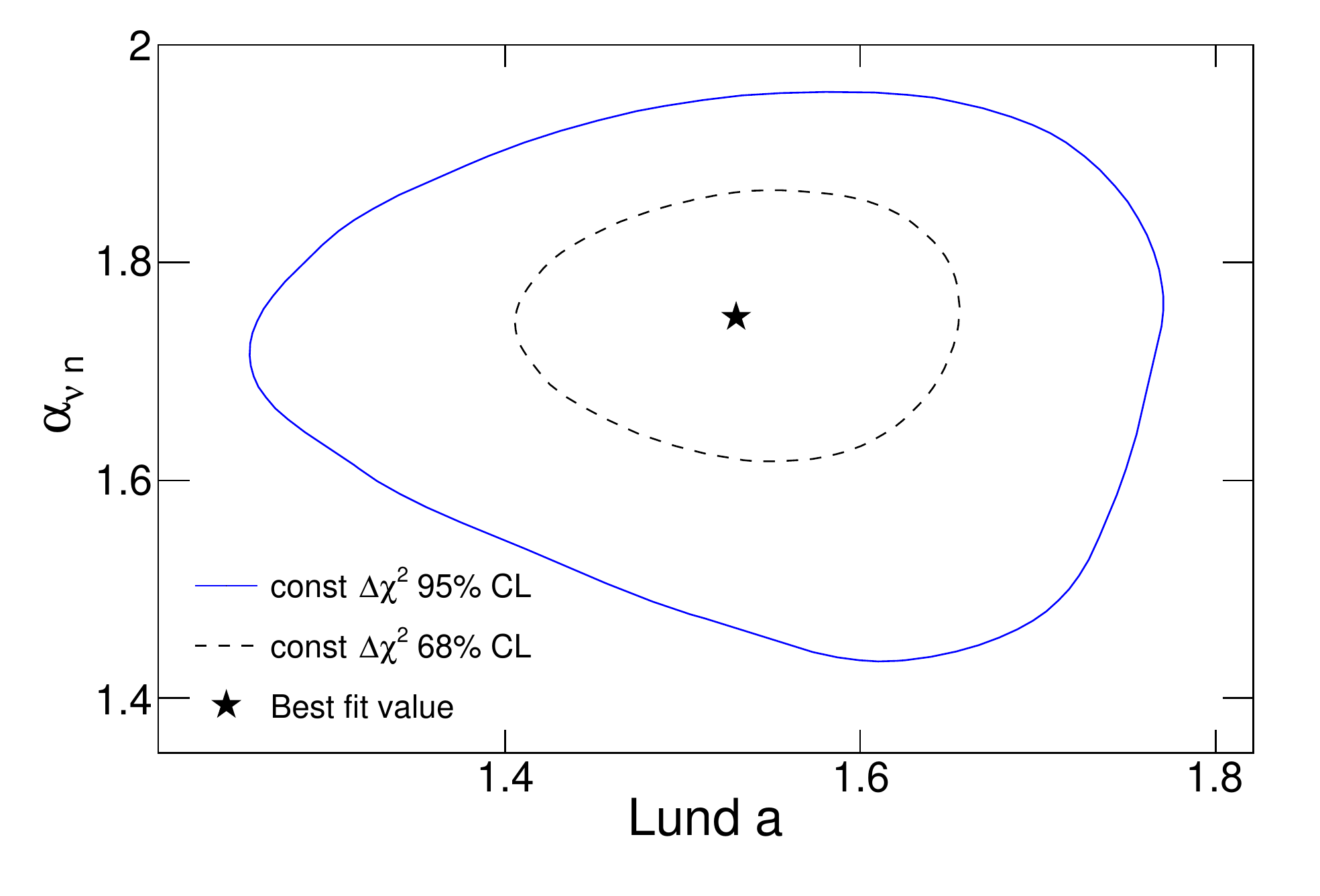}
    \end{subfigure}
    
    \begin{subfigure}{0.45\columnwidth}
        \includegraphics[width=0.97\columnwidth]{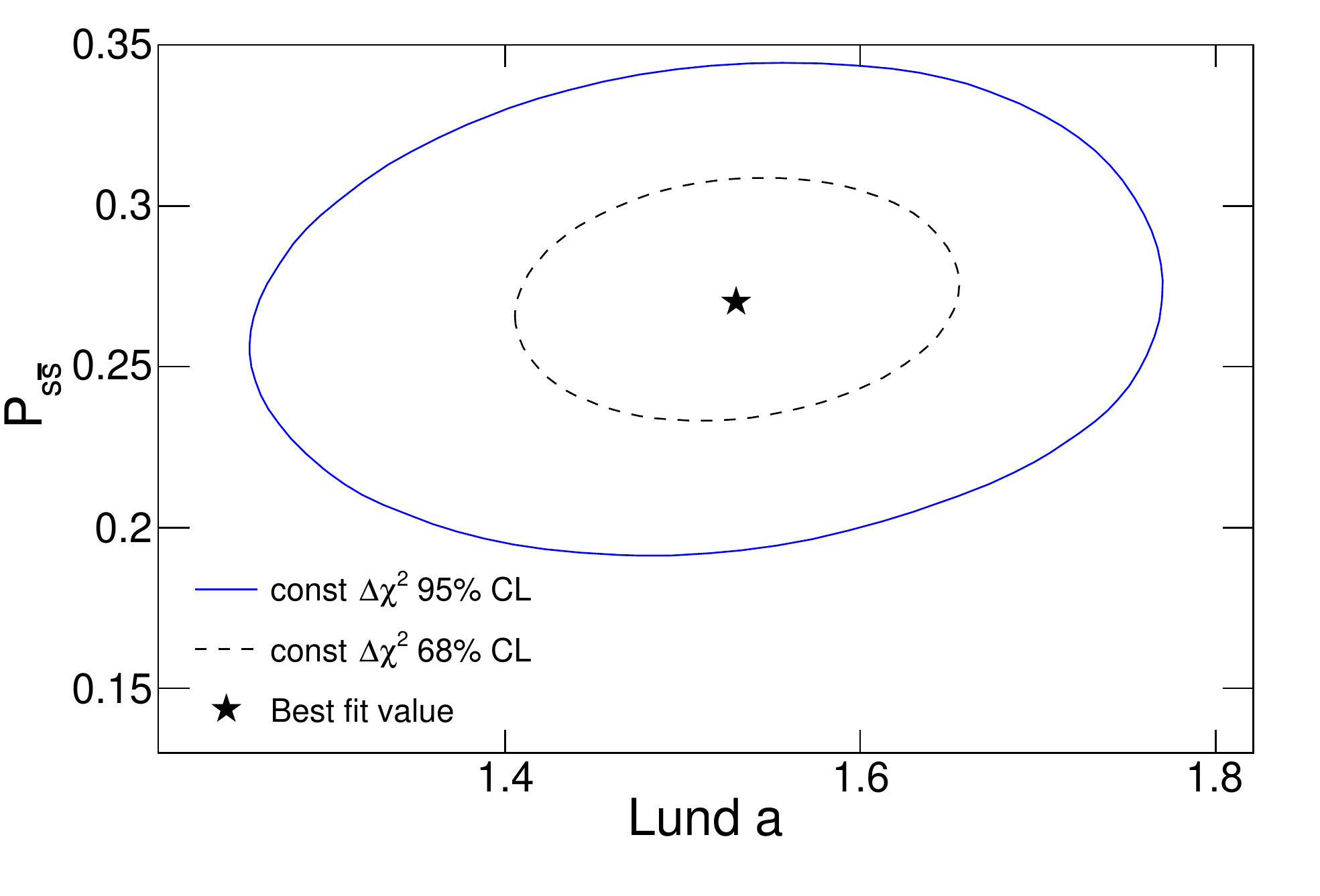}
    \end{subfigure}
    \begin{subfigure}{0.45\columnwidth}
        \includegraphics[width=0.97\columnwidth]{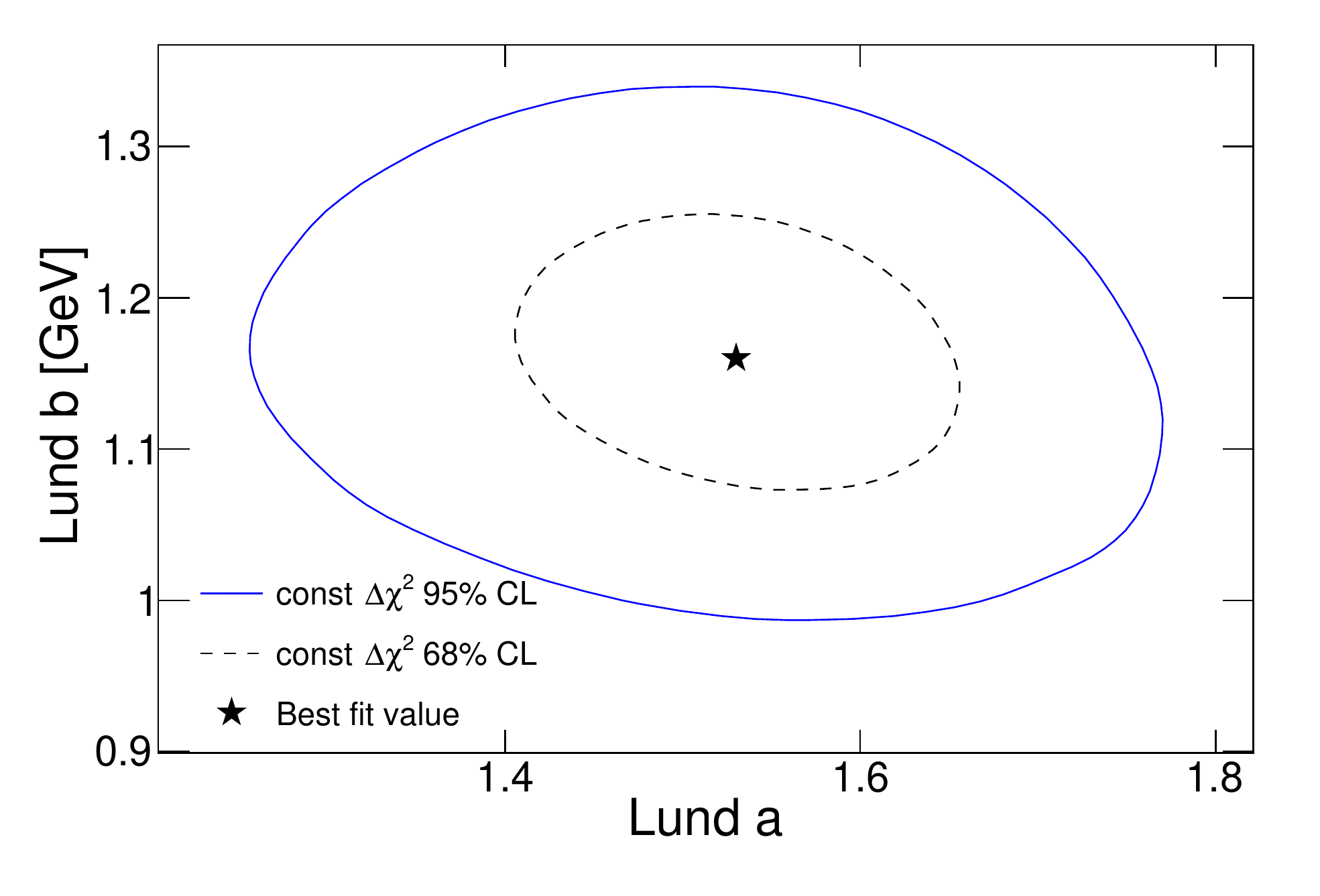}
    \end{subfigure}
    \caption{Joint $\Delta\chi^2_{\text{contour}}(\theta_i,\theta_j)$ function obtained fixing the two parameters under study and minimizing $\Delta\chi^2_{contour}(\boldsymbol{\theta})$ respect the other parameters in the \emph{2021} GENIE global tune. The 95\% and 68\% contour lines are shown as well as the best fit values for the global tune.}
    \label{fig:GlobalContours}
\end{figure*}

For each parameter, the corresponding uncertainty is obtained with the profiling method under the condition $\Delta\chi^2_{\text{profile}} (\theta_i) < 1$. 
The profiles are calculated by fixing the value of the parameter under study $\theta_i$ to a desired value and minimizing the quantity $\Delta\chi^2(\boldsymbol{\theta}) = \chi^2(\boldsymbol{\theta}) - \chi^2_{\min}$ with respect to all others parameters that were allowed to float in the fit. 
The constant $\chi^2_{\min}$ corresponds to the global minimum value of $\chi^2(\boldsymbol{\theta})$. 
Some parameters have a good Gaussian behaviour and a symmetric profile. 
For some others this is not true and this gives rise to asymmetric uncertainties for the parameters.
Example of a symmetric parameter profile compared to the non-Gaussian ones is shown in Fig.~\ref{fig:GlobalProfiles}.
The contours for some pairs of the AGKY parameters are shown in Fig.~\ref{fig:GlobalContours}.  

The fit covariance matrix can be propagated back to the GENIE predictions giving a posterior confidence belt for the prediction associated to the tune. 
As an example, a comparison of the global tune prediction and the associated posterior confidence belt is shown in Fig.~\ref{fig:ProfessorPred}. 

\subsection{The \emph{2021} GENIE AGKY $^{2}$H tune}
\label{subsec:AGKY2HTune}   

For the datasets included in the deuterium only tune, the $\chi^2$ associated to the \emph{2010} GENIE AGKY prediction is $\chi^2_{\text{2010}}=230/52 \text{ DoF}$.
After the tune, the total $\chi^2_{2021(^2\text{H})}$ is $37/52 \text{ DoF}$ that corresponds to a p-value of $0.94$.
Being the deuterium only goodness of fit so much better than the global tune confirms the high tension between H and $^2$H datasets.

Surprisingly, the deuterium only tune shows a better agreement than the global tune when all neutrino-induced hadronization data are considered, see Tab.~\ref{tab:summarychi2}.
Yet, this does not imply that the deuterium only fit is a better tune, it simply reinforces that the discarded dataset are not compatible with the data used in the fit. 

The tension between hydrogen and deuterium data was already observed in other studies where a modified KNO-based model was tuned to averaged charged multiplicity data from bubble chamber experiments~\cite{Kuzmin_2013}. 
Those studies suggest that the origin of tensions between H and $^2$H could be due to rescattering effects on deuterium.
As explained in Sec.~\ref{sec:dataanalysesBBCH}, the bubble chamber experiments claim that rescattering effects have a smaller effect on neutron samples as a consequence of the classification into $\nu_\mu$ on $p$ or $\nu_\mu$ on $n$ events.
This is a consequence of the neutron re-interaction with the proton from the deuterium, which is then kicked out and, therefore, miss-identified as a $\nu_\mu$p event.
If the disagreements were only due to rescattering, the global tune would have a better agreement than the deuterium only tune on $\nu_\mu$ and $\bar{\nu}_\mu$ on neutron data.
However, a better agreement of the global tune on neutron samples is not observed.

\subsection{AGKY global and deuterium only tunes impact on other neutrino-induced hadronization observables}
\label{subsec:other_observables}

The analyses procedure discussed in this paper focuses on the description of the charged averaged multiplicity.
However, as discussed in Sec.~\ref{sec:data_review_lowW}, different observables are linked with the shower particle content description.
In this section, the effect of the \emph{2021} global tune on different hadronization observables is discussed. 

A wider comparison against all available hadronization observables for the G00\_00a\_00\_00a AGKY predictions is reported in~\cite{Yang_2009}.
Some information provided by these observables is included in the tune using priors, see Sec.~\ref{sec:Likelihood}.
The agreement of the \emph{2021} GENIE AGKY global tune with these observables is not compromised. 

\begin{figure}
    \centering
    \includegraphics[width=\textwidth]{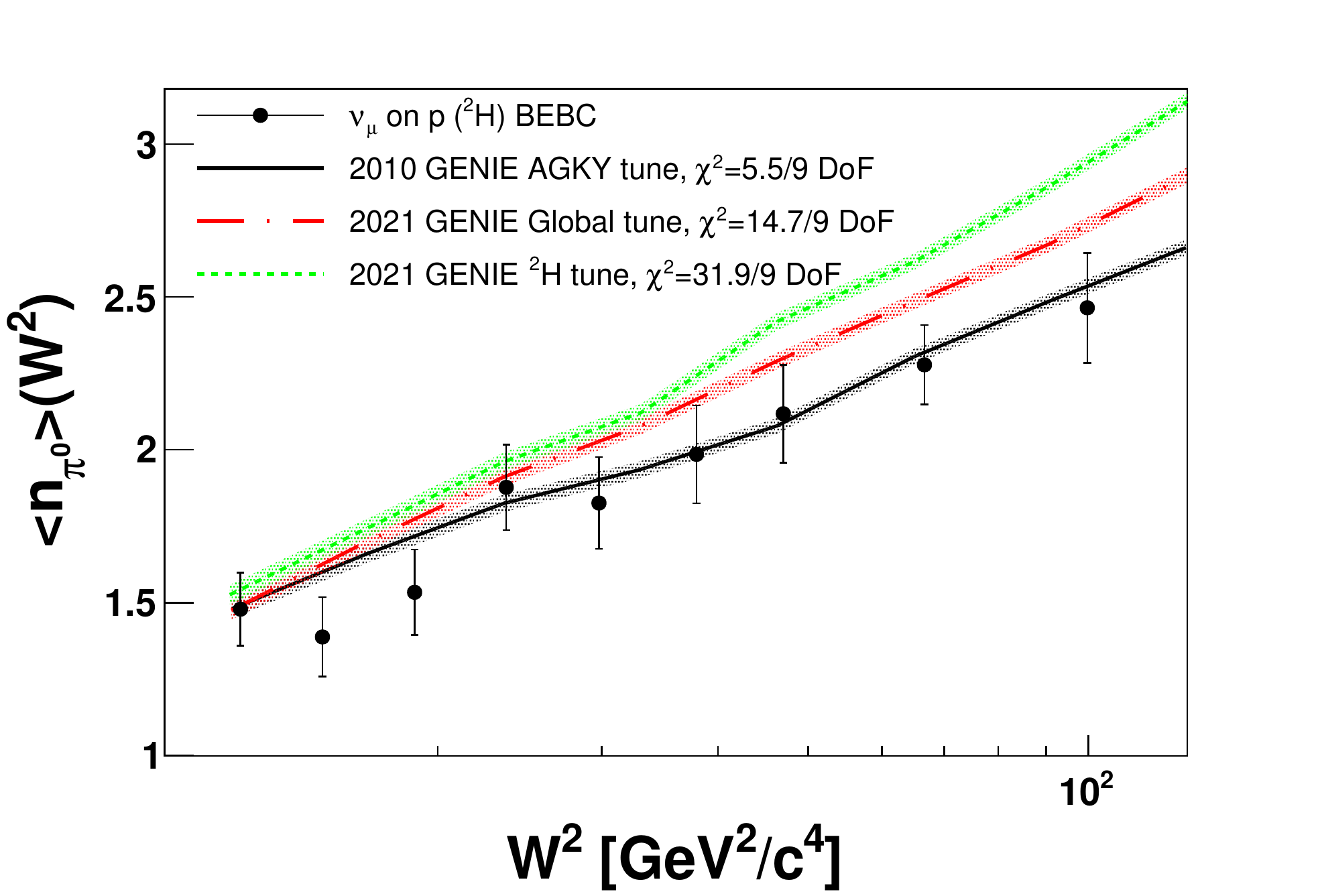}
    \caption{Comparison of the predicted $\langle n_{\pi^{0}}\rangle$ against neutrino-induced hadronization data on $\nu_\mu$ interactions on p from the \ac{BEBC} bubble chamber experiment filled with $^2$H~\cite{Barlag1982,Jongejans1989}. 
    The predictions shown correspond to the \emph{2010} GENIE AGKY (black), the \emph{2021} GENIE AGKY global (red) and the \emph{2021} GENIE AGKY $^{2}$H (green) tunes. }
    \label{fig:neutralpion_comparison}
\end{figure}

\begin{figure}
    \centering
    \begin{subfigure}{\textwidth}
        \includegraphics[width=\textwidth]{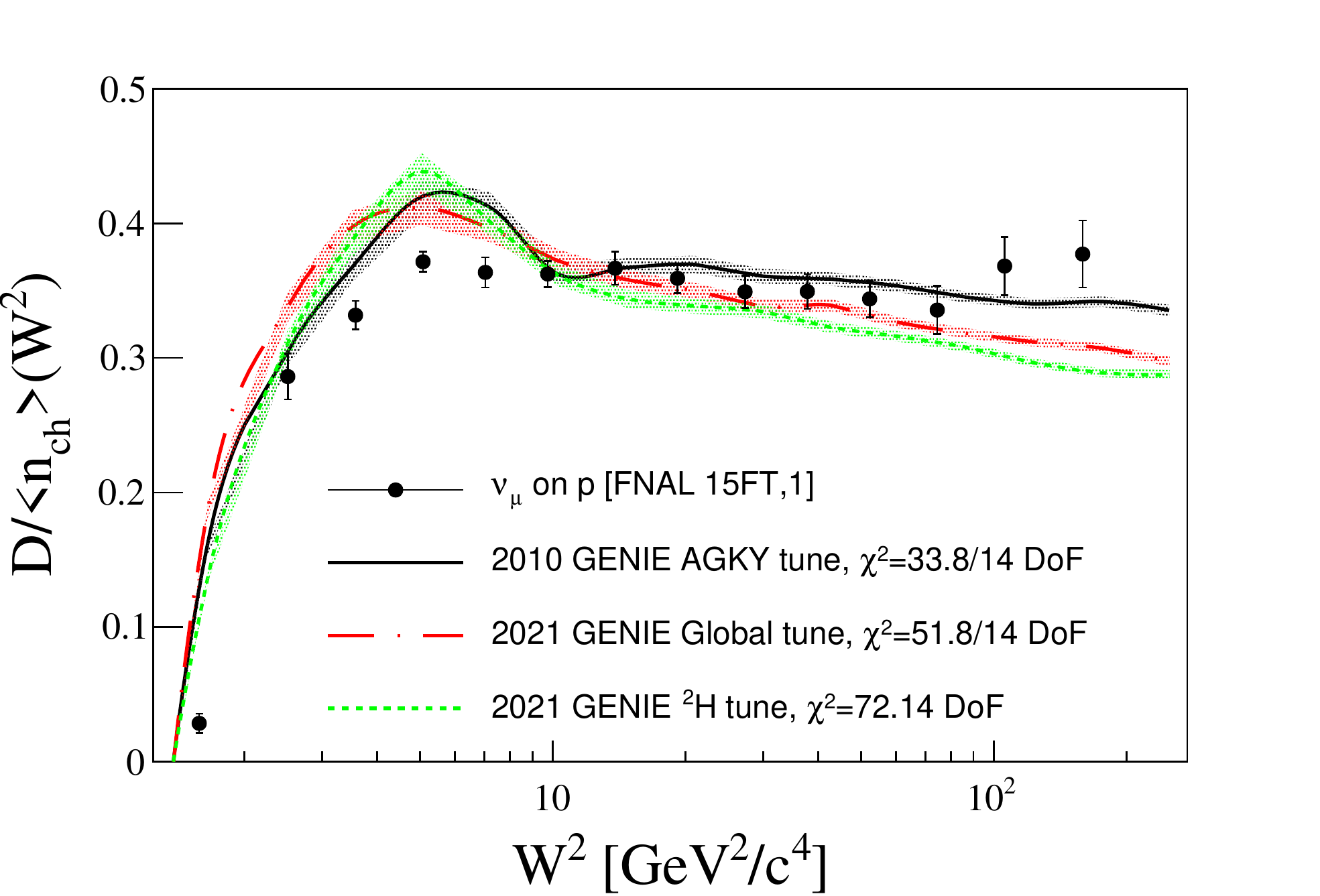}
        \caption{Comparison against $\nu_\mu$ on p data. }
    \end{subfigure}
    
    \begin{subfigure}{\textwidth}
        \includegraphics[width=\textwidth]{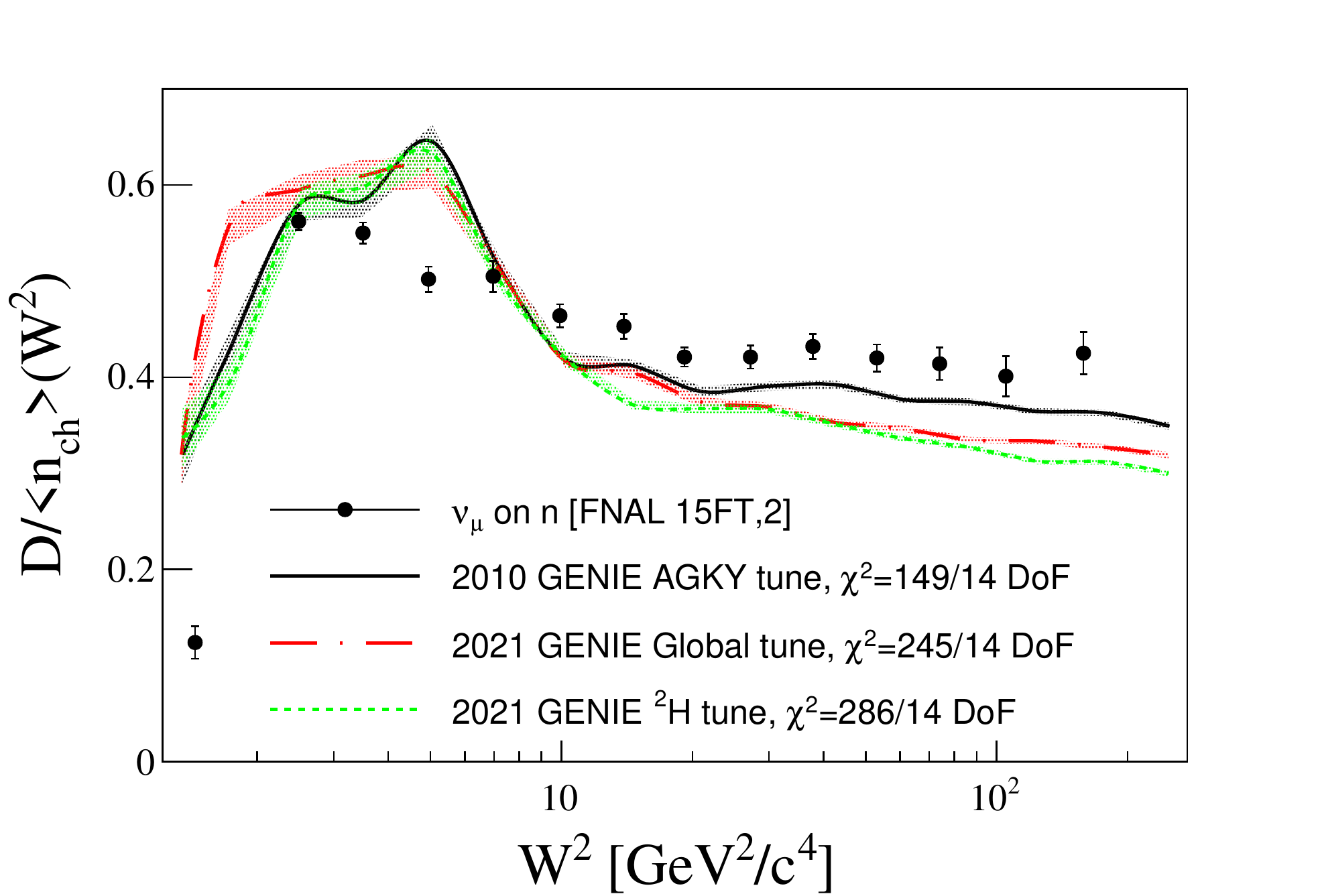}
        \caption{Comparison against $\nu_\mu$ on n data.}
    \end{subfigure}
    \caption{Comparison of the predicted $D/\langle n_{\text{ch}}\rangle$ against neutrino-induced hadrionzation data on $\nu_\mu$ interactions on p (a) and n (b) from the FNAL 15 ft bubble chamber experiment filled with $^2$H~\cite{PhysRevD.27.47}. 
    The predictions shown correspond to the \emph{2010} GENIE AGKY (black), the \emph{2021} GENIE AGKY global (red) and the \emph{2021} GENIE AGKY $^{2}$H (green) tunes.  }
    \label{fig:dispersion_comparsion}
\end{figure}

There are, however, other observables that show tensions with the averaged charged multiplicity data.
The neutral pion averaged multiplicity is related with the charged hadron multiplicity via Eq.~\ref{eq:total_multiplicity}: an increase on the charged averaged multiplicity is equivalent to a higher neutral pion averaged multiplicity.
This result is incompatible with the data, as demonstrated in Fig.~\ref{fig:neutralpion_comparison}.
Another example is the dispersion observable, defined as $D = \sqrt{ \langle n^2 \rangle - \langle n \rangle ^2 } $.
The comparison of data on the ratio of $D/\langle n_{\text{ch}} \rangle$ vs $W^2$ for the different tunes is shown in Fig.~\ref{fig:dispersion_comparsion}.
In this case, the disagreement also increases with the invariant mass.

The tension between charged averaged multiplicity with $\langle n_{\pi^0} \rangle$ and dispersion data was already observed when using the HERMES parameterisation described in Sec.~\ref{subsec:HERMESTUNE}.
The origin of these tensions is beyond the scope of this paper as we aim for a better description of the charged averaged multiplicity data only.
The further understanding of the connection between the different observables would require to repeat the analyses procedure of this paper including other hadronization related observables.
Yet, it is important to understand how the \emph{2021} GENIE AGKY global tune impacts other hadronization related observables.

\subsection{\emph{2021} GENIE AGKY global tune impact at the SIS region}
\begin{figure}
    \centering
    \begin{subfigure}{\textwidth}
        \includegraphics[width=0.9\textwidth]{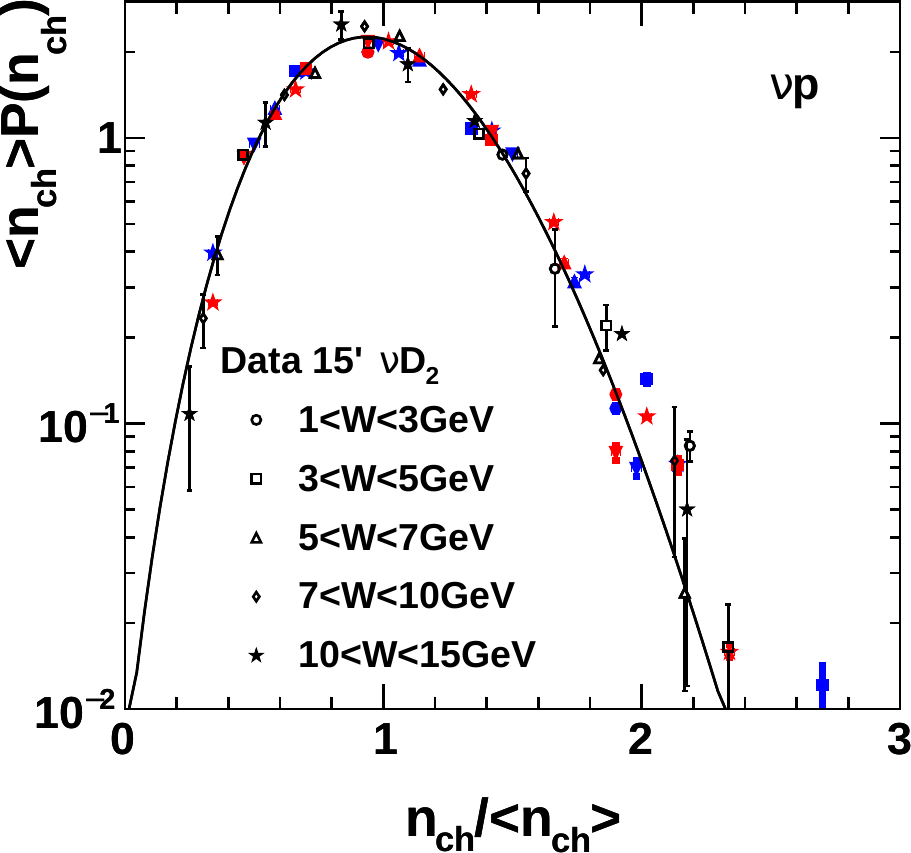}
        \caption{$\nu_\mu p$ KNO scaling distribution.}
    \end{subfigure}
    
    \begin{subfigure}{\textwidth}
        \includegraphics[width=0.9\textwidth]{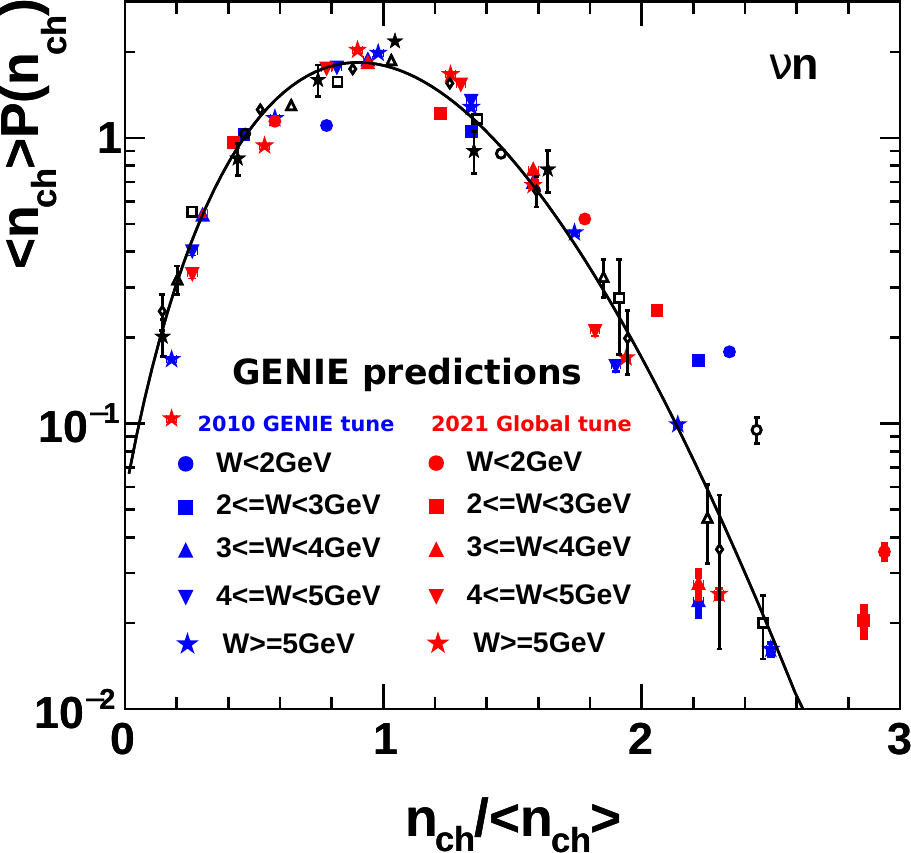}
        \caption{$\nu_\mu n$ KNO scaling distribution.}
    \end{subfigure}
    \caption{Comparison of the KNO scaling distributions for neutrino interactions on deuterium against the predictions for \emph{2010} GENIE tune (blue) and the \emph{2021} GENIE global tune (red). 
    The solid line is the best fit result of the Levi function to FNAL 15 ft bubble chamber data~\cite{PhysRevD.27.47}. 
    The $W$ range used for each data and predicted point is specified in the legend of Fig.~\ref{fig:KNO_comparison} (a) and (b).}
    \label{fig:KNO_comparison}
\end{figure}

Other non-hadronization observables can be affected by this tune. 
The main impact is on the description of the \ac{SIS} region in GENIE, since it is linked with final state multiplicities~\cite{mypaper_1}. 
In GENIE, the \ac{SIS} is modeled applying scaling factors to the \ac{DIS} cross section.
These factors depend on the multiplicity of the process. 
Hence, variations on the final state multiplicity probabilities (Eq.~\ref{eq:Phadro}) change the scaling applied to the \ac{DIS} cross section, affecting the \ac{DIS} contribution to the \ac{SIS}.
The $P^{\text{had}}_{n}$ probability distributions for the \emph{2010} GENIE AGKY tune and for the AGKY global tune are shown in Fig.~\ref{fig:KNO_comparison}.

\begin{figure}
    \centering
    \begin{subfigure}{\textwidth}
        \includegraphics[width=\textwidth]{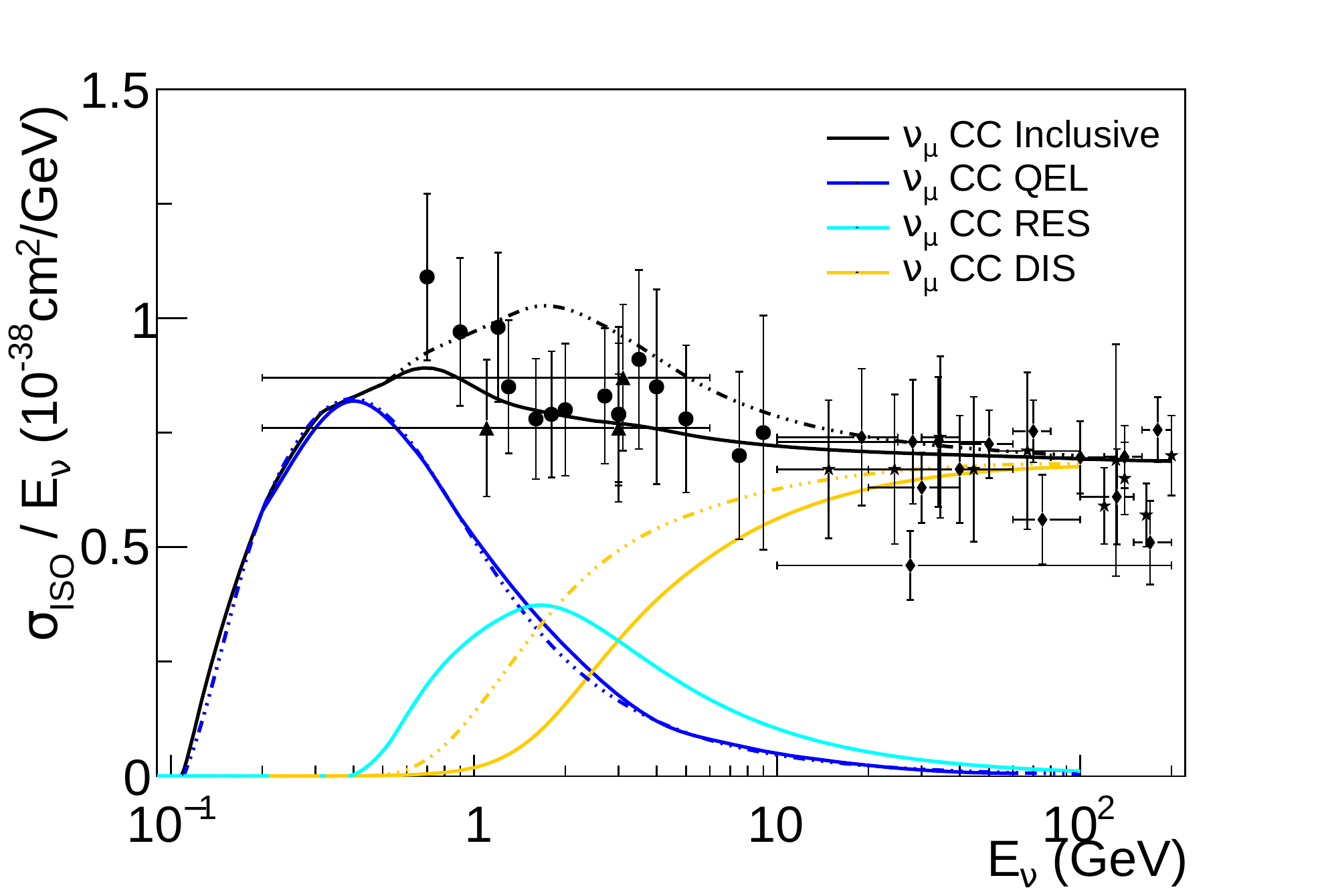}
        \caption{$\nu_\mu$ CC inclusive cross section.}
    \end{subfigure}
    
    \begin{subfigure}{\textwidth}
        \includegraphics[width=\textwidth]{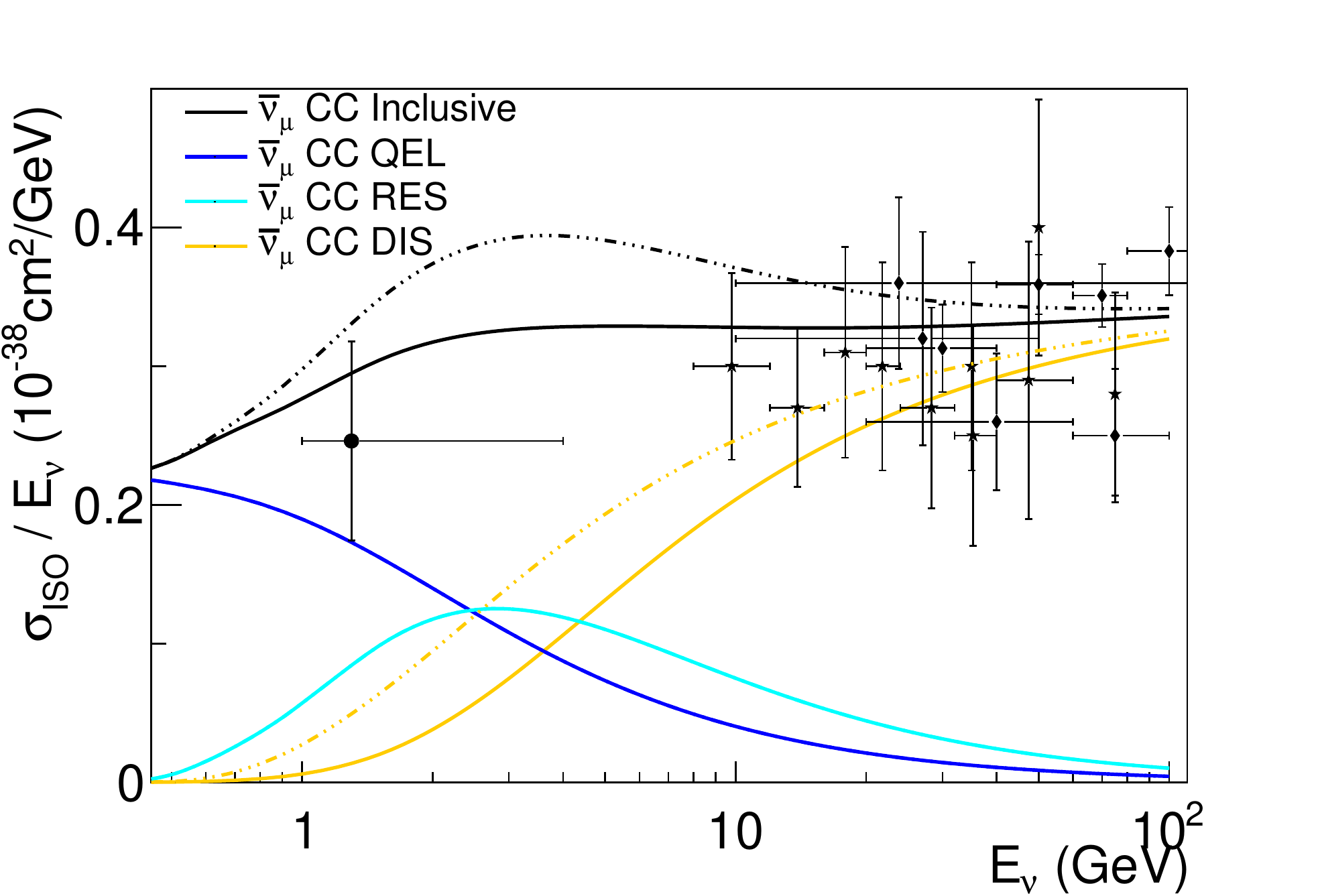}
        \caption{$\bar{\nu}_\mu$ CC inclusive cross section.}
    \end{subfigure}
     \caption{Comparison of the $\nu_\mu$ and $\bar{\nu}_\mu$ CC inclusive cross section on free nucleon for the \emph{2010} GENIE AGKY tune~\cite{mypaper_1} (continuous lines) and the \emph{2021} GENIE global tune (dashed lines) against hydrogen and deuterium data from  ANL 12 ft ($\bigtriangleup$), BNL 7 ft ($\bullet$), \ac{BEBC} ($\diamond$) and FNAL ($\star$).
     The breakdown of the CC QEL, CC RES and CC DIS contributions is shown for before and after the \emph{2021} GENIE AGKY global tune. }
    \label{fig:freenucleon_impact}
\end{figure}

\begin{figure}
    \centering
    \begin{subfigure}{\textwidth}
        \includegraphics[width=\textwidth]{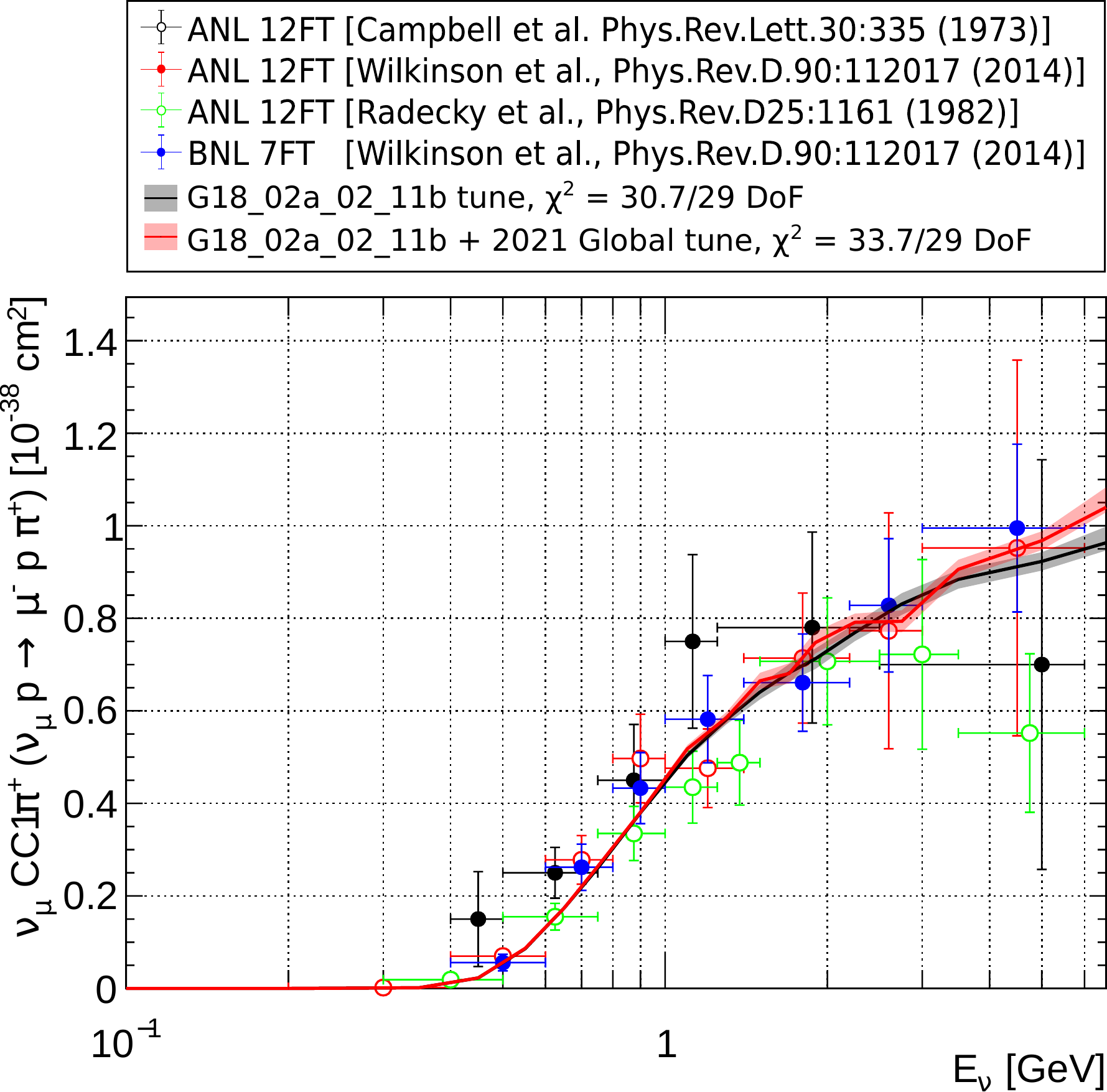}
        \caption{One pion production.}
    \end{subfigure}
    
    \begin{subfigure}{\textwidth}
        \includegraphics[width=\textwidth]{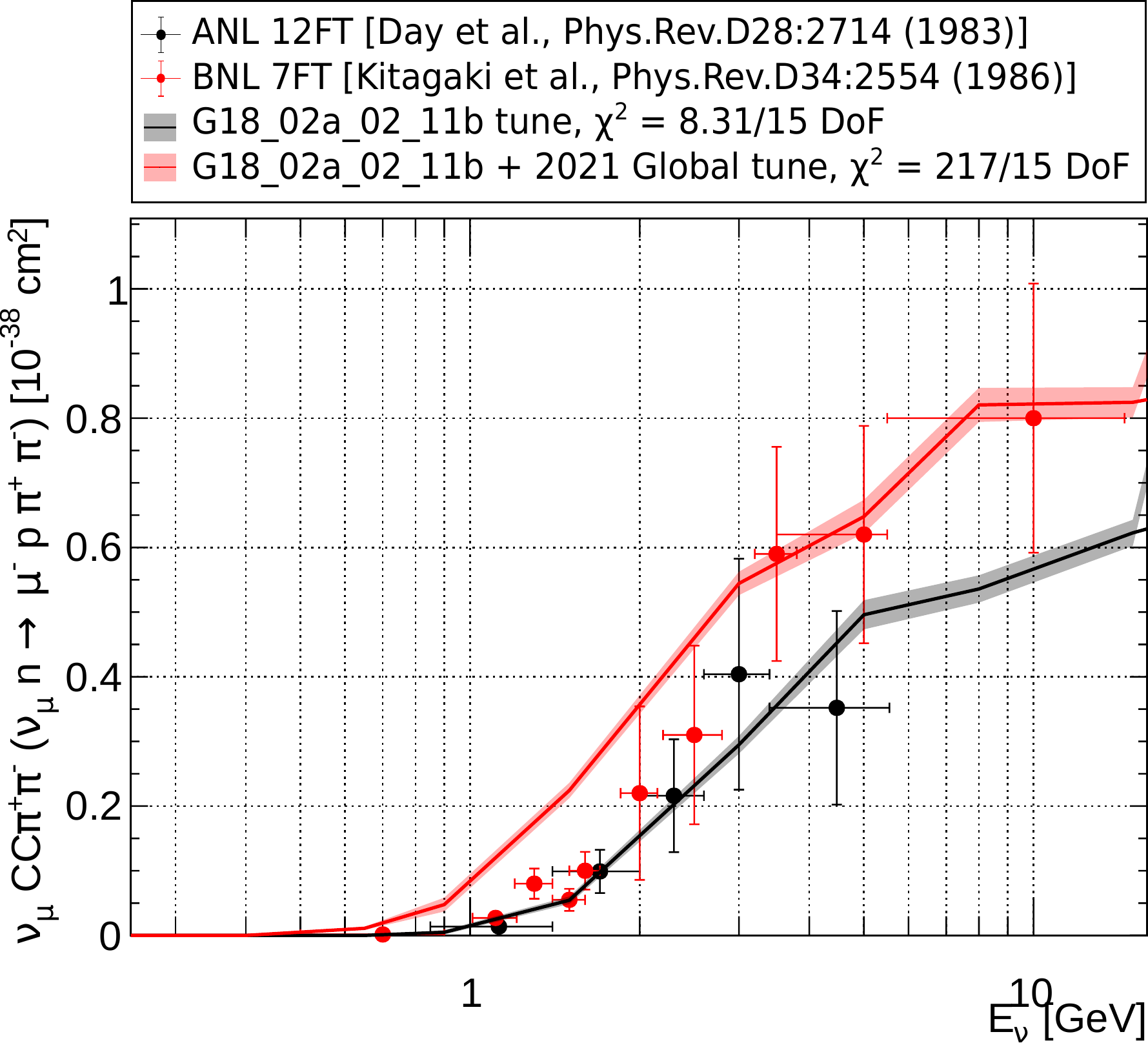}
        \caption{Two pion production.}
    \end{subfigure}
    \caption{Comparison of $\nu_\mu$ CC exclusive cross section data on free nucleon for the \emph{2010} GENIE AGKY tune~\cite{mypaper_1} (black) and the\emph{2021} GENIE global tune (red) against ANL 12 ft and BNL 12 ft data.
    }
    \label{fig:freenucleon_exclusive}
\end{figure}

The impact of the AGKY tune on CC inclusive cross sections is summarised in Fig.~\ref{fig:freenucleon_impact}.
When applying the AGKY global tune to the SIS region, an increase of CC inclusive cross section is observed, for both $\nu_\mu$ and $\bar{\nu}_\mu$.
The exclusive cross sections for different pion multiplicities show that the AGKY tune enhances the 2$\pi$ production whilst the $1\pi$ production barely changes, see Fig.~\ref{fig:freenucleon_exclusive}.
As a consequence, the agreement with inclusive and $\nu_\mu$ CC $\pi^+\pi^-$ data is lost.

Both, the bare nucleon tune \cite{mypaper_1} and the \emph{2021} GENIE global tune, show a preference to increase the two pion production, suggesting that a joint tune could preserve the agreement with inclusive and exclusive data at low-$W$.
This was neglected in previous analyses to minimise the tune's complexity, but this analysis clearly suggests otherwise.
The high-$W$ AGKY parameters do not need anymore refinements. 
On the contrary, the low-$W$ parameters requires a joint tune in order to have a satisfactory result that can be used to extract data driven parameter uncertainties. 

\section{Conclusions}

In this paper, we present the first GENIE tune of the AGKY model~\cite{Andreopoulos:2015wxa,Yang_2009} which was possible thanks to the Professor framework~\cite{Professor}. 
The analysis goal was to improve the GENIE agreement with neutrino charged averaged multiplicity data and to provide with the first data driven constraints on hadronization parameters.
Specifically, we constrained parameters of both low-$W$ empirical model and PYTHIA using data from the \ac{BEBC} and FNAL 15 ft bubble chamber experiments filled with hydrogen and deuterium.

Tensions between hydrogen and deuterium data were observed and two separate tunes were performed:
a global and a deuterium only tune. 
In particular, the \emph{2021} AGKY global tune prediction underpredicts the deuterium data at the PYTHIA region whereas the deuterium only tune overpredicts the hydrogen data.

Further investigations on hadronization samples are needed in order to clarify the origin of this discrepancy. 
A possible solution could come from more recent neutrino experiments that released data on neutrino-induced hadronization. 
This is the case of NOMAD~\cite{Vannucci:2014wna,NOMADData} for $\nu_\mu$ on mainly carbon target, CHORUS for $\nu_\mu$ and $\bar{\nu}_\mu$ on Fuji ET-7B emulsion~\cite{RADICIONI200095,KayisTopaksu:2007pe}, OPERA for $\nu_\mu$ on lead~\cite{giacomelli2006opera,collaboration2017study} and MicroBooNE for $\nu_\mu$ on argon~\cite{MicroBooNEData}. 
But of course these samples include nuclear effects and therefore are not in the scope of this work.

Despite the tensions, the global tune shows a better agreement with the charged averaged multiplicity data and provides the first data driven analysis of this kind using neutrino interactions.
This statistical analysis can be a useful input for proper systematic studies of modern neutrino experiments.
The main effect of the tune is the increase of the averaged charged multiplicity for $W^2>10$ GeV$^2/c^4$, modelled with PYTHIA. 
The low-$W$ region is also affected but constraints due to energy, momentum, charge, baryon number and strangeness conservation laws reduce the available phase space and the effect of the tuning procedure. 

The effect of the \emph{2021} GENIE AGKY global tune at the \ac{SIS} region is an increase on the two pion production cross section, which affects the current agreement with CC inclusive data~\cite{mypaper_1}.
Therefore, we conclude that this tune is more appropriate at higher energies where the contribution of the \ac{SIS} region is not relevant.
The information on the systematic uncertainties coming from the low-$W$ AGKY parameters is still valuable for neutrino experiments interested in the $W<2$ GeV$/c^2$ region. A joint tune of the \ac{SIS} region and hadronization datasets would address this disagreement.

\section{Acknowledgements}

We would like to thank Andy Buckley (University of Glasgow, UK) and Holger Schultz (Institute of Particle Physics Phenomenology, University of Durham, UK)
for their support interfacing the Professor tool with the software products that underpin the GENIE global analyses. 
We would like to thank the CC-IN2P3 Computing Center, as well as the Particle Physics Department at Rutherford Appleton Laboratory for providing computing resources and for their support. 
This work, as well as the ongoing development of several other GENIE physics tunes was enabled through a PhD studentship funded by STFC through LIV.DAT, 
the Liverpool Big Data Science Centre for Doctoral Training (project reference: 2021488).
The initial conceptual and prototyping work for the development of the GENIE / Professor interfaces, as well as for the development of the GENIE global analyses framework that, currently, underpins several analyses, was supported in part through an Associateship Award by the Institute of Particle Physics Phenomenology, University of Durham.

This document was prepared by the GENIE collaboration using the resources of the Fermi National Accelerator Laboratory (Fermilab), a U.S.\ Department of Energy, Office of Science, HEP User Facility. Fermilab is managed by Fermi Research Alliance, LLC (FRA), acting under Contract No. DE-AC02-07CH11359.


\begin{table*}    
    \centering
    \resizebox{\textwidth}{!}{%
    \begin{tabular}{@{\extracolsep\fill}c c c c c c c c c c c c c c} \noalign{\medskip}\hline\hline\noalign{\smallskip}
        & $\alpha_{\nu p}$ & $\alpha_{\nu n}$ & $\alpha_{\bar{\nu} p}$ & $\alpha_{\bar{\nu} n}$ 
        & $\beta{\nu p}$ & $\beta{\nu n}$ & $\beta{\bar{\nu} p}$ & $\beta{\bar{\nu} n}$ 
        & $P_{s\bar{s}}$ & $\langle p_\bot^2 \rangle$ & $E_{\text{CutOff}}$ & Lund $a$ & Lund $b$  \\ \hline\hline\noalign{\smallskip}
        $\alpha_{\nu p}$ &  1.8E-1 & -2.2E-4 & 1.5E-3 & -2.8E-2 & -1.1E-1 & -1.9E-2 & 9.2E-4 & -1.5E-2 & -1.0E-3 & -5.8E-4 & 2.7E-3 & 5.6E-2 & 3.4E-2 \\
        $\alpha_{\nu n}$ & -2.2E-4 & 5.4E-3 & -3.0E-5 & -1.3E-3 & -1.4E-3 & -2.9E-3 & 1.1E-3 & 1.8E-3 & 1.4E-4 & 0.0& -3.0E-5 & 7.2E-4 & 2.4E-3 \\
        $\alpha_{\bar{\nu} p}$ & 1.5E-3 & -3.0E-5 & 6.2E-3 & 5.7E-4 & 2.2E-3 & 7.7E-4 & -2.9E-3 & 1.4E-3 & -8.0E-5 & 1.8E-4 & -6.0E-5 & -3.0E-5 & -3.7E-3 \\
        $\alpha_{\bar{\nu} n}$ & -2.8E-2 & -1.3E-3 & 5.7E-4 & 1.0E-1 & 4.3E-2 & 2.9E-3 & 9.5E-3 & -6.1E-2 & 2.4E-3 & 7.1E-3 & -2.6E-3 & 1.6E-3 & -5.9E-2 \\
        $\beta{\nu p}$ & -1.1E-1 & -1.4E-3 & 2.2E-3 & 4.3E-2 & 1.2E-1 & 3.3E-2 & 4.7E-3 & -6.0E-5 & 7.7E-4 & 1.6E-3 & -3.3E-3 & -3.8E-2 & -4.4E-2 \\
        $\beta{\nu n}$ & -1.9E-2 & -2.9E-3 & 7.7E-4 & 2.9E-3 & 3.3E-2 & 3.2E-2 & 1.1E-2 & 1.1E-2 & 4.9E-4 & -4.2E-3 & -1.1E-3 & -1.8E-3 & -7.5E-3 \\
        $\beta{\bar{\nu} p}$ & 9.2E-4 & 1.1E-3 & -2.9E-3 & 9.5E-3 & 4.7E-3 & 1.1E-2 & 1.3E-2 & -3.0E-3 & -1.6E-3 & -1.3E-3 & -9.9E-4 & 7.2E-3 & 7.7E-4 \\
        $\beta{\bar{\nu} n}$ & -1.5E-2 & 1.8E-3 & 1.4E-3 & -6.1E-2 & -6.0E-5 & 1.1E-2 & -3.0E-3 & 5.2E-2 & -2.4E-3 & -6.1E-3 & 6.3E-4 & -8.5E-3 & 2.9E-2 \\
        $P_{s\bar{s}}$ & -1.0E-3 & 1.4E-4 & -8.0E-5 & 2.4E-3 & 7.7E-4 & 4.9E-4 & -1.6E-3 & -2.4E-3 & 2.3E-3 & -4.8E-4 & -7.0E-5 & 1.2E-4 & -9.6E-4 \\
        $\langle p_\bot^2 \rangle$ & -5.8E-4 & 0.0 & 1.8E-4 & 7.1E-3 & 1.6E-3 & -4.2E-3 & -1.3E-3 & -6.1E-3 & -4.8E-4 & 1.8E-3 & 2.7E-4 & -1.4E-3 & -4.2E-3 \\
        $E_{\text{CutOff}}$  & 2.7E-3 & -3.0E-5 & -6.0E-5 & -2.6E-3 & -3.3E-3 & -1.1E-3 & -9.9E-4 & 6.3E-4 & -7.0E-5 & 2.7E-4 & 2.3E-3 & -7.0E-4 & 9.2E-4 \\
        Lund $a$ & 5.6E-2 & 7.2E-4 & -3.0E-5 & 1.6E-3 & -3.8E-2 & -1.8E-3 & 7.2E-3 & -8.5E-3 & 1.2E-4 & -1.4E-3 & -7.0E-4 & 2.5E-2 & 1.0E-2 \\
        Lund $b$ & 3.4E-2 & 2.4E-3 & -3.7E-3 & -5.9E-2 & -4.4E-2 & -7.5E-3 & 7.7E-4 & 2.9E-2 & -9.6E-4 & -4.2E-3 & 9.2E-4 & 1.0E-2 & 5.0E-2 \\ \noalign{\smallskip}
        \hline\hline
    \end{tabular}
    }
    \caption{Parameter covariance matrix extracted from the \emph{2021} GENIE AGKY global tune.}
    \label{tab:covGlobal}
\end{table*}

\begin{table*}    
    \centering
    \resizebox{\textwidth}{!}{%
    \begin{tabular}{@{\extracolsep\fill}c c c c c c c c c c c c c c} \noalign{\medskip}\hline\hline\noalign{\smallskip}
        & $\alpha_{\nu p}$ & $\alpha_{\nu n}$ & $\alpha_{\bar{\nu} p}$ & $\alpha_{\bar{\nu} n}$ 
        & $\beta{\nu p}$ & $\beta{\nu n}$ & $\beta{\bar{\nu} p}$ & $\beta{\bar{\nu} n}$ 
        & $P_{s\bar{s}}$ & $\langle p_\bot^2 \rangle$ & $E_{\text{CutOff}}$ & Lund $a$ & Lund $b$  \\ \hline\hline\noalign{\smallskip}
        $\alpha_{\nu p}$ & 7.7E-2 & 2.5E-2 & 8.6E-3 & -6.3E-3 & -3.6E-2 & -1.6E-2 & -8.7E-3 & 1.0E-2 & -1.1E-3 & -2.9E-3 & -1.7E-3 & 3.8E-4 & 8.8E-3 \\
        $\alpha_{\nu n}$ & 2.5E-2 & 1.5E-2 & 5.2E-3 & -3.1E-3 & -1.3E-2 & -5.8E-3 & -4.8E-3 & 4.4E-3 & 1.4E-3 & -1.6E-3 & -7.8E-4 & 7.9E-4 & 7.2E-3 \\
        $\alpha_{\bar{\nu} p}$ & 8.6E-3 & 5.2E-3 & 2.4E-2 & -6.9E-3 & 3.0E-4 & -8.1E-3 & -1.8E-2 & -1.9E-3 & 3.4E-3 & 1.5E-3 & -1.2E-4 & -2.9E-3 & -1.2E-3 \\
        $\alpha_{\bar{\nu} n}$ & -6.3E-3 & -3.1E-3 & -6.9E-3 & 8.5E-3 & 1.1E-3 & 5.5E-3 & 4.1E-3 & -3.8E-3 & -7.1E-4 & 3.3E-4 & -5.0E-4 & -7.1E-4 & 2.8E-3 \\
        $\beta{\nu p}$ & -3.6E-2 & -1.3E-2 & 3.0E-4 & 1.1E-3 & 2.2E-2 & 7.4E-3 & 7.2E-4 & -4.5E-3 & -2.0E-5 & 1.0E-3 & 1.9E-3 & -4.6E-3 & -6.2E-3 \\
        $\beta{\nu n}$ & -1.6E-2 & -5.8E-3 & -8.1E-3 & 5.5E-3 & 7.4E-3 & 1.2E-2 & 7.2E-3 & -2.1E-3 & -1.1E-3 & -6.0E-5 & -1.5E-3 & 3.3E-3 & -1.8E-3 \\
        $\beta{\bar{\nu} p}$ & -8.7E-3 & -4.8E-3 & -1.8E-2 & 4.1E-3 & 7.2E-4 & 7.2E-3 & 1.6E-2 & 1.7E-3 & -2.8E-3 & -8.0E-4 & -4.2E-4 & 2.6E-3 & -4.9E-4 \\
        $\beta{\bar{\nu} n}$ & 1.0E-2 & 4.4E-3 & -1.9E-3 & -3.8E-3 & -4.5E-3 & -2.1E-3 & 1.7E-3 & 7.1E-3 & -4.3E-4 & -6.6E-4 & 1.8E-4 & 4.4E-3 & -6.1E-4 \\
        $P_{s\bar{s}}$ & -1.1E-3 & 1.4E-3 & 3.4E-3 & -7.1E-4 & -2.0E-5 & -1.1E-3 & -2.8E-3 & -4.3E-4 & 1.4E-3 & 1.6E-4 & -1.9E-4 & 6.3E-4 & 5.3E-4 \\
        $\langle p_\bot^2 \rangle$ & -2.9E-3 & -1.6E-3 & 1.5E-3 & 3.3E-4 & 1.0E-3 & -6.0E-5 & -8.0E-4 & -6.6E-4 & 1.6E-4 & 9.9E-4 & -2.8E-4 & 1.3E-3 & -1.3E-3 \\
        $E_{\text{CutOff}}$  & -1.7E-3 & -7.8E-4 & -1.2E-4 & -5.0E-4 & 1.9E-3 & -1.5E-3 & -4.2E-4 & 1.8E-4 & -1.9E-4 & -2.8E-4 & 1.5E-3 & -2.9E-3 & -2.0E-5 \\
        Lund $a$ &3.8E-4 & 7.9E-4 & -2.9E-3 & -7.1E-4 & -4.6E-3 & 3.3E-3 & 2.6E-3 & 4.4E-3 & 6.3E-4 & 1.3E-3 & -2.9E-3 & 1.5E-2 & -2.5E-3 \\
        Lund $b$ & 8.8E-3 & 7.2E-3 & -1.2E-3 & 2.8E-3 & -6.2E-3 & -1.8E-3 & -4.9E-4 & -6.1E-4 & 5.3E-4 & -1.3E-3 & -2.0E-5 & -2.5E-3 & 8.5E-3 \\ \noalign{\smallskip}
        \hline\hline
    \end{tabular}
    }
    \caption{Parameter covariance matrix extracted from the \emph{2021} GENIE AGKY $^{2}$H tune.}
    \label{tab:cov2H}
\end{table*}

\appendix

\section{Dataset compatibility study}

\begin{figure*}
    \centering
    \includegraphics[width=0.95\textwidth]{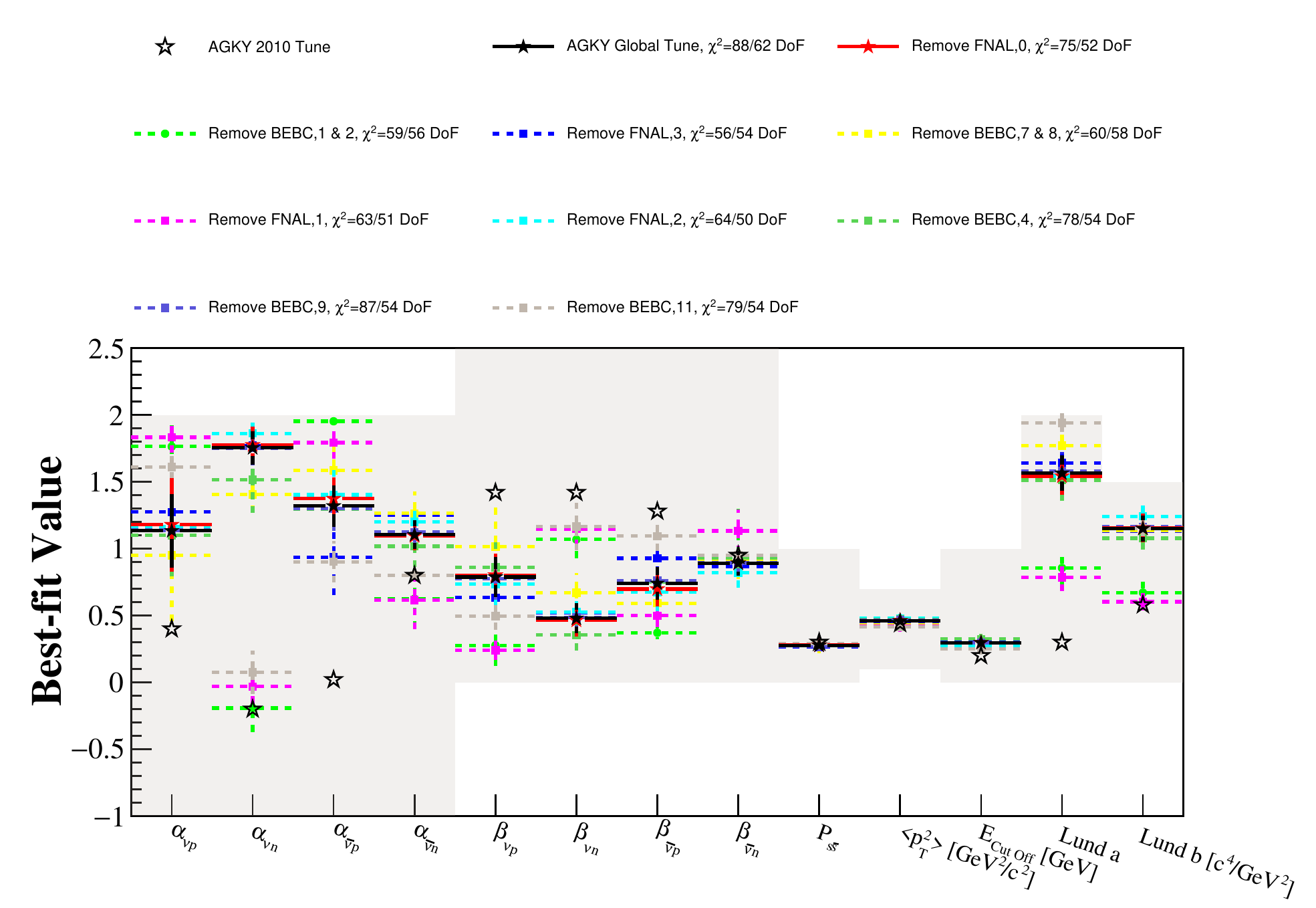}
    \caption{Summary of best-fit values for a series of partial tunes obtained by using all the data from the AGKY 2021 global tune leaving one dataset out at a time. The result of the fit when removing a H dataset is shown in circles, and squares when removing a $^{2}$H dataset. The $\chi^2$ for each tune is listed in the legend. The grey area represents the available range for each parameter. The AGKY \emph{2021} global tune best fit result is shown in filled black stars. The previous GENIE parameters are shown in empty stars (AGKY 2010 tune). 
    The plotted errors are the square roots of the covariance diagonal elements. 
}
    \label{fig:tension_analysis}
\end{figure*}

The tensions highlighted in the paper were investigated to understand if their source could be caused by a specific dataset or analysis procedure. 
In order to do that, we performed a series of tunes using all the data used in the AGKY \emph{2021} global fit leaving one dataset out at a time. 
For each fit, we plotted the parameter best fit value, see Fig.~\ref{fig:tension_analysis}.
Most of the partial fits results are compatible with the AGKY \emph{2021} global tune predicted values.
There are only two datasets in disagreement with the rest: BEBC,1 and 2 (hydrogen dataset) and FNAL,1 (deuterium dataset) \footnote{Please note that BEBC 1 is only one point, which is why we count it together with BEBC 2.}. 
Those datasets use different targets, they come from different experiments, and they were analyzed in different years. 
Hence, the cause of the tension cannot be due to a specific dataset or experiment.

\section*{List of Acronyms}
\begin{acronym}[ICANN]
    \acro  {MC}    [MC]    {Monte Carlo}
    \acro  {CC}    [CC] {Charged-Current}
    \acro  {DIS}   [DIS] {Deep Inelastic Scattering}
    \acro  {SIS}   [SIS] {Shallow Inelastic Scattering}
    \acro  {FSI}   [FSI] {Final State Interactions}
    \acro  {KNO}   [KNO]   {Koba-Nielsen-Olesen scaling law}
    \acro  {AGKY}  [AGKY]  {Andreopoulos-Gallagher-Kehayias-Yang}
    \acro  {BEBC}  [BEBC]  {Big European Bubble Chamber}
    \acro  {CMC}   [CMC]   {Comprehensive Model Configurations}
    \acro  {EMI}   [EMI]   {External Muon Identifier}
    \acro  {LPS}   [LPS]   {Longitudinal Phase Space model}
\end{acronym}


\bibliography{main.bib}

\end{document}